\documentclass[aps,prd,superscriptaddress,twocolumn]{revtex4-1}

\usepackage{wallpaper}
\usepackage{bm}
\usepackage{amsmath}
\usepackage[colorlinks,linkcolor=blue,anchorcolor=blue,citecolor=red]{hyperref}
\usepackage{ulem}
\usepackage{multirow}

\begin{document}

\title{Effects of inner crusts on $g$-mode oscillations in neutron stars}
\author{Hao~Sun}
\affiliation{Center for Gravitation and Cosmology, College of Physical Science and Technology, Yangzhou University, Yangzhou 225009, China}

\author{Jia-Xing~Niu}
\affiliation{Center for Gravitation and Cosmology, College of Physical Science and Technology, Yangzhou University, Yangzhou 225009, China}

\author{Hong-Bo~Li}
\affiliation{Kavli Institute for Astronomy and Astrophysics, Peking University, Beijing 100871, China}

\author{Cheng-Jun Xia}
\email{cjxia@yzu.edu.cn}
\affiliation{Center for Gravitation and Cosmology, College of Physical Science and Technology, Yangzhou University, Yangzhou 225009, China}

\author{Enping Zhou}
\affiliation{Department of Astronomy, School of Physics, Huazhong University of Science and Technology, Wuhan, 430074, China}

\author{Yiqiu Ma}
\affiliation{Center for Gravitational Experiments, Hubei Key Laboratory of Gravitation and Quantum Physics, School of Physics, Huazhong University of Science and Technology, Wuhan, 430074, China}
\affiliation{Department of Astronomy, School of Physics, Huazhong University of Science and Technology, Wuhan, 430074,  China}

\author{Ying-Xun Zhang}
\affiliation{China Institute of Atomic Energy, Beijing 102413, People's Republic of China}
\affiliation{Guangxi Key Laboratory Breeding Base of Nuclear Physics and Technology, Guilin 541004, China}

\date{\today}

\begin{abstract}
In this work we investigate the influence of neutron stars' crusts on the non-radial $g$-mode oscillations and examine their correlations with nuclear matter properties fixed by adopting 10 different relativistic density functionals. At subsaturation densities, neutron star matter takes non-uniform structures and form the crusts. We find that the Brunt-V\"{a}is\"{a}l\"{a} (BV) frequency increases significantly at densities slightly above the neutron drip density (i.e., neutron stars' inner crusts), which leads to crust $g$-mode oscillations with their frequencies insensitive to the adopted density functional. At larger densities, BV frequency increases as well due to the core-crust transitions and emergence of muons, which lead to core $g$-mode oscillations. It is found that the obtained core $g$-mode frequencies generally increase with the slope of nuclear symmetry energy $L$, which eventually intersect with that of the crust $g$ modes adopting large enough $L$. This leads to the avoid-crossing phenomenon for the global $g$ modes that encompass contributions from both the crust and core.
{For neutron stars with fixed masses, the frequency of the global $g_1$ mode generally increases linearly  with $L$, where the Pearson
correlation coefficient ranges from 0.7 to 0.9. The corresponding linear function are then fixed by fitting to the $g_1$ mode frequencies predicted by the 10  relativistic density functionals, which enables the possible measurements of $L$ based on gravitational wave observations.}
In our future study, the effects of the discontinuities in density or shear modulus should be considered, while the temperature, rotation, magnetic field, and superfluid neutron gas in neutron stars could also play important roles.
\end{abstract}

\maketitle

\section{\label{sec:intro}Introduction}
Since the first discovery of neutron stars~\cite{Hewish1968_Nature217-709}, people have been very curious about their internal  compositions and structures. However, due to the difficulties in quantum chromodynamics, the compositions and properties of neutron star matter are still uncertain~\cite{Dutra2012_PRC85-035201, Dutra2014_PRC90-055203, Baym2018_RPP81-056902, Xia2020_PRD102-023031, Li2020_JHEA28-19, Xia2024_PRD110-114009}.  The observational masses and radii of neutron stars could constrain the equation of state (EOS) of neutron star matter to a certain extent~\cite{Antoniadis2013_Science340-1233232, LVC2018_PRL121-161101, Riley2019_ApJL887-L21, Riley2021_ApJL918-L27, Miller2019_ApJL887-L24, Miller2021_ApJL918-L28, Choudhury2024_ApJ971-L20}, but it is difficult to distinguish effectively its internal composition based on these observations alone. With the advent of the multi-messenger era, further observations of neutron star oscillations based on astroseismology seem promising to unveil the internal compositions and structures of neutron stars~\cite{Sagun2020_PRD101-063025, Aerts2021_RMP93-015001, Andersson2021_Universe7-97A, Zhu2023_PRD107-83023, LU2024_SCPMA54-289501, Zhen2024_Symmetry16-2073-8994, Zhang2024_PRD109-063020, Sen2023_Galaxies11-2075-4434, Li2023_PRD108-064005, Zheng2023_PRD107-103048, Zhao2022_PRD106-123002, Constantinou2021_PRD104-123032, Kokkotas2000_AAP366-565, Li:2022qql, Li:2024hzt, Sotani2001_PRD65-024010,  Sotani2024_PRD109-023030}. By carefully analyzing the oscillation frequencies of each mode in neutron stars, it is possible to establish their relations with the thermodynamic quantities of neutron star matter, thereby restricting the EOSs and providing valuable information about the internal structures of neutron stars~\cite{Aerts2021_RMP93-015001}.

There exist various types of oscillations  in neutron stars, i.e., quasi-normal modes (QNMs)~\cite{Kokkotas1999_LRR2-1, Ho2020_PRD101-103009, Yu2017_MNTAS470-350}, where different modes reflect different physical mechanisms. In particular, QNMs can be divided into radial and non-radial oscillations,
{where non-radial oscillations are usually classified according to the source of force that restores the disturbed fluid elements to their equilibrium position. For example, the restoring force for the $p$-mode is provided by the pressure gradient,  $g$-mode by the buoyancy, $r$-mode by the Coriolis force in rotating neutron stars, and $w$-mode by the spacetime itself. For the radial oscillations, there are  fundamental ($f$) and pressure ($p$) modes, whose restoring force is attributed to the pressure gradient.}
In this work, we pay special attention to the non-radial $g$-mode oscillation, which is particularly sensitive to the composition of dense matter. The strong tidal interaction during the merger of binary neutron stars is expected to generate resonance excitations of the $g$-mode~\cite{Lai1994_MNRAS270-611-629, Xu2017_PRD96-083005}, which is expected to be observed by various gravitational wave detectors~\cite{Thorne1967_ApJ-149-591, Punturo2010_CQG27-194002, Regimbau2017_PRL118-151105, Abbott2017_CQG34-044001, Abbott2020_LRR23-3, Maggiore2020_JCAP2020-050}.

The $g$-mode of neutron stars has been discussed in detail in many literatures, including cold neutron stars~\cite{Osaki1973_Apj185-277, Finn1987_MNRAS227-265-293, Kuan2022-MNRAS513-4045}, warm neutron stars~\cite{McDermott1983_ApJ268-837, Krueger2015_PRD92-063009, Sotani2022_PRD105-023007,  Lozano2022_Galaxies10-2075-4434, Lozano2022_Galaxies10-2075-4434}, rotating neutron stars~\cite{Gaertig2009_PRD80-064026, Lovekin2008_ApJ679-2, Kojima1992_PRD46-4289}, hyperon stars~\cite{Tran2023_PRC108-015803, Jaikumar2021_PRD103-123009},  quark stars~\cite{Zhen2024_Symmetry16-2073-8994}, hybrid stars~\cite{Wei2020_ApJ904-187, Constantinou2021_PRD104-123032, Zhao2022_PRD105-103025, Zheng2023_PRD107-103048}, and inverted hybrid stars~\cite{Zhang2024_PRD109-063020}. Most of the investigations focus on the core $g$-mode oscillations in neutron stars and neglect the contributions of crusts, while the study on the crust $g$-mode oscillations is less extensive. For example, previous investigations focus on the effects of the density discontinuity in neutron star crusts~\cite{Finn1987_MNRAS227-265-293, Reisenegger1992_Apj395-240-249}, while the discussion on the $g$-mode containing the inner crust is lacking. Additionally, due to the discontinuous shear modulus at the interface between the fluid core and the solid crust, there exists a crust-core interfacial ($i$) mode~\cite{McDermott1988_ApJ325-725, Tsang2012_PRL108-011102, Pan2020_PRL125-201102, Zhu2023_PRD107-83023}, which is likely to be excited during the inspiral phase of binary neutron star mergers due to its lower frequencies and is thus more observationally feasible. A crust meltdown could also take place in inspiraling binary neutron stars due to the elastic-to-plastic transition~\cite{Tsang2012_PRL108-011102, Pan2020_PRL125-201102}, which may be observed by the ground based gravitational-wave detectors. Since the crust $g$-mode and $i$-mode take place in the vicinity of neutron stars' crusts, the two modes are expected to interact with each other, which alters their frequencies and wave functions, i.e., the avoid-crossing effects. This is expected to affect the tidal excitation of the oscillation modes in a binary neutron star system, where the orbital motion is slightly hindered and affects the waveform of the gravitational waves~\cite{Lai1994_MNRAS270-611-629, Xu2017_PRD96-083005}. In particular, if the frequencies of the oscillation modes are altered, the phase shift of gravitational waves is altered as well, which may be observed by gravitational wave detectors~\cite{Zhu2023_PRD107-83023}. It is thus interesting to examine the crust $g$-mode oscillations in neutron stars and its impacts on the $i$-mode oscillations.

In this work, as a first step, we investigate the $g$-mode oscillations in neutron stars considering the contributions from both the inner crust and core. To examine the impact of nuclear matter properties on $g$-mode oscillations, we adopt 10 unified EOSs predicted by relativistic mean field (RMF) models~\cite{Meng2016}, where the corresponding relativistic density functionals are fixed by reproducing finite nuclei properties, i.e., NL3~\cite{Lalazissis1997_PRC55-540}, PK1~\cite{Long2004_PRC69-034319}, TM1~\cite{Sugahara1994_NPA579-557}, GM1~\cite{Glendenning1991_PRL67-2414}, MTVTC~\cite{Maruyama2005_PRC72-015802}, DD-LZ1~\cite{Wei2020_CPC44-074107}, DD-MEX~\cite{Taninah2020_PLB800-135065}, PKDD~\cite{Long2004_PRC69-034319}, DD-ME2~\cite{Lalazissis2005_PRC71-024312}, TW99~\cite{Typel1999_NPA656-331}. Particularly, the nonuniform structures (droplet, rod, slab, tube, and bubble) of neutron star matter at subsaturation densities are fixed in the framework of Thomas-Fermi approximation (TFA), where the spherical and cylindrical approximations for the Wigner-Seitz cells are employed~\cite{Xia2022_CTP74-095303}. It is found that the crust $g$-mode oscillation plays an important role on the oscillation frequencies of neutron stars and should not be neglected, where the influence of inner crusts on the $g$-mode oscillations in low-mass neutron stars is obvious, and the $g_1$ mode oscillation could be dominated by the inner crust even for massive neutron stars adopting certain functionals. A strong correlation with the slope of symmetry energy and the $g_1$ mode frequency is then identified for neutron stars at fixed masses.

The paper is organized as follows. In Sec.~\ref{sec:the}, we introduce the theoretical framework for relativistic density functional, the non-radial oscillation in neutron stars under hydrostatic equilibrium, and the formulae to estimate the difference between the equilibrium and adiabatic sound velocities. In Sec.~\ref{sec:num}, we present the EOSs fixed by 10 different relativistic density functionals and analysis the origin for the difference between the equilibrium and adiabatic sound velocities and estimate the corresponding Brunt-V\"{a}is\"{a}l\"{a} (BV) frequency. The variation of $g$-mode frequencies with respect to neutron star mass is also illustrated in Sec.~\ref{sec:num}, where the corresponding crust and core $g$-mode frequencies as well as the correlation between the $g$-mode frequencies and nuclear matter properties are examined. Finally, we summarize our work in Sec.~\ref{sec:con}. In this paper, we use the geometric unit $\hbar=c=G=1$, where $c$ and $G$ are the speed of light and gravitational constants, respectively, and the metric signature is $(-, +, +, +)$.

\section{\label{sec:the}THEORETICAL FRAMEWORK}

\subsection{\label{sec:the_EOS} Equation of state}

Since the RMF model~\cite{Meng2016} has been successfully applied to describe finite (hyper) nuclei~\cite{Reinhard1989_RPP52-439, Ring1996_PPNP37-193, Meng2006_PPNP57-470, Paar2007_RPP70-R02, Meng2015_NPP42-093101, Chen2021_SCPMA64-282011, Typel1999_NPA656-331, Vretenar1998_PRC57-R1060, Lu2011_PRC84-014328} and baryonic matter~\cite{Ban2004_PRC69-045805, Weber2007_PPNP59-94, Long2012_PRC85-025806, Sun2012_PRC86-014305, Wang2014_PRC90-055801, Fedoseew2015_PRC91-034307, Gao2017_ApJ849-19}, in this work adopt the RMF model to fix the properties of neutron star matter, which are used to examine the $g$-mode oscillations in neutron stars. More specifically, we adopt the relativistic density functionals illustrated in Ref.~\cite{Xia2022_CTP74-095303}, i.e., those with nonlinear self-couplings (NL3~\cite{Lalazissis1997_PRC55-540}, PK1~\cite{Long2004_PRC69-034319}, TM1~\cite{Sugahara1994_NPA579-557}, GM1~\cite{Glendenning1991_PRL67-2414}, MTVTC~\cite{Maruyama2005_PRC72-015802}) and density-dependent couplings (DD-LZ1~\cite{Wei2020_CPC44-074107}, DD-MEX~\cite{Taninah2020_PLB800-135065}, PKDD~\cite{Long2004_PRC69-034319}, DD-ME2~\cite{Lalazissis2005_PRC71-024312}, TW99~\cite{Typel1999_NPA656-331}). Due to the liquid-gas phase transition of nuclear matter, it is expected that their mixed phases with various inhomogeneous structures, namely nuclear pasta~\cite{Baym1971_ApJ170-299, Negele1973_NPA207-298, Structure1983_PRL50-2066, Hashimoto1984_PTP71-320, Williams1985_NPA435-844}, will appear and form the inner crusts of neutron stars. As density increases, the inhomogeneous nuclear matter will gradually become uniform and form the core of a neutron star.

Here we provide a brief overview for the model, and a more detailed discussion can be found in Ref.~\cite{Xia2022_CTP74-095303}. Various nuclear pasta phases including droplets, rods, slabs, tubes, and bubbles comprised of protons, neutrons, electrons, and muons are considered in our work. The Lagrangian density of RMF models is expressed as
 \begin{eqnarray} \mathcal{L}&=&\sum_{i=n,p}\bar\psi_i[i\gamma^\mu\partial_\mu-\gamma^0(\textsl{g}_{\omega}\omega+\textsl{g}_{\rho}\rho\tau_i+Aq_i)-m_i^{*}]\psi_i \nonumber\\
 		&&+\sum_{l=e,\mu}\bar\psi_l[i\gamma^\mu\partial_\mu-m_l+e\gamma^0A]\psi_l-\frac{1}{4}A_{\mu\nu}A^{\mu\nu}\nonumber\\ 		 &&+\frac{1}{2}\partial_\mu\sigma\partial^\mu\sigma-\frac{1}{2}m_\sigma^2\sigma^2-\frac{1}{4}\omega_{\mu\nu}\omega^{\mu\nu}+\frac{1}{2}m_\omega^2\omega^2 \nonumber\\
 		&&-\frac{1}{4}\rho_{\mu\nu}\rho^{\mu\nu}+\frac{1}{2}m_\rho^2\rho^2+U(\sigma,\omega), \label{equ-1}	
 \end{eqnarray}
where ${\tau}_n=-{\tau}_p=1$ is the third component of isospin for nucleons, $q_p=-q_e=-q_\mu=e$ and $q_n=0$ the charge, $m^*_{n,p}\equiv m_{n,p}+ \textsl{g}_\sigma\sigma$ the effective nucleon mass, $U(\sigma,\omega)$ the nonlinear self-coupling terms and $g_{\sigma,\omega,\rho}(n_\mathrm{b})$ the density-dependent couplings with $n_\mathrm{b}=n_p+n_n$ being the baryon number density. Due to time reversal symmetry, the boson fields $\sigma$, $\omega$, $\rho$ and $A$ take mean values with temporal parts only, and $\omega_{\mu\nu}$, $\rho_{\mu\nu}$, $A_{\mu\nu}$ represents their field tensors.

Carrying out standard variational procedure, the Klein-Gordon equations of various meson and photon fields $(\sigma,\omega,\rho, A)$ can be fixed, i.e.,
\begin{align}
	\label{equ-2}    (-\nabla^2 + m^2_\sigma)\sigma &=-\textsl{g}_\sigma n_s-\textsl{g}_2\sigma^2-\textsl{g}_3\sigma^3, \\
	\label{equ-3}    (-\nabla^2 + m^2_\omega)\omega &=\textsl{g}_\omega n_b-c_3\omega^3, \\
	\label{equ-4}    (-\nabla^2 + m^2_\rho)\rho &=\sum _{i=n,p}\textsl{g}_\rho\tau_i n_i, \\
	\label{equ-5}    -\nabla^2A &=e(n_p -n_e-n_\mu).
\end{align}
These equations (\ref{equ-2})-(\ref{equ-5}) are then solved inside Wigner-Seitz cells adopting spherical and cylindrical approximations, while the density profiles of all particles are fixed iteratively in TFA fulfilling the constancy of chemical potentials, i.e.,
\begin{equation}\label{equ-6}
\mu_i(\Vec{r})=\sqrt{\nu^2_i+m^{*2}_i}+\Sigma^R+g_\omega\omega+g_\rho\tau_i\rho+q_iA=\mathrm{constant}.
\end{equation}
Here $\nu_i$ is the Fermi momentum, which fix the number density $n_i=\nu_i^3/3\pi^2$ of fermion $i$ for cold neutron star matter. The additional ``rearrangement" term
\begin{equation}\label{equ-14}
\Sigma^R = \frac{\mbox{d}g_\sigma}{\mbox{d}n_\mathrm{b}}\sigma n_s+\frac{\mbox{d} g_\omega}{\mbox{d}n_\mathrm{b}}\omega n_\mathrm{b} + \frac{\mbox{d}g_\rho}{\mbox{d}n_\mathrm{b}}\rho \sum_i\tau_i n_i
\end{equation}
arises due to the density-dependent coupling constants~\cite{Lenske1995_PLB345-4}.

\subsection{Non-radial oscillations}

Due to the strong gravitational field inside neutron stars, their structure and dynamic evolution are governed by Einstein's equations in general relativity, i.e.,
\begin{equation}
{R_{\nu\mu} - \frac{1}{2} g_{\nu\mu}R = 8\pi T_{\nu\mu}, \label{eq:1}}
\end{equation}
where $R_{\nu\mu}$ is the Ricci tensor and $R$ the Ricci scalar. The static and spherically symmetric metric describing neutron stars is given by
\begin{equation}
\mbox{d}s^{2} =-\mathrm{e}^{2\Phi } \mbox{d}t^{2} +\mathrm{e}^{2\Lambda}\mbox{d}r^2+r^2(\mbox{d}\theta^{2}+\sin^2\theta \mbox{d}\phi ^2 ),
\label{eq:2}
\end{equation}
where $\Phi$ and $\Lambda$ are metric functions with respect to $r$. Combined with the energy-momentum tensor for a perfect fluid
\begin{equation}
T_{\mu \nu}=(\varepsilon+p)u_{\mu }u_{\nu }+pg_{\mu \nu}, \label{eq:3}
\end{equation}
a mass function $m(r)$ is then defined as $m(r) =r(1-\mathrm{e}^{-2\Lambda})/2$ and satisfies
\begin{equation}
\frac{\mbox{d}m}{\mbox{d}r}=4\pi r^2\varepsilon, \label{eq:4}
\end{equation}
where $\varepsilon$ is the energy density and $p$ the pressure. To fix the distributions of the pressure $p(r)$ and metric function $\Phi(r)$ for the equilibrium structure of a neutron star, we need to solve the Tolman-Oppenheimer-Volkov (TOV) equation, i.e.,
\begin{eqnarray}
 \frac{\mathrm{d} p}{\mathrm{~d} r}&=&-(\varepsilon+p) \frac{\mathrm{d} \Phi}{\mathrm{d} r}, \label{eq:5}\\
 \frac{\mathrm{d} \Phi}{\mathrm{d} r}&=&\frac{m+4 \pi r^{3} p}{r(r-2 m)}, \label{eq:6}
\end{eqnarray}
which is dependent on the EOS of neutron star matter with $p = p(\varepsilon)$.

On the basis of the equilibrium structure of a neutron star, the internal fluid is perturbed, and the Lagrange displacement vector of fluid perturbation can be decomposed into spherical harmonics~\cite{Thorne1967_ApJ-149-591}, i.e.,
\begin{eqnarray}
    \xi^{r}&=&r^{-2}\mathrm{e}^{-\Lambda}W Y^l_m \mathrm{e}^{i\omega t},     \label{eq:8}\\
    \xi^{\theta}&=&-r^{-2}V\partial_{\theta} Y^l_m \mathrm{e}^{i\omega t},     \label{eq:9}\\
    \xi^{\phi}&=&-r^{-2}(\sin\theta)^{-2}V\partial_{\phi} Y^l_m \mathrm{e}^{i\omega t},     \label{eq:10}
\end{eqnarray}
where $Y^l_m(\theta,\phi)$ is the spherical harmonic, $W(r)$ and $V(r)$ the characteristic functions related to radial and angular amplitudes, and $\omega$ the intrinsic frequency of the non-radial oscillation. For a cold, non-rotating, and spherically symmetric neutron star, the non-radial oscillation equation~\cite{Zheng2023_PRD107-103048, Osaki1973_Apj185-277, Zhao2022_PRD105-103025} in the relativistic Cowling approximation~\cite{Cowling1941_MNRAS101-367} is given by
\begin{eqnarray}
    \frac{\mathrm{d} W}{\mathrm{d} r} &=& \frac{\mathrm{d}\varepsilon}{\mathrm{d} p}\left[\omega^2 r^2 \mathrm{e}^{\Lambda-2\Phi}V+\frac{\mathrm{d}\Phi}{\mathrm{d} r}W \right]-l(l+1)\mathrm{e}^{\Lambda}V,     \label{eq:11}\\
    \frac{\mathrm{d} V}{\mathrm{d} r} &=& \mathrm{e}^{\Lambda}(\frac{N^2}{\omega^2} - 1)\frac{W}{r^2}+\frac{N^2}{g \mathrm{e}^{2\Phi-2\Lambda}}V+2gV.     \label{eq:12}
\end{eqnarray}
Note that Cowling approximation neglects the metric perturbations during the fluid oscillations, which significantly simplifies our calculation. The imaginary part of the eigenfrequency is thus absent, where the dampening time of the $g$-mode due to gravitational wave radiation can not be directly estimated. The effectiveness of Cowling approximation has been verified in the literature~\cite{Zhao2022_PRD105-103025, Xu2017_PRD96-083005, Yoshida2002_AAP395-201, Ranea-Sandoval2018_JCAR12-031}, especially for the $g$-mode oscillations with buoyancy as the main restoring force~\cite{Wei2020_ApJ904-187}.
{As illustrated in Ref.~\cite{Zhao2022_PRD105-103025}, the difference between the $g$-mode frequencies obtained in the Cowling approximation and in full general relativity is less than 10\%, which can be further reduced for less massive neutron stars. The deviations arise from Cowling approximation for $f$-mode and $p$-mode, nevertheless, are much larger and reach $\sim$20\%~\cite{Yoshida1997_MNRAS117-122-289, Sotani2020_PRD102-063025}. In such cases, to simplify our calculation, in this work the $g$-mode oscillations  are investigated under Cowling approximation.}

In Eq.~(\ref{eq:12}) $N$ stands for the  Brunt-V\"{a}is\"{a}l\"{a} frequency~\cite{John1980_PUP}, which represents the oscillation frequency of the local fluid unit inside the star, and is defined as
\begin{equation}
    N^2 = g^2\left(\frac{1}{c^2_e}-\frac{1}{c^2_s}\right)\mathrm{e}^{2\Phi-2\Lambda}     \label{eq:13}
\end{equation}
with $g=-(\mbox{d}p/\mbox{d}r)/(p+\varepsilon)$. From Eq.~(\ref{eq:13}), it can be seen that $N^2$ is related to two sound speeds, the equilibrium sound speed $c_e$ and the adiabatic sound speed $c_s$. The square of equilibrium sound velocity $c_{e}^{2}$ is fixed by taking derivative of pressure $p$ with respect to energy density $\varepsilon$ in $\beta$-equilibrium, while the square of the adiabatic sound velocity $c_{s}^{2}$ is obtained by taking derivative of $p$ with respect to $\varepsilon$, holding all the particle fractions $Y_i=n_i/n_\mathrm{b}$ $(i =n, p, e, \mu)$ fixed~\cite{Jaikumar2021_PRD103-123009, Zheng2023_PRD107-103048}, i.e.,
\begin{equation}
    c^2_e =  \frac{\mbox{d}p}{\mbox{d}\varepsilon}, \ \ \ c^2_s = \left.\frac{\mbox{d}p}{\mbox{d}\varepsilon}\right|_{\{Y_i\}}.    \label{eq:14}
\end{equation}
The neutron star described in this paper is comprised of nucleons, which involves four types of particles, namely neutrons, protons, electrons and muons. It should be noted that we have assumed that the time scale of weak reactions of fluid elements in different density gradients is much larger than the time scale of oscillation, so there exist the composition $g$ modes. For the case of rapid nuclear reactions, it was shown that the $g$ modes will be dampened~\cite{Counsell2024_MNRAS531-1721-1729, Pereira2018_ApJ860-12}. The higher order modes are the first to be removed from the oscillation spectrum, where the extent of the removed modes depend on the reaction rate.

For a given equation of state, the non-radial oscillation Eq.~(\ref{eq:11}) and Eq.~(\ref{eq:12}), combined with the boundary conditions for the stellar center $(r=0)$ and surface $(r=R)$, form a Sturm-Liouwe eigenvalue problem. In the core of a neutron star we have
\begin{equation}
   W(0)+lV(0) = 0.     \label{eq:16}
\end{equation}
On the surface of the neutron star at $r=R$, the Lagrange perturbation of pressure must disappear, i.e.,
\begin{equation}
    \omega^2 \mathrm{e}^{\Lambda-2\Phi}V(R)+\frac{1}{R^2}\frac{d\Phi}{dr}(R)W(R)=0.     \label{eq:17}
\end{equation}
{Note that the $g$-modes studied in this work are fixed at  $l=2$ since the quadrupole oscillation mode has a strong coupling with gravitational waves, while the coupling weakens for higher multipolarity  (octupole and higher with $l\geq3$)~\cite{Tran2023_PRC108-015803}.  Additionally, it was shown that only the $l=2$ mode can be coupled to the quadrupole tidal potential~\cite{Xu2017_PRD96-083005}, which is important for the tidal excitation of the $g$-mode oscillations in a binary neutron star system.}
Take $W (0) =1$ and $V (0) = -1/2$ at $r=0$, and then integrate Eqs.~(\ref{eq:11}) and (\ref{eq:12}) from the core to the surface of the neutron star. Once the boundary condition Eq.~(\ref{eq:17}) is satisfied, the discrete eigenvalue $\omega_{i}$ can be obtained, which typically gives
\begin{equation}
    \omega_{gn}<...<\omega_{g1}<\omega_{f}<\omega_{p1}<...<\omega_{pn}.     \label{eq:18}
\end{equation}
{Here $n$ represents the number of nodes for the radial eigenfunction $W(r)$ inside the star. As $n$ approaches to infinity, we have $\lim_{n \to \infty} \omega_{gn}= 0$ and $\lim_{n \to \infty} \omega_{pn}= \infty$.} $\omega_{f}$ is the intrinsic frequency of the fundamental mode, corresponding to the mode without any nodes in the star with $n = 0$; $\omega_{pn}$ is the intrinsic frequency of pressure mode ($p$-mode), taking pressure as the main restoring force. The number of nodes $n$ represents the $n$-order $p$-mode frequency, and the $p$-mode frequency gradually increases with the order $n$;
$\omega_{gn}$ is the intrinsic frequency of gravity mode ($g$-mode), which takes buoyancy as the main restoring force. Different from $p$-mode, the $g$-mode frequency gradually decreases with the order. The $g$-mode produced by chemical stratification is very sensitive to the composition of dense matter. Therefore, they may constrain the EOSs better than $f$-modes and $p$-modes. In particular, the $g$-mode frequency depends on the proton fraction $Y_p$, which is dominated by the symmetry energy~\cite{Xia2022_CTP74-095303}. Thus this work is devoted to the $g$-mode oscillations up to the third order at $l= 2$.

It is worth mentioning that in this work we neglect the $g$-mode oscillations arise from the density discontinuities due to the variations of nuclear species in neutron star outer crusts~\cite{Finn1987_MNRAS227-265-293} and assuming there are no convection~\cite{Reisenegger1992_Apj395-240-249} (zero buoyancy case), i.e., taking $c_{e}^{2}=c_{s}^{2}$ at densities far below the neutron drip density ($n_\mathrm{b}\lesssim 10^{-10}\ \mathrm{fm}^{-3}$) in outer crusts with fixed nuclear species. Note that at larger densities the proton and neutron numbers of nuclei change smoothly and take non-integer numbers, resulting differences between $c_{e}$ and $c_{s}$ at $n_\mathrm{b}\gtrsim 10^{-6}\ \mathrm{fm}^{-3}$. Then the average effects due to variations of nuclear species are effectively considered for $g$-mode oscillations. The effects of density discontinuity at interfaces among various types nuclei shapes (droplet, rod, slab, tube, bubble, and uniform) are insignificant, so that we treat their transitions as continuous, i.e., in the absence of the discontinuous $g$-mode (or $i$-mode)~\cite{Reisenegger1992_Apj395-240-249, Zhu2023_PRD107-83023}. On this basis, we examine the $g$-mode oscillations  of neutron stars including the contributions from the inner crusts. It can be seen from Ref.~\cite{Jaikumar2021_PRD103-123009} that the difference between the two sound speeds leads to the existence of $g$-mode, and the magnitude of the difference generally determines the $g$-mode frequency.

As illustrated in Refs.~\cite{Jaikumar2021_PRD103-123009, Tran2023_PRC108-015803}, for uniform ($n,p,e,\mu$) neutron star matter, a formula on the sound velocity difference was derived and reads
\begin{eqnarray}
   c^2_s-c^2_e =
   \frac{n_\mathrm{b}^{2}}{\mu_{n}}\left[\left.\frac{\partial\left(\mu_{n}-\mu_{e}-\mu_{p}\right)}{\partial n_{b}}\right|_{\{Y_i\}} \frac{\mbox{d} Y_{p}}{\mbox{d} n_\mathrm{b}}
   \notag\right.
   \\
   \phantom{=\;\;}
   \left.  +\left.\frac{\partial\left(\mu_{e}-\mu_{\mu}\right)}{\partial n_{b}}\right|_{\{Y_i\}} \frac{\mbox{d} Y_{\mu}}{\mbox{d} n_\mathrm{b}}\right].
    \label{eq:19}
\end{eqnarray}
For nonuniform matter that comprise neutron stars' crusts, we fix their sound velocities numerically based on Eq.~(\ref{eq:14}), where at fixed particle fractions $Y_i$ and baryon number density $n_\mathrm{b}$ we search for the optimum configuration that minimize the energy of the system as in Ref.~\cite{Okamoto2012_PLB713-284, Okamoto2013_PRC88-025801, Xia2021_PRC103-055812, Xia2022_CTP74-095303, Xia2022_PRC105-045803}.

\section{\label{sec:num}RESULTS AND DISCUSSION}

\begin{figure}
\includegraphics[width=\linewidth]{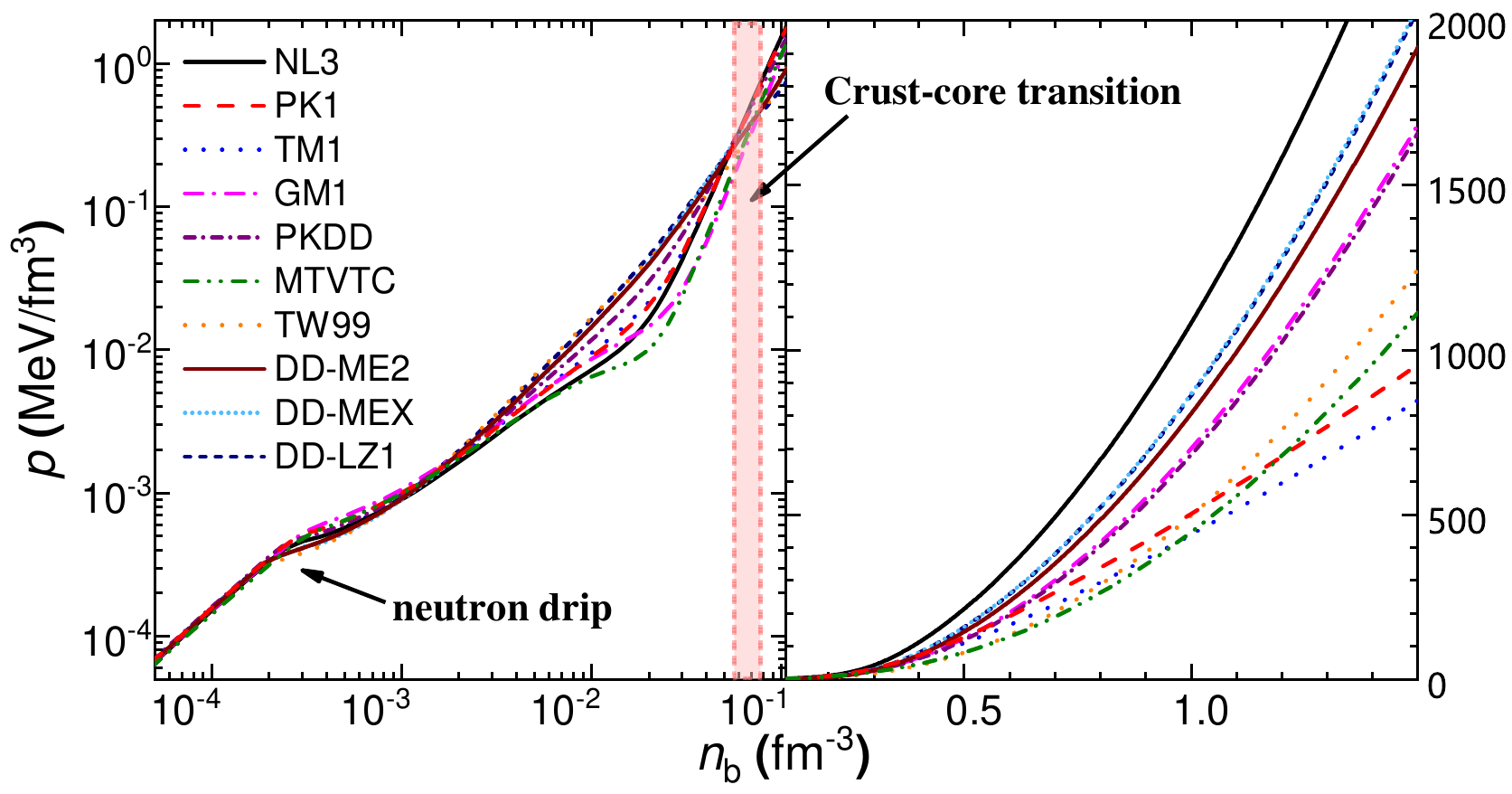}
\caption{\label{Fig:T1}{Pressure of neutron star matter in the crust (left) and core (right) regions as functions of baryon number density, which are predicted by the 10 relativistic density functionals as indicated in Table~\ref{table:prop}. The vertical red band ($0.054 \lesssim n_\mathrm{b} \lesssim 0.074\ \mathrm{fm}^{-3}$) represents the crust-core transition densities, while the neutron drip density lies within $0.0002 \lesssim n_\mathrm{b} \lesssim 0.0003\ \mathrm{fm}^{-3}$. The densities in between the neutron drip and crust-core transition densities correspond to the inner crust matter in neutron stars.}}
\end{figure}

In Fig.~\ref{Fig:T1} we present the pressure as functions of baryon number density obtained with the 10 relativistic density functionals (NL3, PK1, TM1, GM1, MTVTC, PKDD, DD-LZ1, DD-MEX, DD-ME2 and TW99)~\cite{Xia2022_CTP74-095303}. At $n_\mathrm{b} \lesssim 0.0002\ \mathrm{fm}^{-3}$ Fig.~\ref{Fig:T1} shows the EOSs for the outer crusts of neutron stars, where neutron-rich nuclei and electrons form a coulomb lattice. As density increases, neutrons begin to drip out from nuclei and form a neutron gas, the neutron star matter is thus basically a liquid-gas mixed phase of nuclear matter, where the neutron drip density for the 10 models lies within the range {$0.0002 \lesssim n_\mathrm{b} \lesssim 0.0003\ \mathrm{fm}^{-3}$}. At larger densities, the nuclei will eventually deform and form nuclear pasta~\cite{Baym1971_ApJ170-299, Negele1973_NPA207-298, Structure1983_PRL50-2066, Hashimoto1984_PTP71-320, Williams1985_NPA435-844}. As we further increase the density, eventually the nuclear pasta becomes unstable and the neutron star matter is restored into a homogenous state, forming neutron star cores. The vertical red band in Fig.~\ref{Fig:T1} indicates the crust-core transition densities, which lie within the range {$0.054 \lesssim n_\mathrm{b} \lesssim 0.074\ \mathrm{fm}^{-3}$}. In between the neutron drip density and crust-core transition density is the inner crust region for neutron stars, and above these densities are the core regions of neutron stars. In comparison with those in neutron stars' outer crusts, it was shown that the EOSs in neutron stars' inner crusts are very sensitive to the adopted density functionals~\cite{Xia2022_CTP74-095303}. In particular, the relative uncertainty of the EOSs in neutron stars' inner crust increases with density and decreases after reaching the peak value at {$n_\mathrm{b} \approx 0.02\ \mathrm{fm}^{-3}$}, where the EOSs predicted by the density functionals with larger slope of symmetry energy $L$ are generally stiffer at the density {$n_\mathrm{b}\gtrsim 0.02\ \mathrm{fm}^{-3}$}.

\begin{table}[h]
\setlength{\tabcolsep}{3.7pt}
\caption{\label{table:prop} {Saturation properties of nuclear matter predicted by the 10 relativistic density functionals (NL3~\cite{Lalazissis1997_PRC55-540}, PK1~\cite{Long2004_PRC69-034319}, TM1~\cite{Sugahara1994_NPA579-557}, GM1~\cite{Glendenning1991_PRL67-2414}, MTVTC~\cite{Maruyama2005_PRC72-015802}, DD-LZ1~\cite{Wei2020_CPC44-074107}, DD-MEX~\cite{Taninah2020_PLB800-135065}, PKDD~\cite{Long2004_PRC69-034319}, DD-ME2~\cite{Lalazissis2005_PRC71-024312}, TW99~\cite{Typel1999_NPA656-331}). Here $n_0$ represents the saturation density of nuclear matter, $B$ the binding energy, $K$ the incompressibility, $J$ the skewness, $S$ the symmetry energy, $L$ and $K_\mathrm{sym}$ the slope and curvature of  symmetry energy, which correspond to the coefficients in the Taylor series of nuclear binding energy as indicated in Eqs.~(\ref{eq:e0}) and (\ref{eq:esym}).}}
\begin{tabular}{l|ccccccc} \hline \hline
$ $ & $n_0$ & $B$ & $K$ & $J$ & $S$ & $L$ & $K_\mathrm{sym}$ \\
 & fm${}^{-3}$ & MeV & MeV & MeV & MeV & MeV & MeV    \\ \hline
NL3    & 0.148 & $-16.25$  & 271.7 & $204  $ & 37.4 & 118.6 & 101 \\
PK1    & 0.148 & $-16.27$  & 282.7 & $-27.8$ & 37.6 & 115.9 & 55 \\
TM1    & 0.145 & $-16.26$  & 281.2 & $-285 $ & 36.9 & 110.8 & 34 \\
GM1    & 0.153 & $-16.33$  & 300.5 & $-216 $ & 32.5 & 94.0  & 18 \\
PKDD   & 0.150 & $-16.27$  & 262.2 & $-119 $ & 36.8 & 90.2  & $-81 $\\
MTVTC  & 0.153 & $-16.30$  & 239.8 & $-513 $ & 32.5 & 89.6  & $-6.5$ \\
TW99   & 0.153 & $-16.24$  & 240.2 & $-540 $ & 32.8 & 55.3  & $-125$ \\
DD-ME2 & 0.152 & $-16.13$  & 250.8 & $477  $ & 32.3 & 51.2  & $-87 $\\
DD-MEX & 0.152 & $-16.11$  & 267.6 & $874  $ & 32.3 & 49.7  & $-72 $\\
DD-LZ1 & 0.158 & $-16.06$  & 230.7 & $1330 $ & 32.0 & 42.5  & $-20 $\\
\hline
\end{tabular}
\end{table}

The properties of nuclear matter around the saturation density predicted by the 10 relativistic density functionals are indicated in Table~\ref{table:prop}.
{The binding energy for zero-temperature nuclear matter at given baryon number density $n_\mathrm{b}$ and asymmetry parameter $\delta = (n_n-n_p)/n_\mathrm{b}$ can be expressed as
\begin{equation}
e(n_\mathrm{b},\delta)=e_0\left(n_\mathrm{b}\right) +e_\mathrm{sym}\left(n_\mathrm{b}\right)\delta^2,
\end{equation}
where $e_0\left(n_\mathrm{b}\right)$ is the binding energy of symmetric nuclear matter and $e_\mathrm{sym}\left(n_\mathrm{b}\right)$ the symmetry energy. Expanding $e_0\left(n_\mathrm{b}\right)$ and $e_\mathrm{sym}\left(n_\mathrm{b}\right)$ in the Taylor series and omitting the higher order terms, we have
\begin{eqnarray}
e_0 &=&B+\frac{K}{18} \left ( \frac{n_\mathrm{b}}{n_0}-1  \right ) ^{2}+\frac{J}{162} \left ( \frac{n_\mathrm{b}}{n_0}-1  \right )^{3}, \label{eq:e0} \\
e_\mathrm{sym} &=&S+\frac{L}{3} \left ( \frac{n_\mathrm{b}}{n_0}-1  \right )+\frac{K_{\mathrm{sym}}}{18} \left ( \frac{n_\mathrm{b}}{n_0}-1  \right )^{2},
\label{eq:esym}
\end{eqnarray}
where $n_0$ is the nuclear saturation  density, $B$ the binding energy, $K$ the incompressibility, $J$ the skewness, $S$ the symmetry energy, $L$ and $K_\mathrm{sym}$ the slope and curvature of symmetry energy.} It is worth mentioning that the lower-order coefficients are well constrained based on various terrestrial experiments and nuclear theories, e.g., $B\approx -16$ MeV, $K = 240 \pm 20$ MeV~\cite{Shlomo2006_EPJA30-23}, $S = 31.7 \pm 3.2$ MeV and $L = 58.7 \pm 28.1$ MeV~\cite{Li2013_PLB727-276, Oertel2017_RMP89-015007}. Meanwhile, based on various data from astrophysical observations, heavy-ion collisions, and neutron skin thicknesses for $^{208}$Pb  and $^{48}$Ca, it is possible to further constrain the higher-order coefficients despite the large uncertainty~\cite{Farine1997_NPA615-135, Zhang2020_PRC101-034303, Essick2021_PRL127-192701, Xie2021_JPG48-025110, PREX2021_PRL126-172502, CREX2022_PRL129-042501}. In general, as indicated in Table~\ref{table:prop}, the density functionals PKDD, MTVTC, TW99, DD-ME2, DD-MEX, and DD-LZ1 predict nuclear matter properties that are roughly consistent with those constraints, while those with nonlinear self-couplings (NL3, PK1, TM1, GM1) predicts too large incompressibility $K$ and slope $L$ of symmetry energy, resulting too stiff EOSs for neutron star matter.

\begin{figure}
\includegraphics[width=\linewidth]{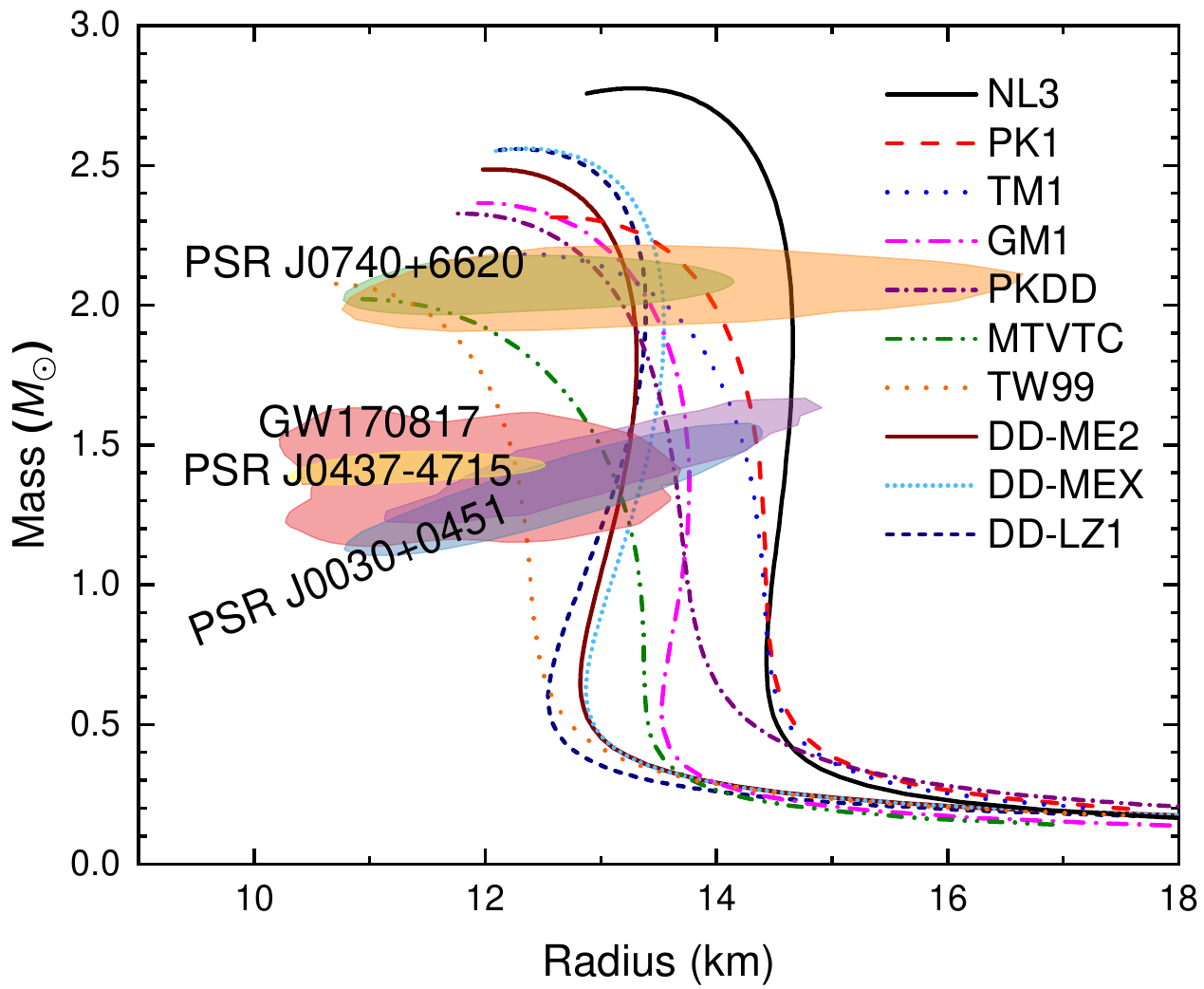}
\caption{\label{Fig:MR} Mass-radius relations of neutron stars predicted by the 10 relativistic density functionals are indicated in Table~\ref{table:prop}. The shaded regions represent the $M$-$R$ constraints from the binary neutron star merger event GW170817 within 90\% credible region~\cite{LVC2018_PRL121-161101}, the observational pulse-profiles in PSR J0030+0451, PSR J0740+6620, and PSR J0437-4715 within 68\% credible region~\cite{Riley2019_ApJL887-L21, Riley2021_ApJL918-L27, Miller2019_ApJL887-L24, Miller2021_ApJL918-L28, Choudhury2024_ApJ971-L20}.}
\end{figure}

Based on the EOSs indicated in Fig.~\ref{Fig:T1}, we then solve the TOV equations~(\ref{eq:5}-\ref{eq:6}) and fix the corresponding equilibrium structures of neutron stars, where their mass-radius relations are presented in Fig.~\ref{Fig:MR}. Evidently, the maximum masses of the neutron stars predicted by the 10 relativistic density functionals exceed two solar masses, which is consistent with the constraints from the mass measurements of massive neutron stars~\cite{Antoniadis2013_Science340-1233232}. Nevertheless, compared with the constraints on neutron stars' radii~\cite{LVC2018_PRL121-161101, Riley2019_ApJL887-L21, Riley2021_ApJL918-L27, Miller2019_ApJL887-L24, Miller2021_ApJL918-L28, Choudhury2024_ApJ971-L20}, several relativistic density functionals (e.g., NL3, PK1, and TM1) predict too large radii, which could be ruled out if the neutron star is comprised entirely of $npe\mu$ matter. Note that there are updates on the radius measurements for PSR J0030+0451 and PSR J0740+6620~\cite{Salmi2024_ApJ976-58, Vinciguerra2024_ApJ961-62}, which generally coincide with previous measurements.

\begin{figure}
\includegraphics[width=\linewidth]{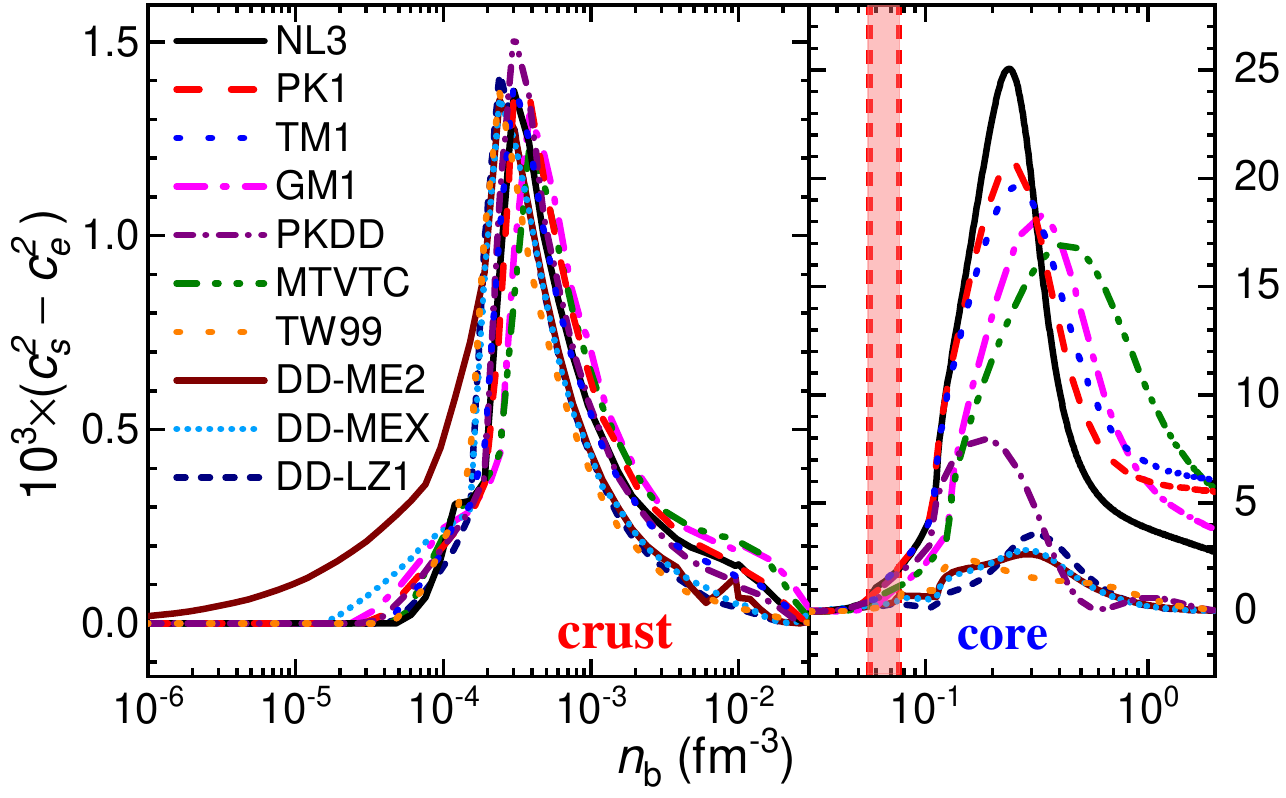}
\caption{\label{Fig:T2}The differences between the adiabatic and equilibrium sound velocities $c^2_s-c^2_e$ of neutron star matter as functions of baryon number density $n_\mathrm{b}$, where the red band indicates the crust-core transition densities.}
\end{figure}

We then present the differences between the adiabatic and equilibrium sound velocities $c^2_s-c^2_e$ in Fig.~\ref{Fig:T2}, which are obtained based on the 10 relativistic density functionals indicated in Table~\ref{table:prop}. The light red band in the figure corresponds to the crust-core phase transition densities.  Evidently, in neutron star crusts, the obtained values for $c^2_s-c^2_e$ are nonzero, which arise due to the variations of nuclear species and the emergence of neutron gas outside of the nuclei (above neutron drip densities at $n_\mathrm{b} \gtrsim 0.0002\ \mathrm{fm}^{-3}$). The differences reach their peak values at $n_\mathrm{b} \approx 0.0003\ \mathrm{fm}^{-3}$, and then decrease with density. Note that the values of $c^2_s-c^2_e$ fixed by various density functionals share a similar trend in neutron star crusts, which suggest the similar origin. At larger densities, neutron star matter becomes uniform, and $c^2_s-c^2_e$ eventually increases, where the slope increases with the slope of symmetry energy $L$. This becomes more evident at densities $n_\mathrm{b} \gtrsim 0.1\ \mathrm{fm}^{-3}$ with the emergence of muons, where $c^2_s-c^2_e$ increases quickly with density and reaches its peak at $0.2 \lesssim n_\mathrm{b} \lesssim 0.5\ \mathrm{fm}^{-3}$. Based on the saturation properties of nuclear matter indicated in Table~\ref{table:prop}, it is found that the peak value is strongly correlated with the slope of symmetric energy $L$. In particular, if the relativistic density functionals (DD-LZ1, DD-MEX, DD-ME2, and TW99) are adopted, the peak values for $c^2_s-c^2_e$ are small and comparable to those in neutron star crusts.

\begin{figure}
\includegraphics[width=\linewidth]{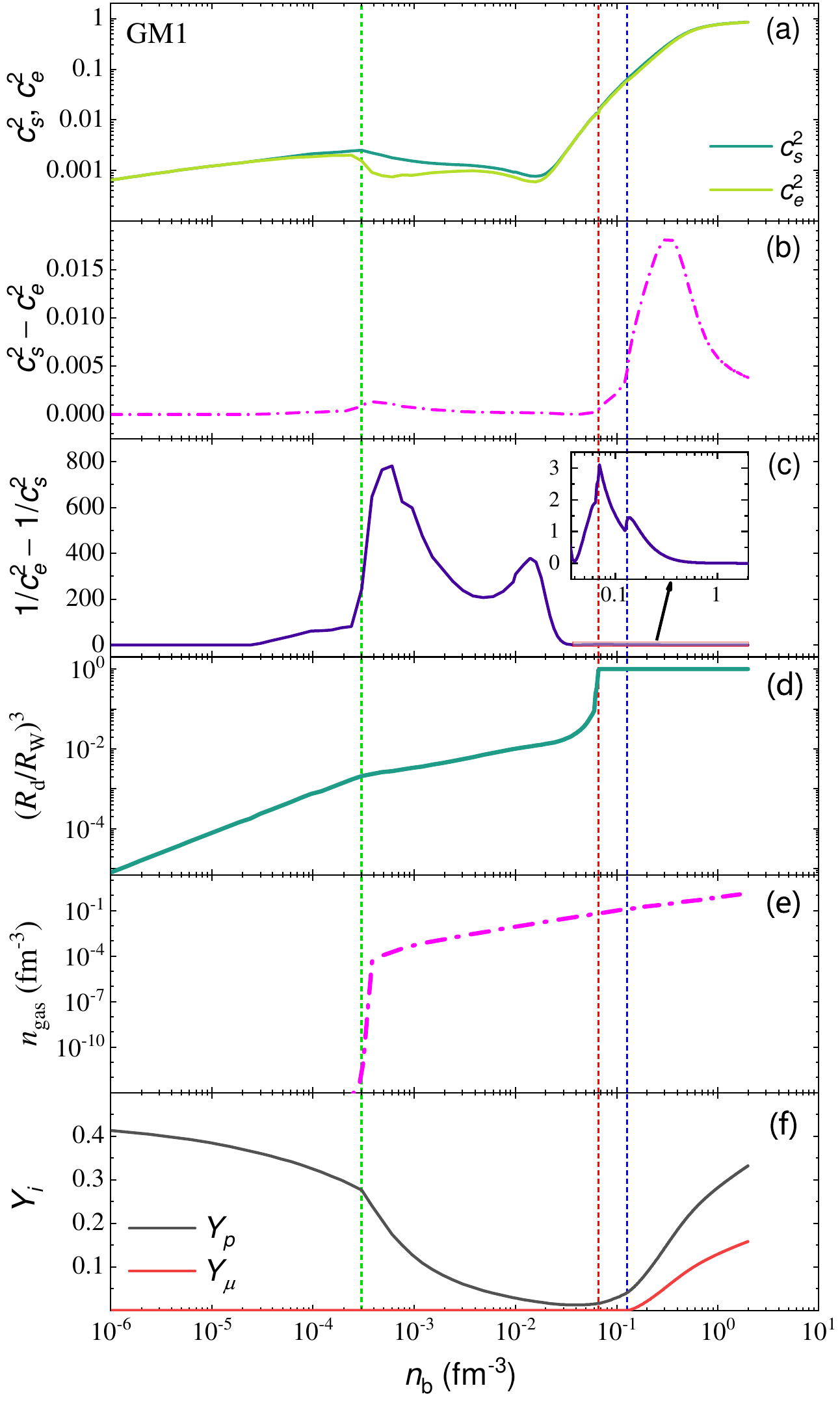}
\caption{\label{Fig:T3} The square of sound velocities ($c^2_e$ and $c^2_s$) and their difference ($c^2_s-c^2_e$, $1/c^2_s-1/c^2_e$), the filling factor $(R_d/R_W)^3$, the neutron gas density $n_\mathrm{gas}$, and the proton and muon fractions $Y_{p,\mu}$ as functions of baryon number density predicted by the relativistic density functional GM1. The vertical green, red, and blue dashed lines indicate the neutron drip density, the crustal-core phase transition density, and the onset density for muons.}
\end{figure}

To better understand the mechanism for the variations of sound velocity differences near neutron drip densities and onset densities of muons, in Fig.~\ref{Fig:T3} we present the sound velocities obtained with the relativistic density functional GM1, while other models give similar behaviors as indicated in Fig.~\ref{Fig:T2}. In Fig.~\ref{Fig:T3}(a), we can see that both $c^2_e$ and $c^2_s$ alter their trends after neutrons start to drip out, while $c^2_e$ decreases faster with density. This can be identified easily in Fig.~\ref{Fig:T3}(b) on the evolution of the sound velocity difference $c^2_s-c^2_e$. {Note that $c^2_e$ becomes discontinuous with the emergence of new degrees of freedom. For example, as indicated by the green, red, and blue dashed vertical lines in Fig.~\ref{Fig:T3}, $c^2_e$ is discontinuous at densities where neutron drip (the neutron gas densities $n_\mathrm{gas}>0$), core-crust transition (the filling factor $(R_d/R_W)^3\rightarrow 1$ with $R_d$ and $R_W$ being the sizes of droplets and Wigner-Seitz cells), and onset of muons (muon fraction $Y_\mu>0$) take place.} Unlike the equilibrium sound velocity, adiabatic sound velocity is always continuous as a function of density, which can be understood according to the definition of the adiabatic sound velocity $c_s$ in Eq.~(\ref{eq:14}), i.e., the $c_s$ is continues since there is no sudden changes of composition as the particle fractions are fixed in the derivative.

Figure~\ref{Fig:T3}(c) shows {the difference of the squared reciprocals} of the two sound speeds, which determines the local oscillation frequency, namely the BV frequency according to Eq.~(\ref{eq:13}). We can see clearly that the inner crust has a very high value for $1/c^2_e-1/c^2_s$ after neutron drip (indicated by the vertical green line), which is caused by the emergence of neutron gas outside of nuclei. At larger densities, there exist two spikes located at the red and blue dashed lines, which are caused by the core-crust transition ($n_b\approx0.07\ \mathrm{fm}^{-3}$) and the onset of muons ($n_b\approx0.12\ \mathrm{fm}^{-3}$)~\cite{Jaikumar2021_PRD103-123009}. It is found that each spike in the BV frequency corresponds to the appearance of a new degree of freedom. This is attributed to variations of the equilibrium sound velocity, which is particularly sensitive to the composition of matter inside neutron stars and thus a significant change in the BV frequency takes place with a variation in the degrees of freedom.

\begin{figure*}
\includegraphics[width=0.8\linewidth]{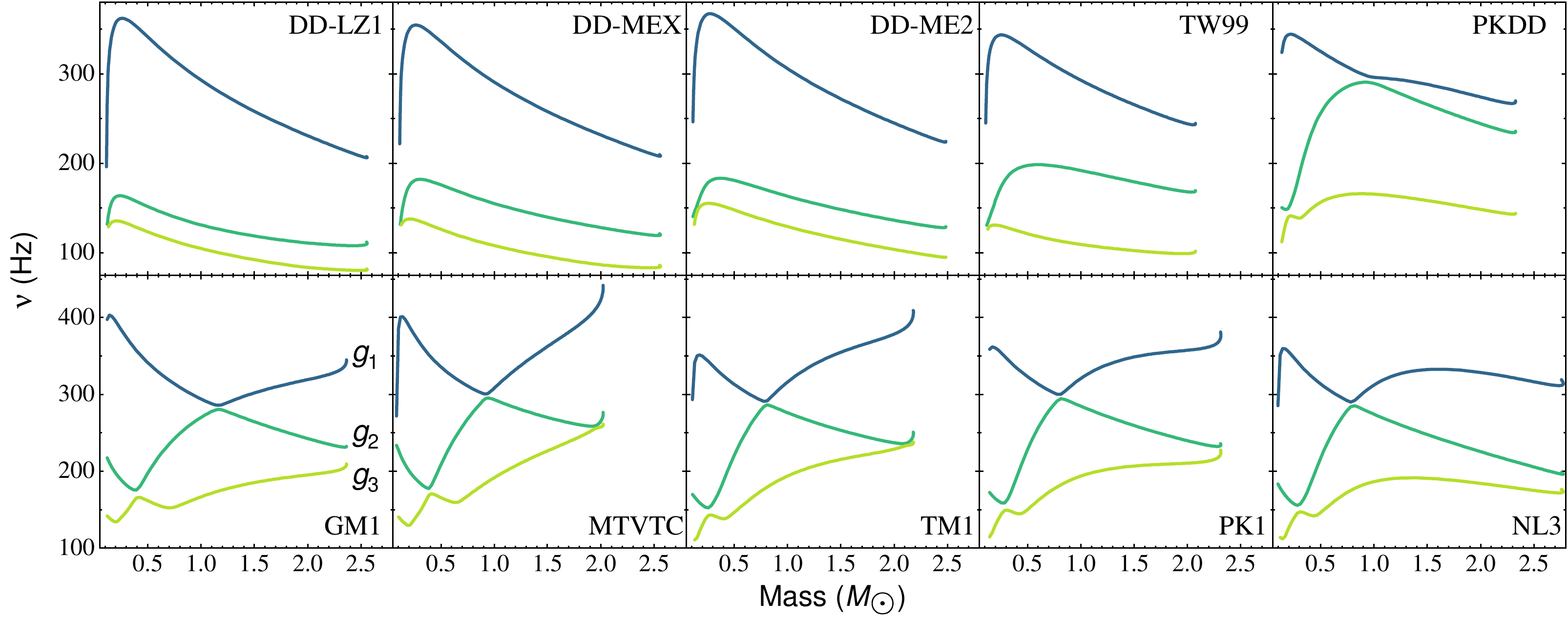}
\caption{\label{Fig:T4} Frequencies of the first three $g$-modes as functions of the stellar mass predicted by the 10 relativistic density functionals. }
\end{figure*}

In Fig.~\ref{Fig:T4} we then present the frequencies of the first three $g$ modes ($g_1$, $g_2$, and $g_3$) as functions of neutron star's mass. It can be easily identified that the trends of the $g$-mode frequencies are distinctively different for those predicted by the density functionals with density-dependent couplings and nonlinear self-couplings except for PKDD, which behave as an intermediate state. For nonlinear self-coupling models, the avoid-crossing phenomenon can be easily identified, which is also observed in other types of oscillation modes, e.g., the $f$ and $p$ modes. Gondek et al. explained in detail the origin of the avoid-crossing phenomenons in $f$ and $p$ modes~\cite{Gondek1999_AAP344-117}, which arise due to the differences in the EOSs and oscillation properties between the core and crust in a neutron star. This is also the case for the $g$-mode oscillations examined in this work, where the avoid-crossing phenomenon emerges as we increase neutron star's mass. The avoid-crossing phenomenon is not observed for density-dependent coupling models except for PKDD, where the first three $g$-mode frequencies decrease with neutron star's mass at $M>0.5\ M_{\odot}$, and rise slightly when approaching to the maximum mass. In general, the avoid-crossing phenomenon of PKDD is less significant than that of nonlinear self-coupling models.

\begin{figure}
\centering
\includegraphics[width=\linewidth]{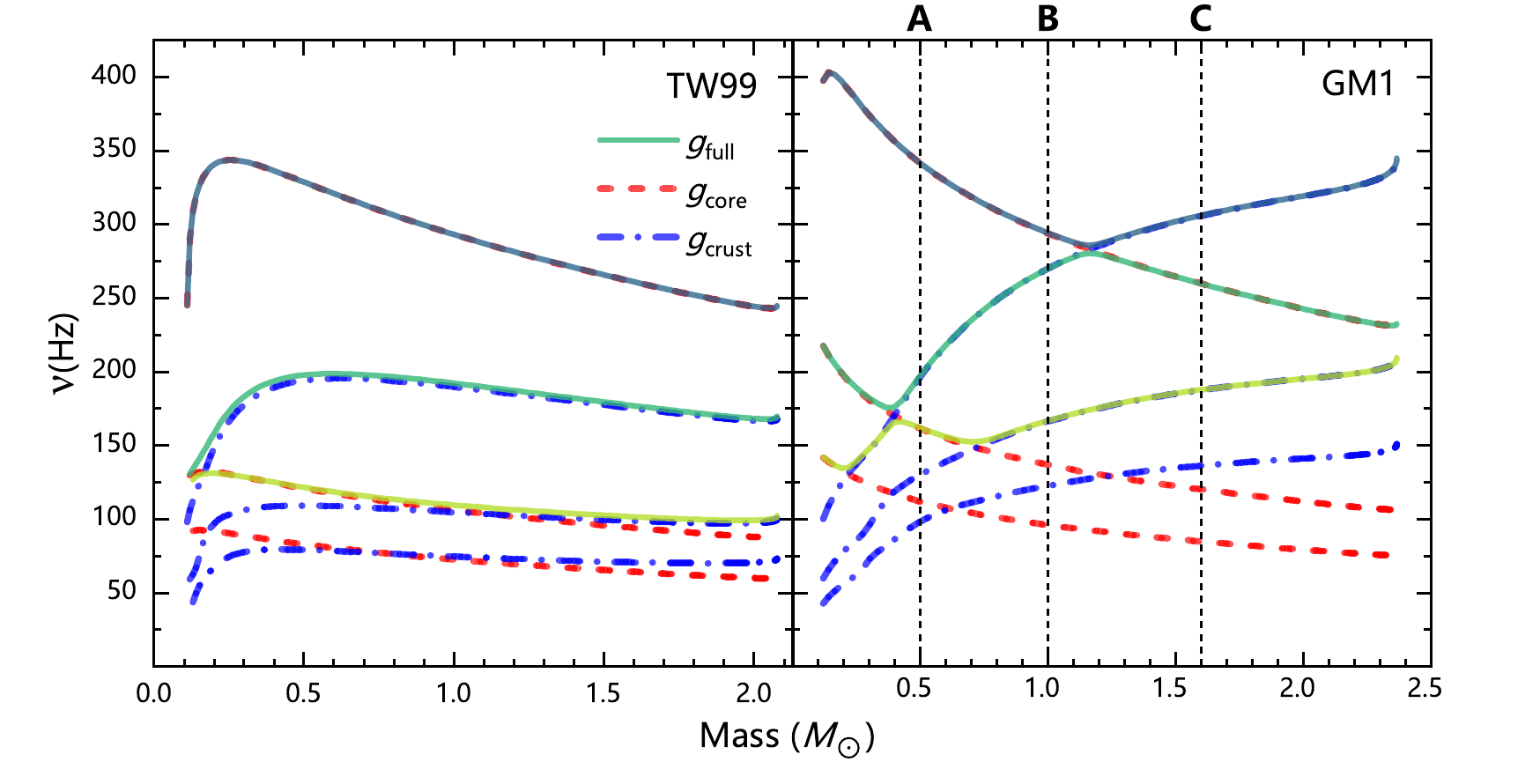}
\caption{\label{Fig:T5} {Frequencies of the first three global $g$-modes (solid), core $g$-modes (red dashed), and crustal $g$-modes (blue dash-dotted) as functions of  neutron star's mass, where the relativistic density functionals TW99 (left) and GM1 (right) are adopted. In the right panel,  three neutron stars with different masses, A (0.5$M_{\odot}$), B (1$M_{\odot}$) and C (1.6$M_{\odot}$), are selected to study their oscillation energy distributions for the global $g_1$, $g_2$ and $g_3$ modes, which are presented in Fig.~\ref{Fig:T6}.} }
\end{figure}%

\begin{figure}
\centering
\includegraphics[width=\linewidth]{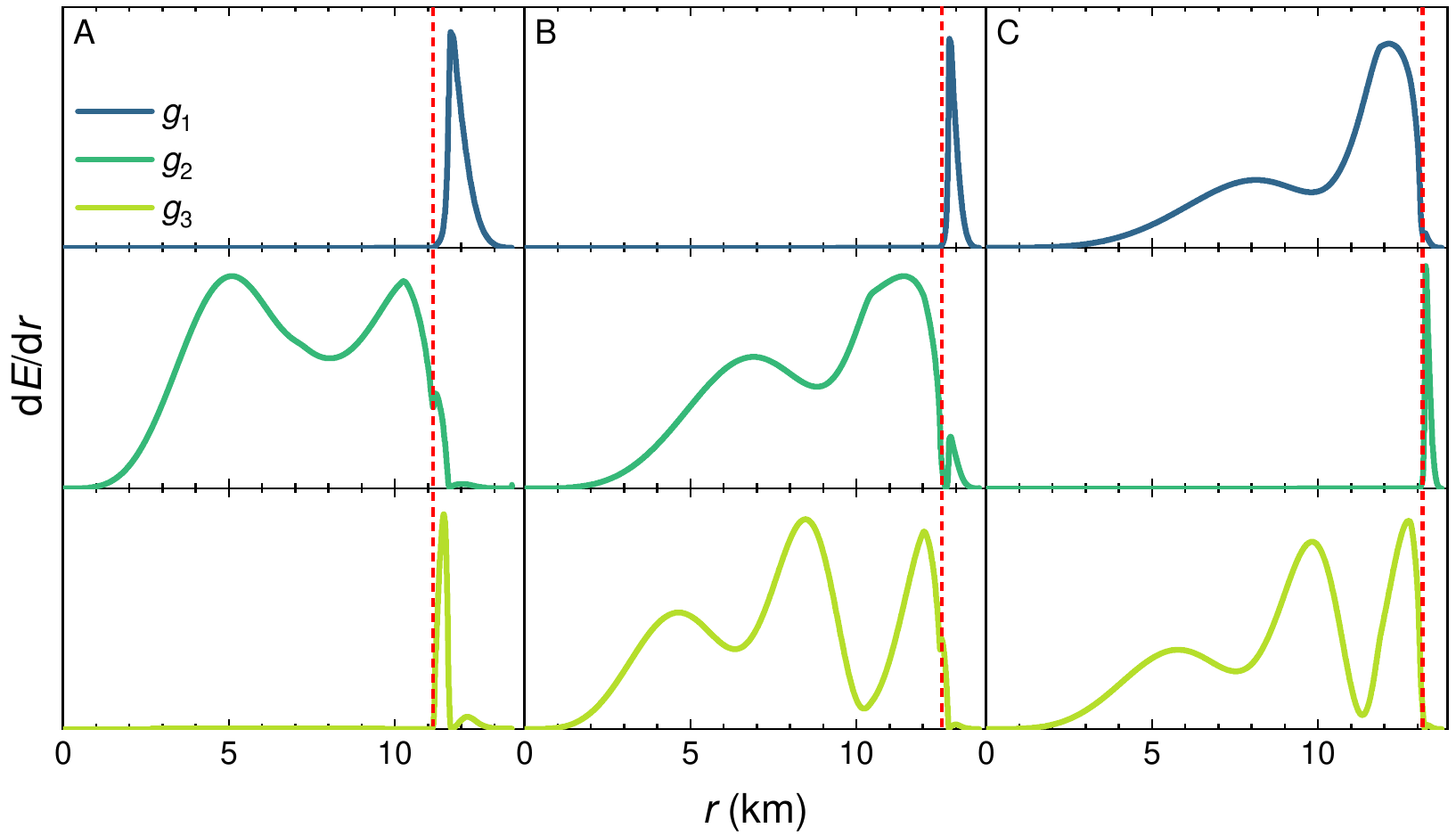}
\caption{\label{Fig:T6}Radial distributions of the oscillation energies for $g_1$, $g_2$ and $g_3$ modes of neutron stars at A, B, and C in the right panel of Fig.~\ref{Fig:T5}. The red dotted lines indicate to the crust-core boundaries.}
\end{figure}

To verify if the avoid-crossing phenomenons in $g$-mode oscillations are attributed to the different oscillatory properties of neutron star's core and crust, we take the density functionals TW99 and GM1 for further study, which represent the density-dependent coupling and nonlinear self-coupling models, respectively. By taking $c_s = c_e$ in the crust or core regions, we can then examine the $g$-mode oscillations in neutron star's core and crust separately, while the squared deviations of the two sound velocities in Fig.~\ref{Fig:T2} are adopted to fix the $g$-mode frequencies. As shown in Fig.~\ref{Fig:T5}, we present the obtained frequencies of the first two $g$ modes ($g_1$ and $g_2$) due to oscillations in neutron stars' crusts (red dashed) and cores (blue dash-dotted), as well as the first three $g$ modes ($g_1$, $g_2$, and $g_3$) arise from the entire star. We can see that these $g$-mode frequencies overlap at certain neutron star masses, which supports the conjecture that the $g$-mode oscillations are dominated by the oscillations in either the core or the crust.

In the right panel of Fig.~\ref{Fig:T5}, the frequencies of the crust $g$ modes decrease with neutron star's mass, while that of the core $g$ mode increase. The intersection of the two $g$ modes forms the so-called avoid-crossing phenomenon from the perspective of the global $g$ mode. This phenomenon was explained by Reisenegger and Goldreich~\cite{Reisenegger1992_Apj395-240-249}, where the core and crust can be interpreted as a pair of weakly coupled resonant cavities, in which the mode of one cavity is barely affected by the existence of the other cavity. However, Reisenegger and Goldreich studied the $g$ mode oscillations in the outer crust of a neutron star, while in this work we focus on the $g$ mode oscillations in the inner crust and core. As shown in Fig.~\ref{Fig:T3}(c), the BV frequencies are concentrated in the inner crust and core. Note that there is also avoid-crossing phenomenon in the left panel of Fig.~\ref{Fig:T5}. The global $g_1$ mode predicted by TW99 is overlapped with the crust $g_1$ mode almost completely, while the global $g_2$ mode is overlapped with the core $g_1$ mode, and the global $g_3$ mode arises due to the coupling between the crust $g_2$ mode and the core $g_2$ mode.

To show this more clearly, we select three neutron stars with different masses at A (0.5$M_{\odot}$), B (1$M_{\odot}$) and C  (1.6$M_{\odot}$) in the right panel of Fig.~\ref{Fig:T5} and examine the distribution of the oscillation energies corresponding to the three $g$ modes. Based on the eigenfunctions $W (r)$ and $V(r)$ obtained by solving Eq.~(\ref{eq:11}) and Eq.~(\ref{eq:12}) under the Cowling approximation, the radial distributions of the energy generated by the $g$-mode oscillations of neutron stars are fixed with~\cite{Reisenegger1992_Apj395-240-249, McDermott1983_ApJ268-837, Zheng2023_PRD107-103048}
\begin{equation}
\frac{\mbox{d} E}{\mbox{d} r}=\frac{\omega^{2}}{2}(p+\varepsilon) \mathrm{e}^{\Lambda-\Phi} \left[\frac{W^{2}}{r^{2}}+\ell(\ell+1) V^{2}\right].
    \label{eq:20}
\end{equation}
The obtained results are then presented in Fig.~\ref{Fig:T6}, where the radial distribution of the internal oscillation energies for $g_1$, $g_2$ and $g_3$ modes of neutron stars at $M=$ 0.5 $M_{\odot}$, 1 $M_{\odot}$ and 1.6 $M_{\odot}$ predicted by  relativistic density functional GM1 are indicated. The vertical red dotted lines indicate the boundary between the inner crust and core of the neutron stars. As we increase the mass of neutron stars, the thickness of the crust decreases. It is evident that when oscillation energy is concentrated in the crust, the $g$-mode oscillation is dominated by the crust, and vice versa for the core. This further confirms the previous conclusion that the core and the inner crust can be interpreted as a pair of weakly coupled resonant cavities, which leads to the avoid-crossing phenomenon in neutron stars' $g$ modes.

\begin{figure}
\includegraphics[width=\linewidth]{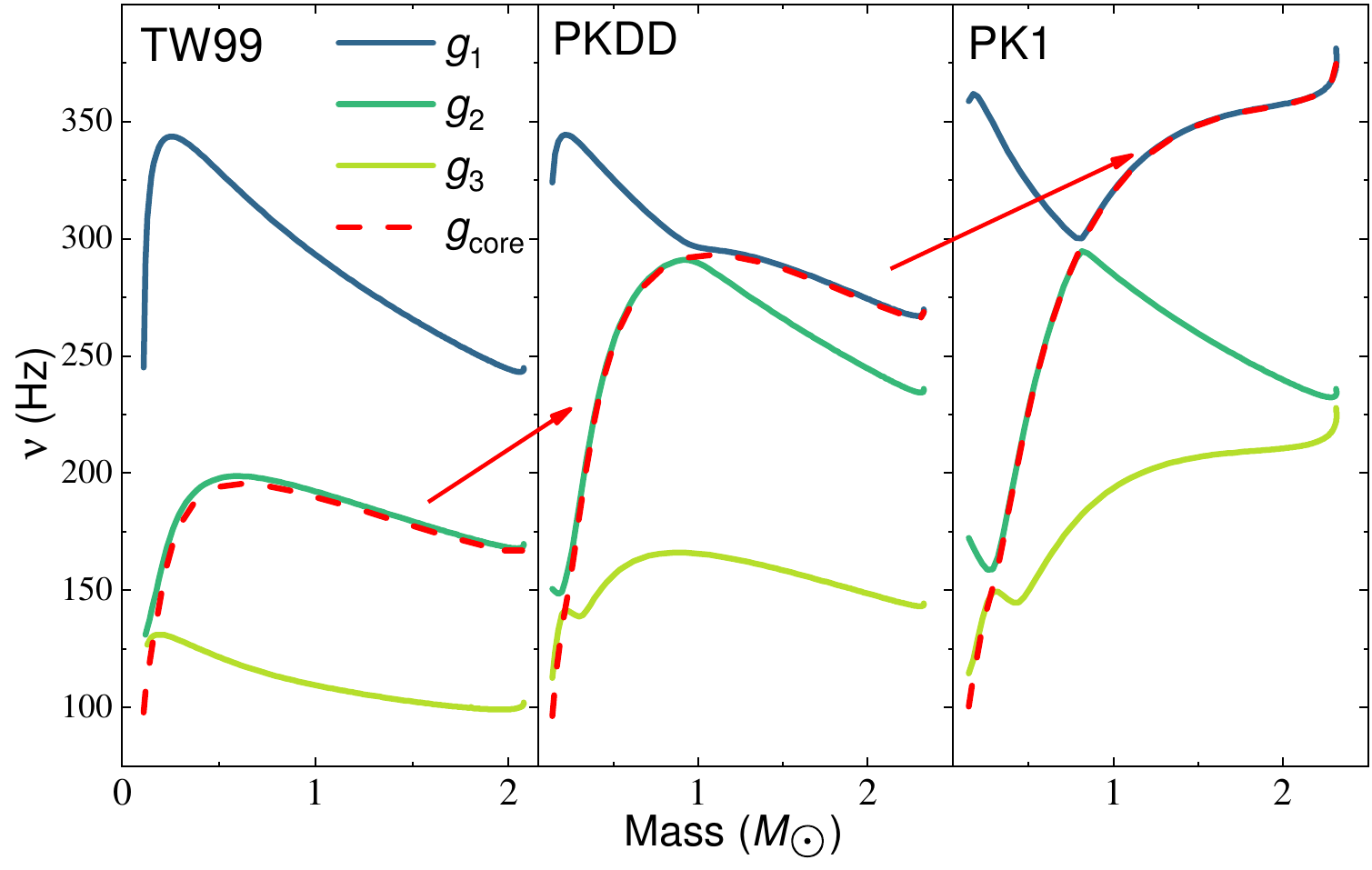}
\caption{\label{Fig:T7} Frequencies of the core $g_1$ mode (red dashed) and the global $g_1$, $g_2$, and $g_3$ modes (solid) predicted by three relativistic density functionals TW99, PKDD, and PK1, where the corresponding slopes of symmetry energy are $L = 55.3$, 90.2, and 115.9 MeV. {The two red arrows indicate the evolution of the core $g_1$ mode as $L$ increases from 55.3 MeV to 115.9 MeV.}}
\end{figure}

\begin{figure}
\includegraphics[width=\linewidth]{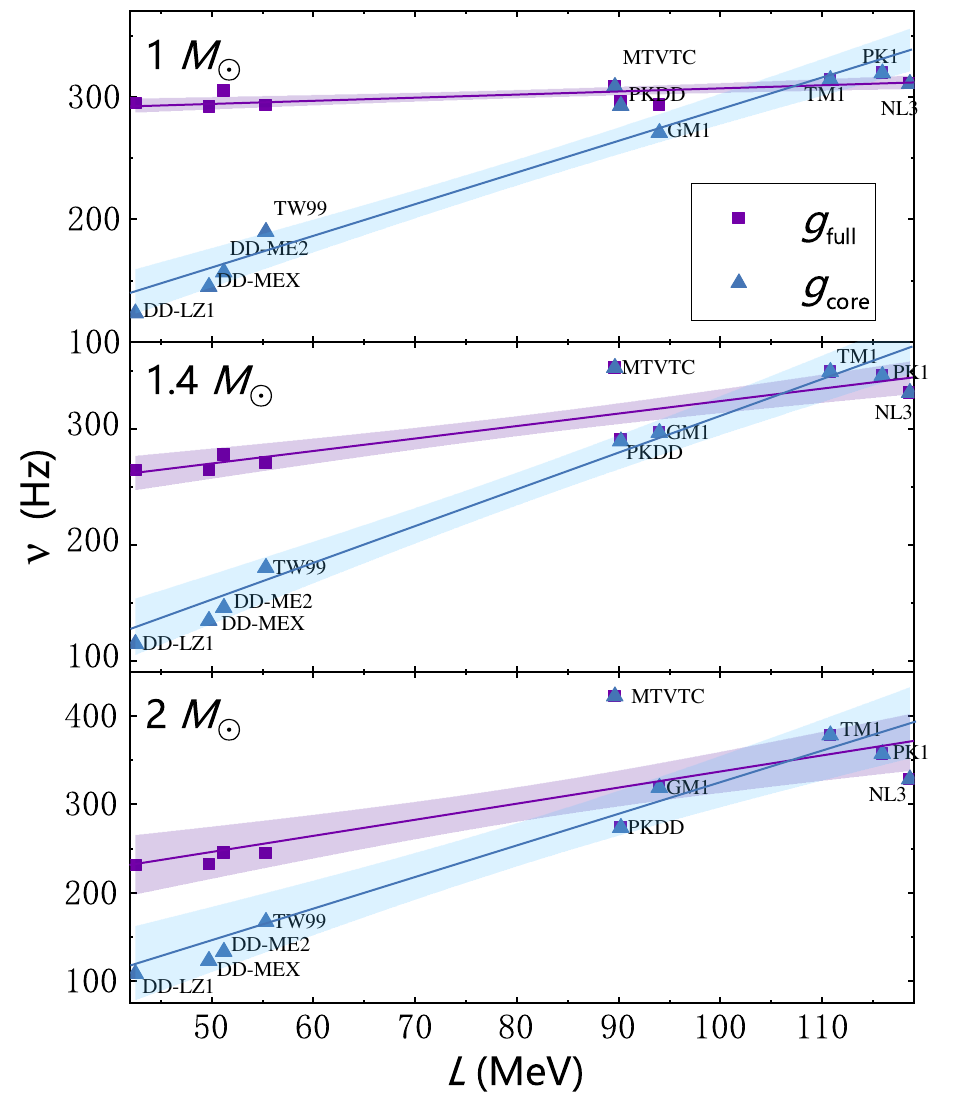}
\caption{\label{Fig:T8} Frequencies of the global and core $g_1$ modes predicted by the 10 density functionals with different slopes of symmetry energy $L$. {The solid lines are linear fits for these discrete points as indicated in Eq.~(\ref{eq:Lfit}) with the coefficients and Pearson correlation coefficients listed in Table~\ref{table:prop2}. The purple and blue regions are 80\% confidence bands for the linear fit of the global and core $g_1$-modes with respect to the slope of symmetric energy $L$.}}
\end{figure}

Finally, we investigate the impact of nuclear matter properties on the $g$-mode oscillations in neutron stars. From Fig.~\ref{Fig:T4} and Fig.~\ref{Fig:T5}, it can be identified that the crust $g$-mode frequencies as functions of neutron star mass share similar trends, while the core $g$-mode frequencies behave differently and are sensitive to the adopted density functionals. To clarify this, as indicated in Fig.~\ref{Fig:T7}, we then examine the $g$-mode frequencies predicted by three relativistic density functionals TW99, PKDD, and PK1. The core $g_1$ mode is indicated by the red dashed curve, while the global $g_1$, $g_2$, and $g_3$ modes are given by the solid curves. It is found that the frequencies of the core $g_1$ mode {for neutron stars with masses $M\gtrsim 1\ M_{\odot}$ generally} increase with the slope of symmetry energy $L$ for nuclear matter, where $L$ increases as we adopt TW99, PKDD, and PK1 sequentially. {In particular, linear correlations between the frequencies of $g$-mode oscillations and $L$ are identified, which will be addressed in Fig.~\ref{Fig:T8}.} Meanwhile, according to Fig.~\ref{Fig:T2}, the square deviations of the sound velocities in neutron stars' cores also increase as we adopt the density functionals with larger $L$. In fact, the restoring force of $g$ mode is originated from the tendency towards chemical equilibrium, so that the corresponding frequency is related to the frequency of local chemical oscillations, which is proportional to the difference between the equilibrium and adiabatic sound velocities~\cite{Zhao2022_PRD105-103025}.

As indicated by the red arrows in Fig.~\ref{Fig:T7}, the frequencies of the core $g_1$ mode increase with $L$. Since the variation on the frequencies of  crust $g$ modes is relatively small when we adopt different relativistic density functionals, the frequencies of the core $g_1$ mode as functions of neutron star mass will eventually intersect with that of the crust $g$ modes for neutron stars at certain masses. The two $g$-mode oscillations are then coupled in the neutron star at masses close to the intersection points, where the avoid-crossing phenomenon takes place. Note that if we adopt the density functionals with small $L$ such as TW99, the square deviations of the sound velocities in neutron stars' cores are small and comparable to that in the crusts, which predicts relatively low frequencies for the core $g$ mode. In such cases, the crust and core $g_1$ modes do not intersect at any neutron star masses, so that the avoid-crossing phenomenon between the crust and core $g_1$ modes never takes place. This well explains the overall difference in the trends of the $g$-mode frequencies predicted by the dense-dependent coupling model and the nonlinear self-coupling model in Fig.~\ref{Fig:T4}, while the PKDD model can be regarded as a transition stage in between.

According to Fig.~\ref{Fig:T2}, the sound velocity differences at the cores of neutron stars have a strong correlation with the slope of symmetric energy $L$, so that the $g$-mode frequency could also have a correlation with $L$, which enables us to constrain the slope of symmetric energy $L$ based on the observations of $g$-mode oscillations in neutron stars. As shown in Fig.~\ref{Fig:T8}, we plot the frequencies of the global and core $g_1$ modes predicted by the 10 density functionals for neutron stars with masses $M=$ 1 $M_{\odot}$, 1.4 $M_{\odot}$ and 2 $M_{\odot}$, where the correlations with respect to $L$ can be identified. Evidently, the frequencies of core $g_1$ mode show a clear linear correlation with $L$. {To quantify this correlation, we define
\begin{equation}
\nu = a L + b. \label{eq:Lfit}
\end{equation}
The coefficients $a$ and $b$ are then fixed by carrying out linear fits for the frequencies of the global and core $g_1$-modes with respect to the slope of symmetric energy $L$, which are then presented in Table~\ref{table:prop2} along with the corresponding Pearson correlation coefficients $R$. The best-fit line and 80\% confidence bands are presented in Fig.~\ref{Fig:T8} as well.}

\begin{table}[h]
\setlength{\tabcolsep}{3.7pt}
{\caption{\label{table:prop2} Best-fit coefficients $a$ and $b$ for the linear correlation function indicated in Eq.~(\ref{eq:Lfit}). The corresponding Pearson correlation coefficient $R$ is also shown, where $R=1$ indicates a perfect positive linear relationship and $R=0$ no linear relationship.}
\begin{tabular}{c|c|cc|c} \hline \hline
         Mass        & type &  $a$   &    $b$    &   $R$  \\
   $M_{\odot}$       &      & Hz/MeV &    Hz     &        \\ \hline
\multirow{2}*{1}     & full & $0.25$ & $282.42$  & $0.72$ \\ \cline{2-5}
                     & core & $2.59$ & $31.22$   & $0.96$ \\ \hline
\multirow{2}*{1.4}   & full & $1.08$ & $216.22$  & $0.87$ \\ \cline{2-5}
                     & core & $3.17$ & $-5.18$   & $0.95$ \\ \hline
\multirow{2}*{2}     & full & $1.82$ & $154.33$  & $0.79$ \\ \cline{2-5}
                     & core & $3.58$ & $-32.03$  & $0.89$ \\ \hline
\end{tabular}}
\end{table}

The impact of neutron star crusts can be identified by comparing the frequencies of the global and core $g_1$ modes, where the frequencies of global $g_1$ modes become larger, especially for the cases adopting the density functionals with small $L$. The correlation between the global $g_1$ modes and $L$ for neutron stars at $M= 1\ M_{\odot}$ thus becomes less evident {with smaller $R$}, where the crust $g_1$ mode dominates with its frequency $\nu_{g1}\approx 300$ Hz. For neutron stars with larger masses, the $g$-mode oscillations in neutron stars' cores start to take effect, where the frequencies of the global $g_1$ modes increase even before the avoid-crossing phenomenon takes place. In such cases, the correlation between the global $g_1$ modes and $L$ remains, e.g., those in the neutron stars with $M= 1.4\ M_{\odot}$ and 2 $M_{\odot}$ as indicated in Fig.~\ref{Fig:T8}. This provides a theoretical basis for constraining the EOSs of neutron stars' cores based on the future observations of $g$-mode oscillations using various gravitational wave detectors~\cite{Thorne1967_ApJ-149-591, Punturo2010_CQG27-194002, Regimbau2017_PRL118-151105, Abbott2017_CQG34-044001, Abbott2020_LRR23-3, Maggiore2020_JCAP2020-050}.

\section{\label{sec:con}conclusion and prospect}

In this work, we study the influence of neutron stars' crusts on the non-radial $g$-mode oscillations, and examine their correlations with nuclear matter properties by adopting 10 different relativistic density functionals, i.e., those with nonlinear self-couplings (NL3~\cite{Lalazissis1997_PRC55-540}, PK1~\cite{Long2004_PRC69-034319}, TM1~\cite{Sugahara1994_NPA579-557}, GM1~\cite{Glendenning1991_PRL67-2414}, MTVTC~\cite{Maruyama2005_PRC72-015802}) and density-dependent couplings (DD-LZ1~\cite{Wei2020_CPC44-074107}, DD-MEX~\cite{Taninah2020_PLB800-135065}, PKDD~\cite{Long2004_PRC69-034319}, DD-ME2~\cite{Lalazissis2005_PRC71-024312}, TW99~\cite{Typel1999_NPA656-331}). Our findings are listed as follows.
\begin{itemize}
  \item The difference between the adiabatic and equilibrium sound velocities $c^2_s-c^2_e$ increases significantly in neutron stars' inner crusts, which arise due to the variations of nuclear species and the emergence of neutron gas outside of the nuclei above neutron drip densities. This will produce relatively strong $g$-mode oscillations in neutron stars' crusts. The obtained frequencies of crust $g_1$ mode generally reach their peak values $\nu_{g1}\approx 350$ Hz at the neutron star mass $M\approx 0.2$ $M_{\odot}$, then decrease with $M$. The frequencies and trends of crust $g$ modes  vary little with respect to the adopted relativistic density functionals.
  \item The sound velocity difference $c^2_s-c^2_e$ in neutron stars' cores arise due to the core-crust transitions and emergence of muons, which is sensitive to the adopted density functionals and increases with the slope of nuclear symmetry energy $L$. Consequently, the core $g$-mode frequencies show a strong correlation with $L$, which increase with $L$ for neutron stars at fixed masses.
  \item The inner crust and core of a neutron star form two weakly coupled resonators. As the slope of nuclear symmetry energy $L$ increases, the frequencies of the core $g$ modes increase, which eventually intersect with the crust $g$ modes. The avoid-crossing phenomenon then takes place for the global $g$ modes that encompass $g$-mode oscillations in both crust and core. Since the crust $g_1$ mode dominates for neutron stars with $M\lesssim 1$ $M_{\odot}$, {the slope $a$ of the linear correlation in Eq.~(\ref{eq:Lfit}) is rather small, which  increases quickly with $M$ as indicated in Table~\ref{table:prop2}.}  The measurement of the slope of nuclear symmetry energy $L$ based on the future observations of $g$-mode oscillations then becomes viable, e.g., from the gravitational waves emitted by binary neutron star mergers~\cite{Pratten2020_NC11-1, Abbott2017_CQG34-044001}.
\end{itemize}
Note that in our current study, the effects of the discontinuities in density~\cite{Finn1987_MNRAS227-265-293, Reisenegger1992_Apj395-240-249} or shear modulus~\cite{McDermott1988_ApJ325-725, Tsang2012_PRL108-011102, Pan2020_PRL125-201102, Zhu2023_PRD107-83023} were neglected, which could lead to the mode-mode interactions. The effects of temperature, rotation, magnetic field, and superfluid neutron gas of the neutron star could also play important roles, which should be considered in our future study.

\section*{ACKNOWLEDGMENTS}
The authors would like to thank Dr. Yong Gao, Dr. Hao-Jui Kuan, and Dr. Zhiqiang Miao for fruitful discussions. {The authors would also like to thank the anonymous referee for providing helpful comments and suggestions.} This work was supported by the National SKA Program of China (Grant No. 2020SKA0120300), the National Natural Science Foundation of China (Grant No. 12275234, No. 12447148) and the China Postdoctoral Science Foundation (2024M760081).


\begin{thebibliography}{117}%
\makeatletter
\providecommand \@ifxundefined [1]{%
 \@ifx{#1\undefined}
}%
\providecommand \@ifnum [1]{%
 \ifnum #1\expandafter \@firstoftwo
 \else \expandafter \@secondoftwo
 \fi
}%
\providecommand \@ifx [1]{%
 \ifx #1\expandafter \@firstoftwo
 \else \expandafter \@secondoftwo
 \fi
}%
\providecommand \natexlab [1]{#1}%
\providecommand \enquote  [1]{``#1''}%
\providecommand \bibnamefont  [1]{#1}%
\providecommand \bibfnamefont [1]{#1}%
\providecommand \citenamefont [1]{#1}%
\providecommand \href@noop [0]{\@secondoftwo}%
\providecommand \href [0]{\begingroup \@sanitize@url \@href}%
\providecommand \@href[1]{\@@startlink{#1}\@@href}%
\providecommand \@@href[1]{\endgroup#1\@@endlink}%
\providecommand \@sanitize@url [0]{\catcode `\\12\catcode `\$12\catcode
  `\&12\catcode `\#12\catcode `\^12\catcode `\_12\catcode `\%12\relax}%
\providecommand \@@startlink[1]{}%
\providecommand \@@endlink[0]{}%
\providecommand \url  [0]{\begingroup\@sanitize@url \@url }%
\providecommand \@url [1]{\endgroup\@href {#1}{\urlprefix }}%
\providecommand \urlprefix  [0]{URL }%
\providecommand \Eprint [0]{\href }%
\providecommand \doibase [0]{http://dx.doi.org/}%
\providecommand \selectlanguage [0]{\@gobble}%
\providecommand \bibinfo  [0]{\@secondoftwo}%
\providecommand \bibfield  [0]{\@secondoftwo}%
\providecommand \translation [1]{[#1]}%
\providecommand \BibitemOpen [0]{}%
\providecommand \bibitemStop [0]{}%
\providecommand \bibitemNoStop [0]{.\EOS\space}%
\providecommand \EOS [0]{\spacefactor3000\relax}%
\providecommand \BibitemShut  [1]{\csname bibitem#1\endcsname}%
\let\auto@bib@innerbib\@empty
\bibitem [{\citenamefont {Hewish}\ \emph {et~al.}(1968)\citenamefont {Hewish},
  \citenamefont {Bell}, \citenamefont {Pilkington}, \citenamefont {Scott},\
  and\ \citenamefont {Collins}}]{Hewish1968_Nature217-709}%
  \BibitemOpen
  \bibfield  {author} {\bibinfo {author} {\bibfnamefont {A.}~\bibnamefont
  {Hewish}}, \bibinfo {author} {\bibfnamefont {S.~J.}\ \bibnamefont {Bell}},
  \bibinfo {author} {\bibfnamefont {J.~D.~H.}\ \bibnamefont {Pilkington}},
  \bibinfo {author} {\bibfnamefont {P.~F.}\ \bibnamefont {Scott}}, \ and\
  \bibinfo {author} {\bibfnamefont {R.~A.}\ \bibnamefont {Collins}},\ }\href
  {http://www.nature.com/nature/journal/v217/n5130/abs/217709a0.html}
  {\bibfield  {journal} {\bibinfo  {journal} {Nature}\ }\textbf {\bibinfo
  {volume} {217}},\ \bibinfo {pages} {709} (\bibinfo {year}
  {1968})}\BibitemShut {NoStop}%
\bibitem [{\citenamefont {Dutra}\ \emph {et~al.}(2012)\citenamefont {Dutra},
  \citenamefont {Louren\ifmmode~\mbox{\c{c}}\else \c{c}\fi{}o}, \citenamefont
  {S\'a~Martins}, \citenamefont {Delfino}, \citenamefont {Stone},\ and\
  \citenamefont {Stevenson}}]{Dutra2012_PRC85-035201}%
  \BibitemOpen
  \bibfield  {author} {\bibinfo {author} {\bibfnamefont {M.}~\bibnamefont
  {Dutra}}, \bibinfo {author} {\bibfnamefont {O.}~\bibnamefont
  {Louren\ifmmode~\mbox{\c{c}}\else \c{c}\fi{}o}}, \bibinfo {author}
  {\bibfnamefont {J.~S.}\ \bibnamefont {S\'a~Martins}}, \bibinfo {author}
  {\bibfnamefont {A.}~\bibnamefont {Delfino}}, \bibinfo {author} {\bibfnamefont
  {J.~R.}\ \bibnamefont {Stone}}, \ and\ \bibinfo {author} {\bibfnamefont
  {P.~D.}\ \bibnamefont {Stevenson}},\ }\href {\doibase
  10.1103/PhysRevC.85.035201} {\bibfield  {journal} {\bibinfo  {journal} {Phys.
  Rev. C}\ }\textbf {\bibinfo {volume} {85}},\ \bibinfo {pages} {035201}
  (\bibinfo {year} {2012})}\BibitemShut {NoStop}%
\bibitem [{\citenamefont {Dutra}\ \emph {et~al.}(2014)\citenamefont {Dutra},
  \citenamefont {Louren\ifmmode~\mbox{\c{c}}\else \c{c}\fi{}o}, \citenamefont
  {Avancini}, \citenamefont {Carlson}, \citenamefont {Delfino}, \citenamefont
  {Menezes}, \citenamefont {Provid\^encia}, \citenamefont {Typel},\ and\
  \citenamefont {Stone}}]{Dutra2014_PRC90-055203}%
  \BibitemOpen
  \bibfield  {author} {\bibinfo {author} {\bibfnamefont {M.}~\bibnamefont
  {Dutra}}, \bibinfo {author} {\bibfnamefont {O.}~\bibnamefont
  {Louren\ifmmode~\mbox{\c{c}}\else \c{c}\fi{}o}}, \bibinfo {author}
  {\bibfnamefont {S.~S.}\ \bibnamefont {Avancini}}, \bibinfo {author}
  {\bibfnamefont {B.~V.}\ \bibnamefont {Carlson}}, \bibinfo {author}
  {\bibfnamefont {A.}~\bibnamefont {Delfino}}, \bibinfo {author} {\bibfnamefont
  {D.~P.}\ \bibnamefont {Menezes}}, \bibinfo {author} {\bibfnamefont
  {C.}~\bibnamefont {Provid\^encia}}, \bibinfo {author} {\bibfnamefont
  {S.}~\bibnamefont {Typel}}, \ and\ \bibinfo {author} {\bibfnamefont {J.~R.}\
  \bibnamefont {Stone}},\ }\href {\doibase 10.1103/PhysRevC.90.055203}
  {\bibfield  {journal} {\bibinfo  {journal} {Phys. Rev. C}\ }\textbf {\bibinfo
  {volume} {90}},\ \bibinfo {pages} {055203} (\bibinfo {year}
  {2014})}\BibitemShut {NoStop}%
\bibitem [{\citenamefont {Baym}\ \emph {et~al.}(2018)\citenamefont {Baym},
  \citenamefont {Hatsuda}, \citenamefont {Kojo}, \citenamefont {Powell},
  \citenamefont {Song},\ and\ \citenamefont
  {Takatsuka}}]{Baym2018_RPP81-056902}%
  \BibitemOpen
  \bibfield  {author} {\bibinfo {author} {\bibfnamefont {G.}~\bibnamefont
  {Baym}}, \bibinfo {author} {\bibfnamefont {T.}~\bibnamefont {Hatsuda}},
  \bibinfo {author} {\bibfnamefont {T.}~\bibnamefont {Kojo}}, \bibinfo {author}
  {\bibfnamefont {P.~D.}\ \bibnamefont {Powell}}, \bibinfo {author}
  {\bibfnamefont {Y.}~\bibnamefont {Song}}, \ and\ \bibinfo {author}
  {\bibfnamefont {T.}~\bibnamefont {Takatsuka}},\ }\href
  {http://stacks.iop.org/0034-4885/81/i=5/a=056902} {\bibfield  {journal}
  {\bibinfo  {journal} {Rep. Prog. Phys.}\ }\textbf {\bibinfo {volume} {81}},\
  \bibinfo {pages} {056902} (\bibinfo {year} {2018})}\BibitemShut {NoStop}%
\bibitem [{\citenamefont {Xia}\ \emph {et~al.}(2020)\citenamefont {Xia},
  \citenamefont {Maruyama}, \citenamefont {Yasutake}, \citenamefont {Tatsumi},
  \citenamefont {Shen},\ and\ \citenamefont {Togashi}}]{Xia2020_PRD102-023031}%
  \BibitemOpen
  \bibfield  {author} {\bibinfo {author} {\bibfnamefont {C.-J.}\ \bibnamefont
  {Xia}}, \bibinfo {author} {\bibfnamefont {T.}~\bibnamefont {Maruyama}},
  \bibinfo {author} {\bibfnamefont {N.}~\bibnamefont {Yasutake}}, \bibinfo
  {author} {\bibfnamefont {T.}~\bibnamefont {Tatsumi}}, \bibinfo {author}
  {\bibfnamefont {H.}~\bibnamefont {Shen}}, \ and\ \bibinfo {author}
  {\bibfnamefont {H.}~\bibnamefont {Togashi}},\ }\href {\doibase
  10.1103/PhysRevD.102.023031} {\bibfield  {journal} {\bibinfo  {journal}
  {Phys. Rev. D}\ }\textbf {\bibinfo {volume} {102}},\ \bibinfo {pages}
  {023031} (\bibinfo {year} {2020})}\BibitemShut {NoStop}%
\bibitem [{\citenamefont {Li}\ \emph {et~al.}(2020)\citenamefont {Li},
  \citenamefont {Zhu}, \citenamefont {Zhou}, \citenamefont {Dong},
  \citenamefont {Hu},\ and\ \citenamefont {Xia}}]{Li2020_JHEA28-19}%
  \BibitemOpen
  \bibfield  {author} {\bibinfo {author} {\bibfnamefont {A.}~\bibnamefont
  {Li}}, \bibinfo {author} {\bibfnamefont {Z.-Y.}\ \bibnamefont {Zhu}},
  \bibinfo {author} {\bibfnamefont {E.-P.}\ \bibnamefont {Zhou}}, \bibinfo
  {author} {\bibfnamefont {J.-M.}\ \bibnamefont {Dong}}, \bibinfo {author}
  {\bibfnamefont {J.-N.}\ \bibnamefont {Hu}}, \ and\ \bibinfo {author}
  {\bibfnamefont {C.-J.}\ \bibnamefont {Xia}},\ }\href {\doibase
  10.1016/j.jheap.2020.07.001} {\bibfield  {journal} {\bibinfo  {journal}
  {JHEAP}\ }\textbf {\bibinfo {volume} {28}},\ \bibinfo {pages} {19} (\bibinfo
  {year} {2020})}\BibitemShut {NoStop}%
\bibitem [{\citenamefont {Xia}\ \emph {et~al.}(2024)\citenamefont {Xia},
  \citenamefont {Xie},\ and\ \citenamefont {Bakhiet}}]{Xia2024_PRD110-114009}%
  \BibitemOpen
  \bibfield  {author} {\bibinfo {author} {\bibfnamefont {C.-J.}\ \bibnamefont
  {Xia}}, \bibinfo {author} {\bibfnamefont {W.-J.}\ \bibnamefont {Xie}}, \ and\
  \bibinfo {author} {\bibfnamefont {M.}~\bibnamefont {Bakhiet}},\ }\href
  {\doibase 10.1103/PhysRevD.110.114009} {\bibfield  {journal} {\bibinfo
  {journal} {Phys. Rev. D}\ }\textbf {\bibinfo {volume} {110}},\ \bibinfo
  {pages} {114009} (\bibinfo {year} {2024})}\BibitemShut {NoStop}%
\bibitem [{\citenamefont {Antoniadis}\ \emph {et~al.}(2013)\citenamefont
  {Antoniadis}, \citenamefont {Freire}, \citenamefont {Wex}, \citenamefont
  {Tauris}, \citenamefont {Lynch}, \citenamefont {van Kerkwijk}, \citenamefont
  {Kramer}, \citenamefont {Bassa}, \citenamefont {Dhillon}, \citenamefont
  {Driebe}, \citenamefont {Hessels}, \citenamefont {Kaspi}, \citenamefont
  {Kondratiev}, \citenamefont {Langer}, \citenamefont {Marsh}, \citenamefont
  {McLaughlin}, \citenamefont {Pennucci}, \citenamefont {Ransom}, \citenamefont
  {Stairs}, \citenamefont {van Leeuwen}, \citenamefont {Verbiest},\ and\
  \citenamefont {Whelan}}]{Antoniadis2013_Science340-1233232}%
  \BibitemOpen
  \bibfield  {author} {\bibinfo {author} {\bibfnamefont {J.}~\bibnamefont
  {Antoniadis}}, \bibinfo {author} {\bibfnamefont {P.~C.~C.}\ \bibnamefont
  {Freire}}, \bibinfo {author} {\bibfnamefont {N.}~\bibnamefont {Wex}},
  \bibinfo {author} {\bibfnamefont {T.~M.}\ \bibnamefont {Tauris}}, \bibinfo
  {author} {\bibfnamefont {R.~S.}\ \bibnamefont {Lynch}}, \bibinfo {author}
  {\bibfnamefont {M.~H.}\ \bibnamefont {van Kerkwijk}}, \bibinfo {author}
  {\bibfnamefont {M.}~\bibnamefont {Kramer}}, \bibinfo {author} {\bibfnamefont
  {C.}~\bibnamefont {Bassa}}, \bibinfo {author} {\bibfnamefont {V.~S.}\
  \bibnamefont {Dhillon}}, \bibinfo {author} {\bibfnamefont {T.}~\bibnamefont
  {Driebe}}, \bibinfo {author} {\bibfnamefont {J.~W.~T.}\ \bibnamefont
  {Hessels}}, \bibinfo {author} {\bibfnamefont {V.~M.}\ \bibnamefont {Kaspi}},
  \bibinfo {author} {\bibfnamefont {V.~I.}\ \bibnamefont {Kondratiev}},
  \bibinfo {author} {\bibfnamefont {N.}~\bibnamefont {Langer}}, \bibinfo
  {author} {\bibfnamefont {T.~R.}\ \bibnamefont {Marsh}}, \bibinfo {author}
  {\bibfnamefont {M.~A.}\ \bibnamefont {McLaughlin}}, \bibinfo {author}
  {\bibfnamefont {T.~T.}\ \bibnamefont {Pennucci}}, \bibinfo {author}
  {\bibfnamefont {S.~M.}\ \bibnamefont {Ransom}}, \bibinfo {author}
  {\bibfnamefont {I.~H.}\ \bibnamefont {Stairs}}, \bibinfo {author}
  {\bibfnamefont {J.}~\bibnamefont {van Leeuwen}}, \bibinfo {author}
  {\bibfnamefont {J.~P.~W.}\ \bibnamefont {Verbiest}}, \ and\ \bibinfo {author}
  {\bibfnamefont {D.~G.}\ \bibnamefont {Whelan}},\ }\href {\doibase
  10.1126/science.1233232} {\bibfield  {journal} {\bibinfo  {journal}
  {Science}\ }\textbf {\bibinfo {volume} {340}},\ \bibinfo {pages} {1233232}
  (\bibinfo {year} {2013})}\BibitemShut {NoStop}%
\bibitem [{\citenamefont {{LIGO Scientific and Virgo
  Collaborations}}(2018)}]{LVC2018_PRL121-161101}%
  \BibitemOpen
  \bibfield  {author} {\bibinfo {author} {\bibnamefont {{LIGO Scientific and
  Virgo Collaborations}}},\ }\href {\doibase 10.1103/PhysRevLett.121.161101}
  {\bibfield  {journal} {\bibinfo  {journal} {Phys. Rev. Lett.}\ }\textbf
  {\bibinfo {volume} {121}},\ \bibinfo {pages} {161101} (\bibinfo {year}
  {2018})}\BibitemShut {NoStop}%
\bibitem [{\citenamefont {Riley}\ \emph {et~al.}(2019)\citenamefont {Riley},
  \citenamefont {Watts}, \citenamefont {Bogdanov}, \citenamefont {Ray},
  \citenamefont {Ludlam}, \citenamefont {Guillot}, \citenamefont {Arzoumanian},
  \citenamefont {Baker}, \citenamefont {Bilous}, \citenamefont {Chakrabarty},
  \citenamefont {Gendreau}, \citenamefont {Harding}, \citenamefont {Ho},
  \citenamefont {Lattimer}, \citenamefont {Morsink},\ and\ \citenamefont
  {Strohmayer}}]{Riley2019_ApJL887-L21}%
  \BibitemOpen
  \bibfield  {author} {\bibinfo {author} {\bibfnamefont {T.~E.}\ \bibnamefont
  {Riley}}, \bibinfo {author} {\bibfnamefont {A.~L.}\ \bibnamefont {Watts}},
  \bibinfo {author} {\bibfnamefont {S.}~\bibnamefont {Bogdanov}}, \bibinfo
  {author} {\bibfnamefont {P.~S.}\ \bibnamefont {Ray}}, \bibinfo {author}
  {\bibfnamefont {R.~M.}\ \bibnamefont {Ludlam}}, \bibinfo {author}
  {\bibfnamefont {S.}~\bibnamefont {Guillot}}, \bibinfo {author} {\bibfnamefont
  {Z.}~\bibnamefont {Arzoumanian}}, \bibinfo {author} {\bibfnamefont {C.~L.}\
  \bibnamefont {Baker}}, \bibinfo {author} {\bibfnamefont {A.~V.}\ \bibnamefont
  {Bilous}}, \bibinfo {author} {\bibfnamefont {D.}~\bibnamefont {Chakrabarty}},
  \bibinfo {author} {\bibfnamefont {K.~C.}\ \bibnamefont {Gendreau}}, \bibinfo
  {author} {\bibfnamefont {A.~K.}\ \bibnamefont {Harding}}, \bibinfo {author}
  {\bibfnamefont {W.~C.~G.}\ \bibnamefont {Ho}}, \bibinfo {author}
  {\bibfnamefont {J.~M.}\ \bibnamefont {Lattimer}}, \bibinfo {author}
  {\bibfnamefont {S.~M.}\ \bibnamefont {Morsink}}, \ and\ \bibinfo {author}
  {\bibfnamefont {T.~E.}\ \bibnamefont {Strohmayer}},\ }\href {\doibase
  10.3847/2041-8213/ab481c} {\bibfield  {journal} {\bibinfo  {journal}
  {Astrophys. J.}\ }\textbf {\bibinfo {volume} {887}},\ \bibinfo {pages} {L21}
  (\bibinfo {year} {2019})}\BibitemShut {NoStop}%
\bibitem [{\citenamefont {Riley}\ \emph {et~al.}(2021)\citenamefont {Riley},
  \citenamefont {Watts}, \citenamefont {Ray}, \citenamefont {Bogdanov},
  \citenamefont {Guillot}, \citenamefont {Morsink}, \citenamefont {Bilous},
  \citenamefont {Arzoumanian}, \citenamefont {Choudhury}, \citenamefont
  {Deneva}, \citenamefont {Gendreau}, \citenamefont {Harding}, \citenamefont
  {Ho}, \citenamefont {Lattimer}, \citenamefont {Loewenstein}, \citenamefont
  {Ludlam}, \citenamefont {Markwardt}, \citenamefont {Okajima}, \citenamefont
  {Prescod-Weinstein}, \citenamefont {Remillard}, \citenamefont {Wolff},
  \citenamefont {Fonseca}, \citenamefont {Cromartie}, \citenamefont {Kerr},
  \citenamefont {Pennucci}, \citenamefont {Parthasarathy}, \citenamefont
  {Ransom}, \citenamefont {Stairs}, \citenamefont {Guillemot},\ and\
  \citenamefont {Cognard}}]{Riley2021_ApJL918-L27}%
  \BibitemOpen
  \bibfield  {author} {\bibinfo {author} {\bibfnamefont {T.~E.}\ \bibnamefont
  {Riley}}, \bibinfo {author} {\bibfnamefont {A.~L.}\ \bibnamefont {Watts}},
  \bibinfo {author} {\bibfnamefont {P.~S.}\ \bibnamefont {Ray}}, \bibinfo
  {author} {\bibfnamefont {S.}~\bibnamefont {Bogdanov}}, \bibinfo {author}
  {\bibfnamefont {S.}~\bibnamefont {Guillot}}, \bibinfo {author} {\bibfnamefont
  {S.~M.}\ \bibnamefont {Morsink}}, \bibinfo {author} {\bibfnamefont {A.~V.}\
  \bibnamefont {Bilous}}, \bibinfo {author} {\bibfnamefont {Z.}~\bibnamefont
  {Arzoumanian}}, \bibinfo {author} {\bibfnamefont {D.}~\bibnamefont
  {Choudhury}}, \bibinfo {author} {\bibfnamefont {J.~S.}\ \bibnamefont
  {Deneva}}, \bibinfo {author} {\bibfnamefont {K.~C.}\ \bibnamefont
  {Gendreau}}, \bibinfo {author} {\bibfnamefont {A.~K.}\ \bibnamefont
  {Harding}}, \bibinfo {author} {\bibfnamefont {W.~C.~G.}\ \bibnamefont {Ho}},
  \bibinfo {author} {\bibfnamefont {J.~M.}\ \bibnamefont {Lattimer}}, \bibinfo
  {author} {\bibfnamefont {M.}~\bibnamefont {Loewenstein}}, \bibinfo {author}
  {\bibfnamefont {R.~M.}\ \bibnamefont {Ludlam}}, \bibinfo {author}
  {\bibfnamefont {C.~B.}\ \bibnamefont {Markwardt}}, \bibinfo {author}
  {\bibfnamefont {T.}~\bibnamefont {Okajima}}, \bibinfo {author} {\bibfnamefont
  {C.}~\bibnamefont {Prescod-Weinstein}}, \bibinfo {author} {\bibfnamefont
  {R.~A.}\ \bibnamefont {Remillard}}, \bibinfo {author} {\bibfnamefont {M.~T.}\
  \bibnamefont {Wolff}}, \bibinfo {author} {\bibfnamefont {E.}~\bibnamefont
  {Fonseca}}, \bibinfo {author} {\bibfnamefont {H.~T.}\ \bibnamefont
  {Cromartie}}, \bibinfo {author} {\bibfnamefont {M.}~\bibnamefont {Kerr}},
  \bibinfo {author} {\bibfnamefont {T.~T.}\ \bibnamefont {Pennucci}}, \bibinfo
  {author} {\bibfnamefont {A.}~\bibnamefont {Parthasarathy}}, \bibinfo {author}
  {\bibfnamefont {S.}~\bibnamefont {Ransom}}, \bibinfo {author} {\bibfnamefont
  {I.}~\bibnamefont {Stairs}}, \bibinfo {author} {\bibfnamefont
  {L.}~\bibnamefont {Guillemot}}, \ and\ \bibinfo {author} {\bibfnamefont
  {I.}~\bibnamefont {Cognard}},\ }\href {\doibase 10.3847/2041-8213/ac0a81}
  {\bibfield  {journal} {\bibinfo  {journal} {Astrophys. J.}\ }\textbf
  {\bibinfo {volume} {918}},\ \bibinfo {pages} {L27} (\bibinfo {year}
  {2021})}\BibitemShut {NoStop}%
\bibitem [{\citenamefont {Miller}\ \emph {et~al.}(2019)\citenamefont {Miller},
  \citenamefont {Lamb}, \citenamefont {Dittmann}, \citenamefont {Bogdanov},
  \citenamefont {Arzoumanian}, \citenamefont {Gendreau}, \citenamefont
  {Guillot}, \citenamefont {Harding}, \citenamefont {Ho}, \citenamefont
  {Lattimer}, \citenamefont {Ludlam}, \citenamefont {Mahmoodifar},
  \citenamefont {Morsink}, \citenamefont {Ray}, \citenamefont {Strohmayer},
  \citenamefont {Wood}, \citenamefont {Enoto}, \citenamefont {Foster},
  \citenamefont {Okajima}, \citenamefont {Prigozhin},\ and\ \citenamefont
  {Soong}}]{Miller2019_ApJL887-L24}%
  \BibitemOpen
  \bibfield  {author} {\bibinfo {author} {\bibfnamefont {M.~C.}\ \bibnamefont
  {Miller}}, \bibinfo {author} {\bibfnamefont {F.~K.}\ \bibnamefont {Lamb}},
  \bibinfo {author} {\bibfnamefont {A.~J.}\ \bibnamefont {Dittmann}}, \bibinfo
  {author} {\bibfnamefont {S.}~\bibnamefont {Bogdanov}}, \bibinfo {author}
  {\bibfnamefont {Z.}~\bibnamefont {Arzoumanian}}, \bibinfo {author}
  {\bibfnamefont {K.~C.}\ \bibnamefont {Gendreau}}, \bibinfo {author}
  {\bibfnamefont {S.}~\bibnamefont {Guillot}}, \bibinfo {author} {\bibfnamefont
  {A.~K.}\ \bibnamefont {Harding}}, \bibinfo {author} {\bibfnamefont
  {W.~C.~G.}\ \bibnamefont {Ho}}, \bibinfo {author} {\bibfnamefont {J.~M.}\
  \bibnamefont {Lattimer}}, \bibinfo {author} {\bibfnamefont {R.~M.}\
  \bibnamefont {Ludlam}}, \bibinfo {author} {\bibfnamefont {S.}~\bibnamefont
  {Mahmoodifar}}, \bibinfo {author} {\bibfnamefont {S.~M.}\ \bibnamefont
  {Morsink}}, \bibinfo {author} {\bibfnamefont {P.~S.}\ \bibnamefont {Ray}},
  \bibinfo {author} {\bibfnamefont {T.~E.}\ \bibnamefont {Strohmayer}},
  \bibinfo {author} {\bibfnamefont {K.~S.}\ \bibnamefont {Wood}}, \bibinfo
  {author} {\bibfnamefont {T.}~\bibnamefont {Enoto}}, \bibinfo {author}
  {\bibfnamefont {R.}~\bibnamefont {Foster}}, \bibinfo {author} {\bibfnamefont
  {T.}~\bibnamefont {Okajima}}, \bibinfo {author} {\bibfnamefont
  {G.}~\bibnamefont {Prigozhin}}, \ and\ \bibinfo {author} {\bibfnamefont
  {Y.}~\bibnamefont {Soong}},\ }\href {\doibase 10.3847/2041-8213/ab50c5}
  {\bibfield  {journal} {\bibinfo  {journal} {Astrophys. J.}\ }\textbf
  {\bibinfo {volume} {887}},\ \bibinfo {pages} {L24} (\bibinfo {year}
  {2019})}\BibitemShut {NoStop}%
\bibitem [{\citenamefont {Miller}\ \emph {et~al.}(2021)\citenamefont {Miller},
  \citenamefont {Lamb}, \citenamefont {Dittmann}, \citenamefont {Bogdanov},
  \citenamefont {Arzoumanian}, \citenamefont {Gendreau}, \citenamefont
  {Guillot}, \citenamefont {Ho}, \citenamefont {Lattimer}, \citenamefont
  {Loewenstein}, \citenamefont {Morsink}, \citenamefont {Ray}, \citenamefont
  {Wolff}, \citenamefont {Baker}, \citenamefont {Cazeau}, \citenamefont
  {Manthripragada}, \citenamefont {Markwardt}, \citenamefont {Okajima},
  \citenamefont {Pollard}, \citenamefont {Cognard}, \citenamefont {Cromartie},
  \citenamefont {Fonseca}, \citenamefont {Guillemot}, \citenamefont {Kerr},
  \citenamefont {Parthasarathy}, \citenamefont {Pennucci}, \citenamefont
  {Ransom},\ and\ \citenamefont {Stairs}}]{Miller2021_ApJL918-L28}%
  \BibitemOpen
  \bibfield  {author} {\bibinfo {author} {\bibfnamefont {M.~C.}\ \bibnamefont
  {Miller}}, \bibinfo {author} {\bibfnamefont {F.~K.}\ \bibnamefont {Lamb}},
  \bibinfo {author} {\bibfnamefont {A.~J.}\ \bibnamefont {Dittmann}}, \bibinfo
  {author} {\bibfnamefont {S.}~\bibnamefont {Bogdanov}}, \bibinfo {author}
  {\bibfnamefont {Z.}~\bibnamefont {Arzoumanian}}, \bibinfo {author}
  {\bibfnamefont {K.~C.}\ \bibnamefont {Gendreau}}, \bibinfo {author}
  {\bibfnamefont {S.}~\bibnamefont {Guillot}}, \bibinfo {author} {\bibfnamefont
  {W.~C.~G.}\ \bibnamefont {Ho}}, \bibinfo {author} {\bibfnamefont {J.~M.}\
  \bibnamefont {Lattimer}}, \bibinfo {author} {\bibfnamefont {M.}~\bibnamefont
  {Loewenstein}}, \bibinfo {author} {\bibfnamefont {S.~M.}\ \bibnamefont
  {Morsink}}, \bibinfo {author} {\bibfnamefont {P.~S.}\ \bibnamefont {Ray}},
  \bibinfo {author} {\bibfnamefont {M.~T.}\ \bibnamefont {Wolff}}, \bibinfo
  {author} {\bibfnamefont {C.~L.}\ \bibnamefont {Baker}}, \bibinfo {author}
  {\bibfnamefont {T.}~\bibnamefont {Cazeau}}, \bibinfo {author} {\bibfnamefont
  {S.}~\bibnamefont {Manthripragada}}, \bibinfo {author} {\bibfnamefont
  {C.~B.}\ \bibnamefont {Markwardt}}, \bibinfo {author} {\bibfnamefont
  {T.}~\bibnamefont {Okajima}}, \bibinfo {author} {\bibfnamefont
  {S.}~\bibnamefont {Pollard}}, \bibinfo {author} {\bibfnamefont
  {I.}~\bibnamefont {Cognard}}, \bibinfo {author} {\bibfnamefont {H.~T.}\
  \bibnamefont {Cromartie}}, \bibinfo {author} {\bibfnamefont {E.}~\bibnamefont
  {Fonseca}}, \bibinfo {author} {\bibfnamefont {L.}~\bibnamefont {Guillemot}},
  \bibinfo {author} {\bibfnamefont {M.}~\bibnamefont {Kerr}}, \bibinfo {author}
  {\bibfnamefont {A.}~\bibnamefont {Parthasarathy}}, \bibinfo {author}
  {\bibfnamefont {T.~T.}\ \bibnamefont {Pennucci}}, \bibinfo {author}
  {\bibfnamefont {S.}~\bibnamefont {Ransom}}, \ and\ \bibinfo {author}
  {\bibfnamefont {I.}~\bibnamefont {Stairs}},\ }\href {\doibase
  10.3847/2041-8213/ac089b} {\bibfield  {journal} {\bibinfo  {journal}
  {Astrophys. J.}\ }\textbf {\bibinfo {volume} {918}},\ \bibinfo {pages} {L28}
  (\bibinfo {year} {2021})}\BibitemShut {NoStop}%
\bibitem [{\citenamefont {Choudhury}\ \emph {et~al.}(2024)\citenamefont
  {Choudhury}, \citenamefont {Salmi}, \citenamefont {Vinciguerra},
  \citenamefont {Riley}, \citenamefont {Kini}, \citenamefont {Watts},
  \citenamefont {Dorsman}, \citenamefont {Bogdanov}, \citenamefont {Guillot},
  \citenamefont {Ray}, \citenamefont {Reardon}, \citenamefont {Remillard},
  \citenamefont {Bilous}, \citenamefont {Huppenkothen}, \citenamefont
  {Lattimer}, \citenamefont {Rutherford}, \citenamefont {Arzoumanian},
  \citenamefont {Gendreau}, \citenamefont {Morsink},\ and\ \citenamefont
  {Ho}}]{Choudhury2024_ApJ971-L20}%
  \BibitemOpen
  \bibfield  {author} {\bibinfo {author} {\bibfnamefont {D.}~\bibnamefont
  {Choudhury}}, \bibinfo {author} {\bibfnamefont {T.}~\bibnamefont {Salmi}},
  \bibinfo {author} {\bibfnamefont {S.}~\bibnamefont {Vinciguerra}}, \bibinfo
  {author} {\bibfnamefont {T.~E.}\ \bibnamefont {Riley}}, \bibinfo {author}
  {\bibfnamefont {Y.}~\bibnamefont {Kini}}, \bibinfo {author} {\bibfnamefont
  {A.~L.}\ \bibnamefont {Watts}}, \bibinfo {author} {\bibfnamefont
  {B.}~\bibnamefont {Dorsman}}, \bibinfo {author} {\bibfnamefont
  {S.}~\bibnamefont {Bogdanov}}, \bibinfo {author} {\bibfnamefont
  {S.}~\bibnamefont {Guillot}}, \bibinfo {author} {\bibfnamefont {P.~S.}\
  \bibnamefont {Ray}}, \bibinfo {author} {\bibfnamefont {D.~J.}\ \bibnamefont
  {Reardon}}, \bibinfo {author} {\bibfnamefont {R.~A.}\ \bibnamefont
  {Remillard}}, \bibinfo {author} {\bibfnamefont {A.~V.}\ \bibnamefont
  {Bilous}}, \bibinfo {author} {\bibfnamefont {D.}~\bibnamefont
  {Huppenkothen}}, \bibinfo {author} {\bibfnamefont {J.~M.}\ \bibnamefont
  {Lattimer}}, \bibinfo {author} {\bibfnamefont {N.}~\bibnamefont
  {Rutherford}}, \bibinfo {author} {\bibfnamefont {Z.}~\bibnamefont
  {Arzoumanian}}, \bibinfo {author} {\bibfnamefont {K.~C.}\ \bibnamefont
  {Gendreau}}, \bibinfo {author} {\bibfnamefont {S.~M.}\ \bibnamefont
  {Morsink}}, \ and\ \bibinfo {author} {\bibfnamefont {W.~C.~G.}\ \bibnamefont
  {Ho}},\ }\href {\doibase 10.3847/2041-8213/ad5a6f} {\bibfield  {journal}
  {\bibinfo  {journal} {Astrophys. J. Lett.}\ }\textbf {\bibinfo {volume}
  {971}},\ \bibinfo {pages} {L20} (\bibinfo {year} {2024})}\BibitemShut
  {NoStop}%
\bibitem [{\citenamefont {Sagun}\ \emph {et~al.}(2020)\citenamefont {Sagun},
  \citenamefont {Panotopoulos},\ and\ \citenamefont
  {Lopes}}]{Sagun2020_PRD101-063025}%
  \BibitemOpen
  \bibfield  {author} {\bibinfo {author} {\bibfnamefont {V.}~\bibnamefont
  {Sagun}}, \bibinfo {author} {\bibfnamefont {G.}~\bibnamefont {Panotopoulos}},
  \ and\ \bibinfo {author} {\bibfnamefont {I.}~\bibnamefont {Lopes}},\ }\href
  {\doibase 10.1103/PhysRevD.101.063025} {\bibfield  {journal} {\bibinfo
  {journal} {Phys. Rev. D}\ }\textbf {\bibinfo {volume} {101}},\ \bibinfo
  {pages} {063025} (\bibinfo {year} {2020})}\BibitemShut {NoStop}%
\bibitem [{\citenamefont {Aerts}(2021)}]{Aerts2021_RMP93-015001}%
  \BibitemOpen
  \bibfield  {author} {\bibinfo {author} {\bibfnamefont {C.}~\bibnamefont
  {Aerts}},\ }\href {\doibase 10.1103/RevModPhys.93.015001} {\bibfield
  {journal} {\bibinfo  {journal} {Rev. Mod. Phys.}\ }\textbf {\bibinfo {volume}
  {93}},\ \bibinfo {pages} {015001} (\bibinfo {year} {2021})}\BibitemShut
  {NoStop}%
\bibitem [{\citenamefont {{Andersson}}(2021)}]{Andersson2021_Universe7-97A}%
  \BibitemOpen
  \bibfield  {author} {\bibinfo {author} {\bibfnamefont {N.}~\bibnamefont
  {{Andersson}}},\ }\href {\doibase 10.3390/universe7040097} {\bibfield
  {journal} {\bibinfo  {journal} {Universe}\ }\textbf {\bibinfo {volume} {7}},\
  \bibinfo {eid} {97} (\bibinfo {year} {2021})}\BibitemShut {NoStop}%
\bibitem [{\citenamefont {Zhu}\ \emph {et~al.}(2023)\citenamefont {Zhu},
  \citenamefont {Wang}, \citenamefont {Xia}, \citenamefont {Zhou},\ and\
  \citenamefont {Ma}}]{Zhu2023_PRD107-83023}%
  \BibitemOpen
  \bibfield  {author} {\bibinfo {author} {\bibfnamefont {J.}~\bibnamefont
  {Zhu}}, \bibinfo {author} {\bibfnamefont {C.}~\bibnamefont {Wang}}, \bibinfo
  {author} {\bibfnamefont {C.}~\bibnamefont {Xia}}, \bibinfo {author}
  {\bibfnamefont {E.}~\bibnamefont {Zhou}}, \ and\ \bibinfo {author}
  {\bibfnamefont {Y.}~\bibnamefont {Ma}},\ }\href {\doibase
  10.1103/PhysRevD.107.083023} {\bibfield  {journal} {\bibinfo  {journal}
  {Phys. Rev. D}\ }\textbf {\bibinfo {volume} {107}},\ \bibinfo {pages}
  {083023} (\bibinfo {year} {2023})}\BibitemShut {NoStop}%
\bibitem [{\citenamefont {Lu}\ \emph {et~al.}(2024)\citenamefont {Lu},
  \citenamefont {Gao}, \citenamefont {Hu}, \citenamefont {Lai}, \citenamefont
  {Li}, \citenamefont {Lu}, \citenamefont {Shao}, \citenamefont {Wang},
  \citenamefont {Wang}, \citenamefont {Wang}, \citenamefont {Xia},
  \citenamefont {Xu}, \citenamefont {Xu}, \citenamefont {Xu}, \citenamefont
  {Yue}, \citenamefont {Zhao}, \citenamefont {Zheng}, \citenamefont {Zhou},\
  and\ \citenamefont {Zou}}]{LU2024_SCPMA54-289501}%
  \BibitemOpen
  \bibfield  {author} {\bibinfo {author} {\bibfnamefont {R.}~\bibnamefont
  {Lu}}, \bibinfo {author} {\bibfnamefont {Y.}~\bibnamefont {Gao}}, \bibinfo
  {author} {\bibfnamefont {Y.}~\bibnamefont {Hu}}, \bibinfo {author}
  {\bibfnamefont {X.}~\bibnamefont {Lai}}, \bibinfo {author} {\bibfnamefont
  {H.}~\bibnamefont {Li}}, \bibinfo {author} {\bibfnamefont {J.}~\bibnamefont
  {Lu}}, \bibinfo {author} {\bibfnamefont {L.}~\bibnamefont {Shao}}, \bibinfo
  {author} {\bibfnamefont {P.}~\bibnamefont {Wang}}, \bibinfo {author}
  {\bibfnamefont {W.}~\bibnamefont {Wang}}, \bibinfo {author} {\bibfnamefont
  {W.}~\bibnamefont {Wang}}, \bibinfo {author} {\bibfnamefont {C.}~\bibnamefont
  {Xia}}, \bibinfo {author} {\bibfnamefont {H.}~\bibnamefont {Xu}}, \bibinfo
  {author} {\bibfnamefont {R.}~\bibnamefont {Xu}}, \bibinfo {author}
  {\bibfnamefont {S.}~\bibnamefont {Xu}}, \bibinfo {author} {\bibfnamefont
  {H.}~\bibnamefont {Yue}}, \bibinfo {author} {\bibfnamefont {L.}~\bibnamefont
  {Zhao}}, \bibinfo {author} {\bibfnamefont {X.}~\bibnamefont {Zheng}},
  \bibinfo {author} {\bibfnamefont {E.}~\bibnamefont {Zhou}}, \ and\ \bibinfo
  {author} {\bibfnamefont {Y.}~\bibnamefont {Zou}},\ }\href {\doibase
  https://doi.org/10.1360/SSPMA-2023-0424} {\bibfield  {journal} {\bibinfo
  {journal} {Sci. China Phys. Mech. Astron.}\ }\textbf {\bibinfo {volume}
  {54}},\ \bibinfo {pages} {289501} (\bibinfo {year} {2024})}\BibitemShut
  {NoStop}%
\bibitem [{\citenamefont {Zhen}\ \emph {et~al.}(2024)\citenamefont {Zhen},
  \citenamefont {Sun}, \citenamefont {Wei}, \citenamefont {Zheng},\ and\
  \citenamefont {Chen}}]{Zhen2024_Symmetry16-2073-8994}%
  \BibitemOpen
  \bibfield  {author} {\bibinfo {author} {\bibfnamefont {Y.}~\bibnamefont
  {Zhen}}, \bibinfo {author} {\bibfnamefont {T.-T.}\ \bibnamefont {Sun}},
  \bibinfo {author} {\bibfnamefont {J.-B.}\ \bibnamefont {Wei}}, \bibinfo
  {author} {\bibfnamefont {Z.-Y.}\ \bibnamefont {Zheng}}, \ and\ \bibinfo
  {author} {\bibfnamefont {H.}~\bibnamefont {Chen}},\ }\href {\doibase
  10.3390/sym16070807} {\bibfield  {journal} {\bibinfo  {journal} {Symmetry}\
  }\textbf {\bibinfo {volume} {16}},\ \bibinfo {pages} {807} (\bibinfo {year}
  {2024})}\BibitemShut {NoStop}%
\bibitem [{\citenamefont {Zhang}\ \emph {et~al.}(2024)\citenamefont {Zhang},
  \citenamefont {Luo}, \citenamefont {Li}, \citenamefont {Shao},\ and\
  \citenamefont {Xu}}]{Zhang2024_PRD109-063020}%
  \BibitemOpen
  \bibfield  {author} {\bibinfo {author} {\bibfnamefont {C.}~\bibnamefont
  {Zhang}}, \bibinfo {author} {\bibfnamefont {Y.}~\bibnamefont {Luo}}, \bibinfo
  {author} {\bibfnamefont {H.-B.}\ \bibnamefont {Li}}, \bibinfo {author}
  {\bibfnamefont {L.}~\bibnamefont {Shao}}, \ and\ \bibinfo {author}
  {\bibfnamefont {R.}~\bibnamefont {Xu}},\ }\href {\doibase
  10.1103/PhysRevD.109.063020} {\bibfield  {journal} {\bibinfo  {journal}
  {Phys. Rev. D}\ }\textbf {\bibinfo {volume} {109}},\ \bibinfo {pages}
  {063020} (\bibinfo {year} {2024})}\BibitemShut {NoStop}%
\bibitem [{\citenamefont {Sen}\ \emph {et~al.}(2023)\citenamefont {Sen},
  \citenamefont {Kumar}, \citenamefont {Kunjipurayil}, \citenamefont
  {Routaray}, \citenamefont {Ghosh}, \citenamefont {Kalita}, \citenamefont
  {Zhao},\ and\ \citenamefont {Kumar}}]{Sen2023_Galaxies11-2075-4434}%
  \BibitemOpen
  \bibfield  {author} {\bibinfo {author} {\bibfnamefont {S.}~\bibnamefont
  {Sen}}, \bibinfo {author} {\bibfnamefont {S.}~\bibnamefont {Kumar}}, \bibinfo
  {author} {\bibfnamefont {A.}~\bibnamefont {Kunjipurayil}}, \bibinfo {author}
  {\bibfnamefont {P.}~\bibnamefont {Routaray}}, \bibinfo {author}
  {\bibfnamefont {S.}~\bibnamefont {Ghosh}}, \bibinfo {author} {\bibfnamefont
  {P.~J.}\ \bibnamefont {Kalita}}, \bibinfo {author} {\bibfnamefont
  {T.}~\bibnamefont {Zhao}}, \ and\ \bibinfo {author} {\bibfnamefont
  {B.}~\bibnamefont {Kumar}},\ }\href {\doibase 10.3390/galaxies11020060}
  {\bibfield  {journal} {\bibinfo  {journal} {Galaxies}\ }\textbf {\bibinfo
  {volume} {11}},\ \bibinfo {pages} {60} (\bibinfo {year} {2023})}\BibitemShut
  {NoStop}%
\bibitem [{\citenamefont {Li}\ \emph {et~al.}(2023)\citenamefont {Li},
  \citenamefont {Gao}, \citenamefont {Shao},\ and\ \citenamefont
  {Xu}}]{Li2023_PRD108-064005}%
  \BibitemOpen
  \bibfield  {author} {\bibinfo {author} {\bibfnamefont {H.-B.}\ \bibnamefont
  {Li}}, \bibinfo {author} {\bibfnamefont {Y.}~\bibnamefont {Gao}}, \bibinfo
  {author} {\bibfnamefont {L.}~\bibnamefont {Shao}}, \ and\ \bibinfo {author}
  {\bibfnamefont {R.-X.}\ \bibnamefont {Xu}},\ }\href {\doibase
  10.1103/PhysRevD.108.064005} {\bibfield  {journal} {\bibinfo  {journal}
  {Phys. Rev. D}\ }\textbf {\bibinfo {volume} {108}},\ \bibinfo {pages}
  {064005} (\bibinfo {year} {2023})}\BibitemShut {NoStop}%
\bibitem [{\citenamefont {Zheng}\ \emph {et~al.}(2023)\citenamefont {Zheng},
  \citenamefont {Sun}, \citenamefont {Chen}, \citenamefont {Wei}, \citenamefont
  {Burgio},\ and\ \citenamefont {Schulze}}]{Zheng2023_PRD107-103048}%
  \BibitemOpen
  \bibfield  {author} {\bibinfo {author} {\bibfnamefont {Z.-Y.}\ \bibnamefont
  {Zheng}}, \bibinfo {author} {\bibfnamefont {T.-T.}\ \bibnamefont {Sun}},
  \bibinfo {author} {\bibfnamefont {H.}~\bibnamefont {Chen}}, \bibinfo {author}
  {\bibfnamefont {J.-B.}\ \bibnamefont {Wei}}, \bibinfo {author} {\bibfnamefont
  {G.~F.}\ \bibnamefont {Burgio}}, \ and\ \bibinfo {author} {\bibfnamefont
  {H.-J.}\ \bibnamefont {Schulze}},\ }\href {\doibase
  10.1103/PhysRevD.107.103048} {\bibfield  {journal} {\bibinfo  {journal}
  {Phys. Rev. D}\ }\textbf {\bibinfo {volume} {107}},\ \bibinfo {pages}
  {103048} (\bibinfo {year} {2023})}\BibitemShut {NoStop}%
\bibitem [{\citenamefont {Zhao}\ and\ \citenamefont
  {Lattimer}(2022)}]{Zhao2022_PRD106-123002}%
  \BibitemOpen
  \bibfield  {author} {\bibinfo {author} {\bibfnamefont {T.}~\bibnamefont
  {Zhao}}\ and\ \bibinfo {author} {\bibfnamefont {J.~M.}\ \bibnamefont
  {Lattimer}},\ }\href {\doibase 10.1103/PhysRevD.106.123002} {\bibfield
  {journal} {\bibinfo  {journal} {Phys. Rev. D}\ }\textbf {\bibinfo {volume}
  {106}},\ \bibinfo {pages} {123002} (\bibinfo {year} {2022})}\BibitemShut
  {NoStop}%
\bibitem [{\citenamefont {Constantinou}\ \emph {et~al.}(2021)\citenamefont
  {Constantinou}, \citenamefont {Han}, \citenamefont {Jaikumar},\ and\
  \citenamefont {Prakash}}]{Constantinou2021_PRD104-123032}%
  \BibitemOpen
  \bibfield  {author} {\bibinfo {author} {\bibfnamefont {C.}~\bibnamefont
  {Constantinou}}, \bibinfo {author} {\bibfnamefont {S.}~\bibnamefont {Han}},
  \bibinfo {author} {\bibfnamefont {P.}~\bibnamefont {Jaikumar}}, \ and\
  \bibinfo {author} {\bibfnamefont {M.}~\bibnamefont {Prakash}},\ }\href
  {\doibase 10.1103/PhysRevD.104.123032} {\bibfield  {journal} {\bibinfo
  {journal} {Phys. Rev. D}\ }\textbf {\bibinfo {volume} {104}},\ \bibinfo
  {pages} {123032} (\bibinfo {year} {2021})}\BibitemShut {NoStop}%
\bibitem [{\citenamefont {Kokkotas}\ and\ \citenamefont
  {Ruoff}(2001)}]{Kokkotas2000_AAP366-565}%
  \BibitemOpen
  \bibfield  {author} {\bibinfo {author} {\bibfnamefont {K.~D.}\ \bibnamefont
  {Kokkotas}}\ and\ \bibinfo {author} {\bibfnamefont {J.}~\bibnamefont
  {Ruoff}},\ }\href {\doibase 10.1051/0004-6361:20000216} {\bibfield  {journal}
  {\bibinfo  {journal} {Astron. Astrophys.}\ }\textbf {\bibinfo {volume}
  {366}},\ \bibinfo {pages} {565} (\bibinfo {year} {2001})}\BibitemShut
  {NoStop}%
\bibitem [{\citenamefont {Li}\ \emph {et~al.}(2022)\citenamefont {Li},
  \citenamefont {Gao}, \citenamefont {Shao}, \citenamefont {Xu},\ and\
  \citenamefont {Xu}}]{Li:2022qql}%
  \BibitemOpen
  \bibfield  {author} {\bibinfo {author} {\bibfnamefont {H.-B.}\ \bibnamefont
  {Li}}, \bibinfo {author} {\bibfnamefont {Y.}~\bibnamefont {Gao}}, \bibinfo
  {author} {\bibfnamefont {L.}~\bibnamefont {Shao}}, \bibinfo {author}
  {\bibfnamefont {R.-X.}\ \bibnamefont {Xu}}, \ and\ \bibinfo {author}
  {\bibfnamefont {R.}~\bibnamefont {Xu}},\ }\href {\doibase
  10.1093/mnras/stac2622} {\bibfield  {journal} {\bibinfo  {journal} {Mon. Not.
  Roy. Astron. Soc.}\ }\textbf {\bibinfo {volume} {516}},\ \bibinfo {pages}
  {6172} (\bibinfo {year} {2022})},\ \Eprint {http://arxiv.org/abs/2206.09407}
  {arXiv:2206.09407 [gr-qc]} \BibitemShut {NoStop}%
\bibitem [{\citenamefont {Li}\ \emph {et~al.}(2024)\citenamefont {Li},
  \citenamefont {Gao}, \citenamefont {Shao},\ and\ \citenamefont
  {Xu}}]{Li:2024hzt}%
  \BibitemOpen
  \bibfield  {author} {\bibinfo {author} {\bibfnamefont {H.-B.}\ \bibnamefont
  {Li}}, \bibinfo {author} {\bibfnamefont {Y.}~\bibnamefont {Gao}}, \bibinfo
  {author} {\bibfnamefont {L.}~\bibnamefont {Shao}}, \ and\ \bibinfo {author}
  {\bibfnamefont {R.-X.}\ \bibnamefont {Xu}},\ }\href {\doibase
  10.3390/universe10040157} {\bibfield  {journal} {\bibinfo  {journal}
  {Universe}\ }\textbf {\bibinfo {volume} {10}},\ \bibinfo {pages} {157}
  (\bibinfo {year} {2024})}\BibitemShut {NoStop}%
\bibitem [{\citenamefont {Sotani}\ \emph {et~al.}(2001)\citenamefont {Sotani},
  \citenamefont {Tominaga},\ and\ \citenamefont
  {Maeda}}]{Sotani2001_PRD65-024010}%
  \BibitemOpen
  \bibfield  {author} {\bibinfo {author} {\bibfnamefont {H.}~\bibnamefont
  {Sotani}}, \bibinfo {author} {\bibfnamefont {K.}~\bibnamefont {Tominaga}}, \
  and\ \bibinfo {author} {\bibfnamefont {K.-i.}\ \bibnamefont {Maeda}},\ }\href
  {\doibase 10.1103/PhysRevD.65.024010} {\bibfield  {journal} {\bibinfo
  {journal} {Phys. Rev. D}\ }\textbf {\bibinfo {volume} {65}},\ \bibinfo
  {pages} {024010} (\bibinfo {year} {2001})}\BibitemShut {NoStop}%
\bibitem [{\citenamefont {Sotani}(2024)}]{Sotani2024_PRD109-023030}%
  \BibitemOpen
  \bibfield  {author} {\bibinfo {author} {\bibfnamefont {H.}~\bibnamefont
  {Sotani}},\ }\href {\doibase 10.1103/PhysRevD.109.023030} {\bibfield
  {journal} {\bibinfo  {journal} {Phys. Rev. D}\ }\textbf {\bibinfo {volume}
  {109}},\ \bibinfo {pages} {023030} (\bibinfo {year} {2024})}\BibitemShut
  {NoStop}%
\bibitem [{\citenamefont {Kokkotas}\ and\ \citenamefont
  {Schmidt}(1999)}]{Kokkotas1999_LRR2-1}%
  \BibitemOpen
  \bibfield  {author} {\bibinfo {author} {\bibfnamefont {K.~D.}\ \bibnamefont
  {Kokkotas}}\ and\ \bibinfo {author} {\bibfnamefont {B.~G.}\ \bibnamefont
  {Schmidt}},\ }\href {\doibase 10.12942/lrr-1999-2} {\bibfield  {journal}
  {\bibinfo  {journal} {Living Rev. Relativ.}\ }\textbf {\bibinfo {volume}
  {2}},\ \bibinfo {pages} {2} (\bibinfo {year} {1999})}\BibitemShut {NoStop}%
\bibitem [{\citenamefont {Ho}\ \emph {et~al.}(2020)\citenamefont {Ho},
  \citenamefont {Jones}, \citenamefont {Andersson},\ and\ \citenamefont
  {Espinoza}}]{Ho2020_PRD101-103009}%
  \BibitemOpen
  \bibfield  {author} {\bibinfo {author} {\bibfnamefont {W.~C.~G.}\
  \bibnamefont {Ho}}, \bibinfo {author} {\bibfnamefont {D.~I.}\ \bibnamefont
  {Jones}}, \bibinfo {author} {\bibfnamefont {N.}~\bibnamefont {Andersson}}, \
  and\ \bibinfo {author} {\bibfnamefont {C.~M.}\ \bibnamefont {Espinoza}},\
  }\href {\doibase 10.1103/PhysRevD.101.103009} {\bibfield  {journal} {\bibinfo
   {journal} {Phys. Rev. D}\ }\textbf {\bibinfo {volume} {101}},\ \bibinfo
  {pages} {103009} (\bibinfo {year} {2020})}\BibitemShut {NoStop}%
\bibitem [{\citenamefont {Yu}\ and\ \citenamefont
  {Weinberg}(2017)}]{Yu2017_MNTAS470-350}%
  \BibitemOpen
  \bibfield  {author} {\bibinfo {author} {\bibfnamefont {H.}~\bibnamefont
  {Yu}}\ and\ \bibinfo {author} {\bibfnamefont {N.~N.}\ \bibnamefont
  {Weinberg}},\ }\href {\doibase 10.1093/mnras/stx1188} {\bibfield  {journal}
  {\bibinfo  {journal} {Mon. Not. R. Astron. Soc.}\ }\textbf {\bibinfo {volume}
  {470}},\ \bibinfo {pages} {350} (\bibinfo {year} {2017})}\BibitemShut
  {NoStop}%
\bibitem [{\citenamefont {Lai}(1994)}]{Lai1994_MNRAS270-611-629}%
  \BibitemOpen
  \bibfield  {author} {\bibinfo {author} {\bibfnamefont {D.}~\bibnamefont
  {Lai}},\ }\href {\doibase 10.1093/mnras/270.3.611} {\bibfield  {journal}
  {\bibinfo  {journal} {Mon. Not. R. Astron. Soc.}\ }\textbf {\bibinfo {volume}
  {270}},\ \bibinfo {pages} {611} (\bibinfo {year} {1994})}\BibitemShut
  {NoStop}%
\bibitem [{\citenamefont {Xu}\ and\ \citenamefont
  {Lai}(2017)}]{Xu2017_PRD96-083005}%
  \BibitemOpen
  \bibfield  {author} {\bibinfo {author} {\bibfnamefont {W.}~\bibnamefont
  {Xu}}\ and\ \bibinfo {author} {\bibfnamefont {D.}~\bibnamefont {Lai}},\
  }\href {\doibase 10.1103/PhysRevD.96.083005} {\bibfield  {journal} {\bibinfo
  {journal} {Phys. Rev. D}\ }\textbf {\bibinfo {volume} {96}},\ \bibinfo
  {pages} {083005} (\bibinfo {year} {2017})}\BibitemShut {NoStop}%
\bibitem [{\citenamefont {{Thorne}}\ and\ \citenamefont
  {{Campolattaro}}(1967)}]{Thorne1967_ApJ-149-591}%
  \BibitemOpen
  \bibfield  {author} {\bibinfo {author} {\bibfnamefont {K.~S.}\ \bibnamefont
  {{Thorne}}}\ and\ \bibinfo {author} {\bibfnamefont {A.}~\bibnamefont
  {{Campolattaro}}},\ }\href {\doibase 10.1086/149288} {\bibfield  {journal}
  {\bibinfo  {journal} {Astrophys. J.}\ ,\ \bibinfo {pages} {591}} (\bibinfo
  {year} {1967})}\BibitemShut {NoStop}%
\bibitem{Punturo2010_CQG27-194002}%
  \BibitemOpen
  \bibfield  {author} {\bibinfo {author} {\bibfnamefont {M.}~\bibnamefont
  {Punturo}}, \bibinfo {author} {\bibfnamefont {M.}~\bibnamefont
  {Abernathy}}, \bibinfo {author} {\bibfnamefont {F.}~\bibnamefont
  {Acernese}}, et al.,\ }
  \href {\doibase 10.1088/0264-9381/27/19/194002} {\bibfield
  {journal} {\bibinfo  {journal} {Class. Quant. Grav.}\ }\textbf {\bibinfo
  {volume} {27}},\ \bibinfo {pages} {194002} (\bibinfo {year}
  {2010})}\BibitemShut {NoStop}%
\bibitem [{\citenamefont {Regimbau}\ \emph {et~al.}(2017)\citenamefont
  {Regimbau}, \citenamefont {Evans}, \citenamefont {Christensen}, \citenamefont
  {Katsavounidis}, \citenamefont {Sathyaprakash},\ and\ \citenamefont
  {Vitale}}]{Regimbau2017_PRL118-151105}%
  \BibitemOpen
  \bibfield  {author} {\bibinfo {author} {\bibfnamefont {T.}~\bibnamefont
  {Regimbau}}, \bibinfo {author} {\bibfnamefont {M.}~\bibnamefont {Evans}},
  \bibinfo {author} {\bibfnamefont {N.}~\bibnamefont {Christensen}}, \bibinfo
  {author} {\bibfnamefont {E.}~\bibnamefont {Katsavounidis}}, \bibinfo {author}
  {\bibfnamefont {B.}~\bibnamefont {Sathyaprakash}}, \ and\ \bibinfo {author}
  {\bibfnamefont {S.}~\bibnamefont {Vitale}},\ }\href {\doibase
  10.1103/PhysRevLett.118.151105} {\bibfield  {journal} {\bibinfo  {journal}
  {Phys. Rev. Lett.}\ }\textbf {\bibinfo {volume} {118}},\ \bibinfo {pages}
  {151105} (\bibinfo {year} {2017})}\BibitemShut {NoStop}%
\bibitem [{\citenamefont {Abbott}\ \emph {et~al.}(2017)\citenamefont {Abbott},
  \citenamefont {Abbott}, \citenamefont {Abbott},\ and\ \citenamefont
  {et~al.}}]{Abbott2017_CQG34-044001}%
  \BibitemOpen
  \bibfield  {author} {\bibinfo {author} {\bibfnamefont {B.~P.}\ \bibnamefont
  {Abbott}}, \bibinfo {author} {\bibfnamefont {R.}~\bibnamefont {Abbott}},
  \bibinfo {author} {\bibfnamefont {T.~D.}\ \bibnamefont {Abbott}}, et~al.,\ }\href {\doibase
  10.1088/1361-6382/aa51f4} {\bibfield  {journal} {\bibinfo  {journal} {Class.
  Quant. Grav.}\ }\textbf {\bibinfo {volume} {34}},\ \bibinfo {pages} {044001}
  (\bibinfo {year} {2017})}\BibitemShut {NoStop}%
\bibitem [{\citenamefont {Abbott}\ \emph {et~al.}(2020)\citenamefont {Abbott},
  \citenamefont {Abbott}, \citenamefont {Abbott},\ and\ \citenamefont
  {et~al.}}]{Abbott2020_LRR23-3}%
  \BibitemOpen
  \bibfield  {author} {\bibinfo {author} {\bibfnamefont {B.~P.}\ \bibnamefont
  {Abbott}}, \bibinfo {author} {\bibfnamefont {R.}~\bibnamefont {Abbott}},
  \bibinfo {author} {\bibfnamefont {T.~D.}\ \bibnamefont {Abbott}}, et~al.,\ }\href {\doibase
  10.1007/s41114-020-00026-9} {\bibfield  {journal} {\bibinfo  {journal}
  {Living Rev. Relativ.}\ }\textbf {\bibinfo {volume} {23}},\ \bibinfo {pages}
  {3} (\bibinfo {year} {2020})}\BibitemShut {NoStop}%
\bibitem [{\citenamefont {Maggiore}\ \emph {et~al.}(2020)\citenamefont
  {Maggiore}, \citenamefont {Broeck}, \citenamefont {Bartolo}, \citenamefont
  {Belgacem}, \citenamefont {Bertacca}, \citenamefont {Bizouard}, \citenamefont
  {Branchesi}, \citenamefont {Clesse}, \citenamefont {Foffa}, \citenamefont
  {Garc\'{\i}a-Bellido}, \citenamefont {Grimm}, \citenamefont {Harms},
  \citenamefont {Hinderer}, \citenamefont {Matarrese}, \citenamefont {Palomba},
  \citenamefont {Peloso}, \citenamefont {Ricciardone},\ and\ \citenamefont
  {Sakellariadou}}]{Maggiore2020_JCAP2020-050}%
  \BibitemOpen
  \bibfield  {author} {\bibinfo {author} {\bibfnamefont {M.}~\bibnamefont
  {Maggiore}}, \bibinfo {author} {\bibfnamefont {C.~V.~D.}\ \bibnamefont
  {Broeck}}, \bibinfo {author} {\bibfnamefont {N.}~\bibnamefont {Bartolo}},
  \bibinfo {author} {\bibfnamefont {E.}~\bibnamefont {Belgacem}}, \bibinfo
  {author} {\bibfnamefont {D.}~\bibnamefont {Bertacca}}, \bibinfo {author}
  {\bibfnamefont {M.~A.}\ \bibnamefont {Bizouard}}, \bibinfo {author}
  {\bibfnamefont {M.}~\bibnamefont {Branchesi}}, \bibinfo {author}
  {\bibfnamefont {S.}~\bibnamefont {Clesse}}, \bibinfo {author} {\bibfnamefont
  {S.}~\bibnamefont {Foffa}}, \bibinfo {author} {\bibfnamefont
  {J.}~\bibnamefont {Garc\'{\i}a-Bellido}}, \bibinfo {author} {\bibfnamefont
  {S.}~\bibnamefont {Grimm}}, \bibinfo {author} {\bibfnamefont
  {J.}~\bibnamefont {Harms}}, \bibinfo {author} {\bibfnamefont
  {T.}~\bibnamefont {Hinderer}}, \bibinfo {author} {\bibfnamefont
  {S.}~\bibnamefont {Matarrese}}, \bibinfo {author} {\bibfnamefont
  {C.}~\bibnamefont {Palomba}}, \bibinfo {author} {\bibfnamefont
  {M.}~\bibnamefont {Peloso}}, \bibinfo {author} {\bibfnamefont
  {A.}~\bibnamefont {Ricciardone}}, \ and\ \bibinfo {author} {\bibfnamefont
  {M.}~\bibnamefont {Sakellariadou}},\ }\href {\doibase
  10.1088/1475-7516/2020/03/050} {\bibfield  {journal} {\bibinfo  {journal} {J.
  Cosmol. Astropart. Phys.}\ }\textbf {\bibinfo {volume} {2020}},\ \bibinfo
  {pages} {050} (\bibinfo {year} {2020})}\BibitemShut {NoStop}%
\bibitem [{\citenamefont {Osaki}\ and\ \citenamefont
  {Hansen}(1973)}]{Osaki1973_Apj185-277}%
  \BibitemOpen
  \bibfield  {author} {\bibinfo {author} {\bibfnamefont {Y.}~\bibnamefont
  {Osaki}}\ and\ \bibinfo {author} {\bibfnamefont {C.~J.}\ \bibnamefont
  {Hansen}},\ }\href {https://api.semanticscholar.org/CorpusID:119818255}
  {\bibfield  {journal} {\bibinfo  {journal} {Astrophys. J.}\ }\textbf
  {\bibinfo {volume} {185}},\ \bibinfo {pages} {277} (\bibinfo {year}
  {1973})}\BibitemShut {NoStop}%
\bibitem [{\citenamefont {Finn}(1987)}]{Finn1987_MNRAS227-265-293}%
  \BibitemOpen
  \bibfield  {author} {\bibinfo {author} {\bibfnamefont {L.~S.}\ \bibnamefont
  {Finn}},\ }\href {\doibase 10.1093/mnras/227.2.265} {\bibfield  {journal}
  {\bibinfo  {journal} {Mon. Not. R. Astron. Soc.}\ }\textbf {\bibinfo {volume}
  {227}},\ \bibinfo {pages} {265} (\bibinfo {year} {1987})}\BibitemShut
  {NoStop}%
\bibitem [{\citenamefont {Kuan}\ \emph {et~al.}(2022)\citenamefont {Kuan},
  \citenamefont {Kr\"{u}ger}, \citenamefont {Suvorov},\ and\ \citenamefont
  {Kokkotas}}]{Kuan2022-MNRAS513-4045}%
  \BibitemOpen
  \bibfield  {author} {\bibinfo {author} {\bibfnamefont {H.-J.}\ \bibnamefont
  {Kuan}}, \bibinfo {author} {\bibfnamefont {C.~J.}\ \bibnamefont {Kr\"{u}ger}},
  \bibinfo {author} {\bibfnamefont {A.~G.}\ \bibnamefont {Suvorov}}, \ and\
  \bibinfo {author} {\bibfnamefont {K.~D.}\ \bibnamefont {Kokkotas}},\ }\href
  {\doibase 10.1093/mnras/stac1101} {\bibfield  {journal} {\bibinfo  {journal}
  {Mon. Not. R. Astron. Soc.}\ }\textbf {\bibinfo {volume} {513}},\ \bibinfo
  {pages} {4045} (\bibinfo {year} {2022})}\BibitemShut {NoStop}%
\bibitem [{\citenamefont {{McDermott}}\ \emph {et~al.}(1983)\citenamefont
  {{McDermott}}, \citenamefont {{van Horn}},\ and\ \citenamefont
  {{Scholl}}}]{McDermott1983_ApJ268-837}%
  \BibitemOpen
  \bibfield  {author} {\bibinfo {author} {\bibfnamefont {P.~N.}\ \bibnamefont
  {{McDermott}}}, \bibinfo {author} {\bibfnamefont {H.~M.}\ \bibnamefont {{van
  Horn}}}, \ and\ \bibinfo {author} {\bibfnamefont {J.~F.}\ \bibnamefont
  {{Scholl}}},\ }\href {\doibase 10.1086/161006} {\bibfield  {journal}
  {\bibinfo  {journal} {Astrophys. J.}\ }\textbf {\bibinfo {volume} {268}},\
  \bibinfo {pages} {837} (\bibinfo {year} {1983})}\BibitemShut {NoStop}%
\bibitem [{\citenamefont {Kr\"uger}\ \emph {et~al.}(2015)\citenamefont
  {Kr\"uger}, \citenamefont {Ho},\ and\ \citenamefont
  {Andersson}}]{Krueger2015_PRD92-063009}%
  \BibitemOpen
  \bibfield  {author} {\bibinfo {author} {\bibfnamefont {C.~J.}\ \bibnamefont
  {Kr\"uger}}, \bibinfo {author} {\bibfnamefont {W.~C.~G.}\ \bibnamefont {Ho}},
  \ and\ \bibinfo {author} {\bibfnamefont {N.}~\bibnamefont {Andersson}},\
  }\href {\doibase 10.1103/PhysRevD.92.063009} {\bibfield  {journal} {\bibinfo
  {journal} {Phys. Rev. D}\ }\textbf {\bibinfo {volume} {92}},\ \bibinfo
  {pages} {063009} (\bibinfo {year} {2015})}\BibitemShut {NoStop}%
\bibitem [{\citenamefont {Sotani}\ and\ \citenamefont
  {Dohi}(2022)}]{Sotani2022_PRD105-023007}%
  \BibitemOpen
  \bibfield  {author} {\bibinfo {author} {\bibfnamefont {H.}~\bibnamefont
  {Sotani}}\ and\ \bibinfo {author} {\bibfnamefont {A.}~\bibnamefont {Dohi}},\
  }\href {\doibase 10.1103/PhysRevD.105.023007} {\bibfield  {journal} {\bibinfo
   {journal} {Phys. Rev. D}\ }\textbf {\bibinfo {volume} {105}},\ \bibinfo
  {pages} {023007} (\bibinfo {year} {2022})}\BibitemShut {NoStop}%
\bibitem [{\citenamefont {Lozano}\ \emph {et~al.}(2022)\citenamefont {Lozano},
  \citenamefont {Tran},\ and\ \citenamefont
  {Jaikumar}}]{Lozano2022_Galaxies10-2075-4434}%
  \BibitemOpen
  \bibfield  {author} {\bibinfo {author} {\bibfnamefont {N.}~\bibnamefont
  {Lozano}}, \bibinfo {author} {\bibfnamefont {V.}~\bibnamefont {Tran}}, \ and\
  \bibinfo {author} {\bibfnamefont {P.}~\bibnamefont {Jaikumar}},\ }\href
  {\doibase 10.3390/galaxies10040079} {\bibfield  {journal} {\bibinfo
  {journal} {Galaxies}\ }\textbf {\bibinfo {volume} {10}},\ \bibinfo {pages}
  {79} (\bibinfo {year} {2022})}\BibitemShut {NoStop}%
\bibitem [{\citenamefont {Gaertig}\ and\ \citenamefont
  {Kokkotas}(2009)}]{Gaertig2009_PRD80-064026}%
  \BibitemOpen
  \bibfield  {author} {\bibinfo {author} {\bibfnamefont {E.}~\bibnamefont
  {Gaertig}}\ and\ \bibinfo {author} {\bibfnamefont {K.~D.}\ \bibnamefont
  {Kokkotas}},\ }\href {\doibase 10.1103/PhysRevD.80.064026} {\bibfield
  {journal} {\bibinfo  {journal} {Phys. Rev. D}\ }\textbf {\bibinfo {volume}
  {80}},\ \bibinfo {pages} {064026} (\bibinfo {year} {2009})}\BibitemShut
  {NoStop}%
\bibitem [{\citenamefont {Lovekin}\ and\ \citenamefont
  {Deupree}(2008)}]{Lovekin2008_ApJ679-2}%
  \BibitemOpen
  \bibfield  {author} {\bibinfo {author} {\bibfnamefont {C.~C.}\ \bibnamefont
  {Lovekin}}\ and\ \bibinfo {author} {\bibfnamefont {R.~G.}\ \bibnamefont
  {Deupree}},\ }\href {\doibase 10.1086/587615} {\bibfield  {journal} {\bibinfo
   {journal} {Astrophys. J.}\ }\textbf {\bibinfo {volume} {679}},\ \bibinfo
  {pages} {1499-1508} (\bibinfo {year} {2008})}\BibitemShut {NoStop}%
\bibitem [{\citenamefont {Kojima}(1992)}]{Kojima1992_PRD46-4289}%
  \BibitemOpen
  \bibfield  {author} {\bibinfo {author} {\bibfnamefont {Y.}~\bibnamefont
  {Kojima}},\ }\href {\doibase 10.1103/PhysRevD.46.4289} {\bibfield  {journal}
  {\bibinfo  {journal} {Phys. Rev. D}\ }\textbf {\bibinfo {volume} {46}},\
  \bibinfo {pages} {4289} (\bibinfo {year} {1992})}\BibitemShut {NoStop}%
\bibitem [{\citenamefont {Tran}\ \emph {et~al.}(2023)\citenamefont {Tran},
  \citenamefont {Ghosh}, \citenamefont {Lozano}, \citenamefont {Chatterjee},\
  and\ \citenamefont {Jaikumar}}]{Tran2023_PRC108-015803}%
  \BibitemOpen
  \bibfield  {author} {\bibinfo {author} {\bibfnamefont {V.}~\bibnamefont
  {Tran}}, \bibinfo {author} {\bibfnamefont {S.}~\bibnamefont {Ghosh}},
  \bibinfo {author} {\bibfnamefont {N.}~\bibnamefont {Lozano}}, \bibinfo
  {author} {\bibfnamefont {D.}~\bibnamefont {Chatterjee}}, \ and\ \bibinfo
  {author} {\bibfnamefont {P.}~\bibnamefont {Jaikumar}},\ }\href {\doibase
  10.1103/PhysRevC.108.015803} {\bibfield  {journal} {\bibinfo  {journal}
  {Phys. Rev. C}\ }\textbf {\bibinfo {volume} {108}},\ \bibinfo {pages}
  {015803} (\bibinfo {year} {2023})}\BibitemShut {NoStop}%
\bibitem [{\citenamefont {Jaikumar}\ \emph {et~al.}(2021)\citenamefont
  {Jaikumar}, \citenamefont {Semposki}, \citenamefont {Prakash},\ and\
  \citenamefont {Constantinou}}]{Jaikumar2021_PRD103-123009}%
  \BibitemOpen
  \bibfield  {author} {\bibinfo {author} {\bibfnamefont {P.}~\bibnamefont
  {Jaikumar}}, \bibinfo {author} {\bibfnamefont {A.}~\bibnamefont {Semposki}},
  \bibinfo {author} {\bibfnamefont {M.}~\bibnamefont {Prakash}}, \ and\
  \bibinfo {author} {\bibfnamefont {C.}~\bibnamefont {Constantinou}},\ }\href
  {\doibase 10.1103/PhysRevD.103.123009} {\bibfield  {journal} {\bibinfo
  {journal} {Phys. Rev. D}\ }\textbf {\bibinfo {volume} {103}},\ \bibinfo
  {pages} {123009} (\bibinfo {year} {2021})}\BibitemShut {NoStop}%
\bibitem [{\citenamefont {Wei}\ \emph {et~al.}(2020{\natexlab{a}})\citenamefont
  {Wei}, \citenamefont {Salinas}, \citenamefont {Kl\"ahn}, \citenamefont
  {Jaikumar},\ and\ \citenamefont {Barry}}]{Wei2020_ApJ904-187}%
  \BibitemOpen
  \bibfield  {author} {\bibinfo {author} {\bibfnamefont {W.}~\bibnamefont
  {Wei}}, \bibinfo {author} {\bibfnamefont {M.}~\bibnamefont {Salinas}},
  \bibinfo {author} {\bibfnamefont {T.}~\bibnamefont {Kl\"ahn}}, \bibinfo
  {author} {\bibfnamefont {P.}~\bibnamefont {Jaikumar}}, \ and\ \bibinfo
  {author} {\bibfnamefont {M.}~\bibnamefont {Barry}},\ }\href {\doibase
  10.3847/1538-4357/abbe02} {\bibfield  {journal} {\bibinfo  {journal}
  {Astrophys. J.}\ }\textbf {\bibinfo {volume} {904}},\ \bibinfo {pages} {187}
  (\bibinfo {year} {2020}{\natexlab{a}})}\BibitemShut {NoStop}%
\bibitem [{\citenamefont {Zhao}\ \emph {et~al.}(2022)\citenamefont {Zhao},
  \citenamefont {Constantinou}, \citenamefont {Jaikumar},\ and\ \citenamefont
  {Prakash}}]{Zhao2022_PRD105-103025}%
  \BibitemOpen
  \bibfield  {author} {\bibinfo {author} {\bibfnamefont {T.}~\bibnamefont
  {Zhao}}, \bibinfo {author} {\bibfnamefont {C.}~\bibnamefont {Constantinou}},
  \bibinfo {author} {\bibfnamefont {P.}~\bibnamefont {Jaikumar}}, \ and\
  \bibinfo {author} {\bibfnamefont {M.}~\bibnamefont {Prakash}},\ }\href
  {\doibase 10.1103/PhysRevD.105.103025} {\bibfield  {journal} {\bibinfo
  {journal} {Phys. Rev. D}\ }\textbf {\bibinfo {volume} {105}},\ \bibinfo
  {pages} {103025} (\bibinfo {year} {2022})}\BibitemShut {NoStop}%
\bibitem [{\citenamefont {Reisenegger}\ and\ \citenamefont
  {Goldreich}(1992)}]{Reisenegger1992_Apj395-240-249}%
  \BibitemOpen
  \bibfield  {author} {\bibinfo {author} {\bibfnamefont {A.}~\bibnamefont
  {Reisenegger}}\ and\ \bibinfo {author} {\bibfnamefont {P.}~\bibnamefont
  {Goldreich}},\ }\href {\doibase 10.1086/171645} {\bibfield  {journal}
  {\bibinfo  {journal} {Astrophys. J.}\ }\textbf {\bibinfo {volume} {395}},\
  \bibinfo {pages} {240} (\bibinfo {year} {1992})}\BibitemShut {NoStop}%
\bibitem [{\citenamefont {{McDermott}}\ \emph {et~al.}(1988)\citenamefont
  {{McDermott}}, \citenamefont {{van Horn}},\ and\ \citenamefont
  {{Hansen}}}]{McDermott1988_ApJ325-725}%
  \BibitemOpen
  \bibfield  {author} {\bibinfo {author} {\bibfnamefont {P.~N.}\ \bibnamefont
  {{McDermott}}}, \bibinfo {author} {\bibfnamefont {H.~M.}\ \bibnamefont {{van
  Horn}}}, \ and\ \bibinfo {author} {\bibfnamefont {C.~J.}\ \bibnamefont
  {{Hansen}}},\ }\href {\doibase 10.1086/166044} {\bibfield  {journal}
  {\bibinfo  {journal} {Astrophys. J.}\ }\textbf {\bibinfo {volume} {325}},\
  \bibinfo {pages} {725} (\bibinfo {year} {1988})}\BibitemShut {NoStop}%
\bibitem [{\citenamefont {Tsang}\ \emph {et~al.}(2012)\citenamefont {Tsang},
  \citenamefont {Read}, \citenamefont {Hinderer}, \citenamefont {Piro},\ and\
  \citenamefont {Bondarescu}}]{Tsang2012_PRL108-011102}%
  \BibitemOpen
  \bibfield  {author} {\bibinfo {author} {\bibfnamefont {D.}~\bibnamefont
  {Tsang}}, \bibinfo {author} {\bibfnamefont {J.~S.}\ \bibnamefont {Read}},
  \bibinfo {author} {\bibfnamefont {T.}~\bibnamefont {Hinderer}}, \bibinfo
  {author} {\bibfnamefont {A.~L.}\ \bibnamefont {Piro}}, \ and\ \bibinfo
  {author} {\bibfnamefont {R.}~\bibnamefont {Bondarescu}},\ }\href {\doibase
  10.1103/PhysRevLett.108.011102} {\bibfield  {journal} {\bibinfo  {journal}
  {Phys. Rev. Lett.}\ }\textbf {\bibinfo {volume} {108}},\ \bibinfo {pages}
  {011102} (\bibinfo {year} {2012})}\BibitemShut {NoStop}%
\bibitem [{\citenamefont {Pan}\ \emph {et~al.}(2020)\citenamefont {Pan},
  \citenamefont {Lyu}, \citenamefont {Bonga}, \citenamefont {Ortiz},\ and\
  \citenamefont {Yang}}]{Pan2020_PRL125-201102}%
  \BibitemOpen
  \bibfield  {author} {\bibinfo {author} {\bibfnamefont {Z.}~\bibnamefont
  {Pan}}, \bibinfo {author} {\bibfnamefont {Z.}~\bibnamefont {Lyu}}, \bibinfo
  {author} {\bibfnamefont {B.}~\bibnamefont {Bonga}}, \bibinfo {author}
  {\bibfnamefont {N.}~\bibnamefont {Ortiz}}, \ and\ \bibinfo {author}
  {\bibfnamefont {H.}~\bibnamefont {Yang}},\ }\href {\doibase
  10.1103/PhysRevLett.125.201102} {\bibfield  {journal} {\bibinfo  {journal}
  {Phys. Rev. Lett.}\ }\textbf {\bibinfo {volume} {125}},\ \bibinfo {pages}
  {201102} (\bibinfo {year} {2020})}\BibitemShut {NoStop}%
\bibitem [{\citenamefont {Meng}(2016)}]{Meng2016}%
  \BibitemOpen
  \bibinfo {editor} {\bibfnamefont {J.}~\bibnamefont {Meng}},\ ed.,\ \href
  {\doibase 10.1142/9872} { {\bibinfo {title} {{Relativistic Density
  Functional for Nuclear Structure}}}},\ \bibinfo {series} {International
  Review of Nuclear Physics}, Vol.~\bibinfo {volume} {10}\ (\bibinfo
  {publisher} {World Scientific Pub Co Pte Lt},\ \bibinfo {year}
  {2016})\BibitemShut {NoStop}%
\bibitem [{\citenamefont {Lalazissis}\ \emph {et~al.}(1997)\citenamefont
  {Lalazissis}, \citenamefont {K\"onig},\ and\ \citenamefont
  {Ring}}]{Lalazissis1997_PRC55-540}%
  \BibitemOpen
  \bibfield  {author} {\bibinfo {author} {\bibfnamefont {G.~A.}\ \bibnamefont
  {Lalazissis}}, \bibinfo {author} {\bibfnamefont {J.}~\bibnamefont {K\"onig}},
  \ and\ \bibinfo {author} {\bibfnamefont {P.}~\bibnamefont {Ring}},\ }\href
  {\doibase 10.1103/PhysRevC.55.540} {\bibfield  {journal} {\bibinfo  {journal}
  {Phys. Rev. C}\ }\textbf {\bibinfo {volume} {55}},\ \bibinfo {pages} {540}
  (\bibinfo {year} {1997})}\BibitemShut {NoStop}%
\bibitem [{\citenamefont {Long}\ \emph {et~al.}(2004)\citenamefont {Long},
  \citenamefont {Meng}, \citenamefont {Giai},\ and\ \citenamefont
  {Zhou}}]{Long2004_PRC69-034319}%
  \BibitemOpen
  \bibfield  {author} {\bibinfo {author} {\bibfnamefont {W.}~\bibnamefont
  {Long}}, \bibinfo {author} {\bibfnamefont {J.}~\bibnamefont {Meng}}, \bibinfo
  {author} {\bibfnamefont {N.~V.}\ \bibnamefont {Giai}}, \ and\ \bibinfo
  {author} {\bibfnamefont {S.-G.}\ \bibnamefont {Zhou}},\ }\href {\doibase
  10.1103/PhysRevC.69.034319} {\bibfield  {journal} {\bibinfo  {journal} {Phys.
  Rev. C}\ }\textbf {\bibinfo {volume} {69}},\ \bibinfo {pages} {034319}
  (\bibinfo {year} {2004})}\BibitemShut {NoStop}%
\bibitem [{\citenamefont {Sugahara}\ and\ \citenamefont
  {Toki}(1994)}]{Sugahara1994_NPA579-557}%
  \BibitemOpen
  \bibfield  {author} {\bibinfo {author} {\bibfnamefont {Y.}~\bibnamefont
  {Sugahara}}\ and\ \bibinfo {author} {\bibfnamefont {H.}~\bibnamefont
  {Toki}},\ }\href {\doibase https://doi.org/10.1016/0375-9474(94)90923-7}
  {\bibfield  {journal} {\bibinfo  {journal} {Nucl. Phys. A}\ }\textbf
  {\bibinfo {volume} {579}},\ \bibinfo {pages} {557} (\bibinfo {year}
  {1994})}\BibitemShut {NoStop}%
\bibitem [{\citenamefont {Glendenning}\ and\ \citenamefont
  {Moszkowski}(1991)}]{Glendenning1991_PRL67-2414}%
  \BibitemOpen
  \bibfield  {author} {\bibinfo {author} {\bibfnamefont {N.~K.}\ \bibnamefont
  {Glendenning}}\ and\ \bibinfo {author} {\bibfnamefont {S.~A.}\ \bibnamefont
  {Moszkowski}},\ }\href {\doibase 10.1103/PhysRevLett.67.2414} {\bibfield
  {journal} {\bibinfo  {journal} {Phys. Rev. Lett.}\ }\textbf {\bibinfo
  {volume} {67}},\ \bibinfo {pages} {2414} (\bibinfo {year}
  {1991})}\BibitemShut {NoStop}%
\bibitem [{\citenamefont {Maruyama}\ \emph {et~al.}(2005)\citenamefont
  {Maruyama}, \citenamefont {Tatsumi}, \citenamefont {Voskresensky},
  \citenamefont {Tanigawa},\ and\ \citenamefont
  {Chiba}}]{Maruyama2005_PRC72-015802}%
  \BibitemOpen
  \bibfield  {author} {\bibinfo {author} {\bibfnamefont {T.}~\bibnamefont
  {Maruyama}}, \bibinfo {author} {\bibfnamefont {T.}~\bibnamefont {Tatsumi}},
  \bibinfo {author} {\bibfnamefont {D.~N.}\ \bibnamefont {Voskresensky}},
  \bibinfo {author} {\bibfnamefont {T.}~\bibnamefont {Tanigawa}}, \ and\
  \bibinfo {author} {\bibfnamefont {S.}~\bibnamefont {Chiba}},\ }\href
  {\doibase 10.1103/PhysRevC.72.015802} {\bibfield  {journal} {\bibinfo
  {journal} {Phys. Rev. C}\ }\textbf {\bibinfo {volume} {72}},\ \bibinfo
  {pages} {015802} (\bibinfo {year} {2005})}\BibitemShut {NoStop}%
\bibitem [{\citenamefont {Wei}\ \emph {et~al.}(2020{\natexlab{b}})\citenamefont
  {Wei}, \citenamefont {Zhao}, \citenamefont {Wang}, \citenamefont {Geng},
  \citenamefont {Sun}, \citenamefont {Niu},\ and\ \citenamefont
  {Long}}]{Wei2020_CPC44-074107}%
  \BibitemOpen
  \bibfield  {author} {\bibinfo {author} {\bibfnamefont {B.}~\bibnamefont
  {Wei}}, \bibinfo {author} {\bibfnamefont {Q.}~\bibnamefont {Zhao}}, \bibinfo
  {author} {\bibfnamefont {Z.-H.}\ \bibnamefont {Wang}}, \bibinfo {author}
  {\bibfnamefont {J.}~\bibnamefont {Geng}}, \bibinfo {author} {\bibfnamefont
  {B.-Y.}\ \bibnamefont {Sun}}, \bibinfo {author} {\bibfnamefont {Y.-F.}\
  \bibnamefont {Niu}}, \ and\ \bibinfo {author} {\bibfnamefont {W.-H.}\
  \bibnamefont {Long}},\ }\href {\doibase 10.1088/1674-1137/44/7/074107}
  {\bibfield  {journal} {\bibinfo  {journal} {Chin. Phys. C}\ }\textbf
  {\bibinfo {volume} {44}},\ \bibinfo {pages} {074107} (\bibinfo {year}
  {2020}{\natexlab{b}})}\BibitemShut {NoStop}%
\bibitem [{\citenamefont {Taninah}\ \emph {et~al.}(2020)\citenamefont
  {Taninah}, \citenamefont {Agbemava}, \citenamefont {Afanasjev},\ and\
  \citenamefont {Ring}}]{Taninah2020_PLB800-135065}%
  \BibitemOpen
  \bibfield  {author} {\bibinfo {author} {\bibfnamefont {A.}~\bibnamefont
  {Taninah}}, \bibinfo {author} {\bibfnamefont {S.}~\bibnamefont {Agbemava}},
  \bibinfo {author} {\bibfnamefont {A.}~\bibnamefont {Afanasjev}}, \ and\
  \bibinfo {author} {\bibfnamefont {P.}~\bibnamefont {Ring}},\ }\href {\doibase
  https://doi.org/10.1016/j.physletb.2019.135065} {\bibfield  {journal}
  {\bibinfo  {journal} {Phys. Lett. B}\ }\textbf {\bibinfo {volume} {800}},\
  \bibinfo {pages} {135065} (\bibinfo {year} {2020})}\BibitemShut {NoStop}%
\bibitem [{\citenamefont {Lalazissis}\ \emph {et~al.}(2005)\citenamefont
  {Lalazissis}, \citenamefont {Nik\ifmmode \check{s}\else
  \v{s}\fi{}i\ifmmode~\acute{c}\else \'{c}\fi{}}, \citenamefont {Vretenar},\
  and\ \citenamefont {Ring}}]{Lalazissis2005_PRC71-024312}%
  \BibitemOpen
  \bibfield  {author} {\bibinfo {author} {\bibfnamefont {G.~A.}\ \bibnamefont
  {Lalazissis}}, \bibinfo {author} {\bibfnamefont {T.}~\bibnamefont
  {Nik\ifmmode \check{s}\else \v{s}\fi{}i\ifmmode~\acute{c}\else \'{c}\fi{}}},
  \bibinfo {author} {\bibfnamefont {D.}~\bibnamefont {Vretenar}}, \ and\
  \bibinfo {author} {\bibfnamefont {P.}~\bibnamefont {Ring}},\ }\href {\doibase
  10.1103/PhysRevC.71.024312} {\bibfield  {journal} {\bibinfo  {journal} {Phys.
  Rev. C}\ }\textbf {\bibinfo {volume} {71}},\ \bibinfo {pages} {024312}
  (\bibinfo {year} {2005})}\BibitemShut {NoStop}%
\bibitem [{\citenamefont {Typel}\ and\ \citenamefont
  {Wolter}(1999)}]{Typel1999_NPA656-331}%
  \BibitemOpen
  \bibfield  {author} {\bibinfo {author} {\bibfnamefont {S.}~\bibnamefont
  {Typel}}\ and\ \bibinfo {author} {\bibfnamefont {H.}~\bibnamefont {Wolter}},\
  }\href {\doibase https://doi.org/10.1016/S0375-9474(99)00310-3} {\bibfield
  {journal} {\bibinfo  {journal} {Nucl. Phys. A}\ }\textbf {\bibinfo {volume}
  {656}},\ \bibinfo {pages} {331} (\bibinfo {year} {1999})}\BibitemShut
  {NoStop}%
\bibitem [{\citenamefont {Xia}\ \emph {et~al.}(2022{\natexlab{a}})\citenamefont
  {Xia}, \citenamefont {Maruyama}, \citenamefont {Li}, \citenamefont {Sun},
  \citenamefont {Long},\ and\ \citenamefont {Zhang}}]{Xia2022_CTP74-095303}%
  \BibitemOpen
  \bibfield  {author} {\bibinfo {author} {\bibfnamefont {C.-J.}\ \bibnamefont
  {Xia}}, \bibinfo {author} {\bibfnamefont {T.}~\bibnamefont {Maruyama}},
  \bibinfo {author} {\bibfnamefont {A.}~\bibnamefont {Li}}, \bibinfo {author}
  {\bibfnamefont {B.~Y.}\ \bibnamefont {Sun}}, \bibinfo {author} {\bibfnamefont
  {W.-H.}\ \bibnamefont {Long}}, \ and\ \bibinfo {author} {\bibfnamefont
  {Y.-X.}\ \bibnamefont {Zhang}},\ }\href {\doibase 10.1088/1572-9494/ac71fd}
  {\bibfield  {journal} {\bibinfo  {journal} {Commun. Theor. Phys.}\ }\textbf
  {\bibinfo {volume} {74}},\ \bibinfo {pages} {095303} (\bibinfo {year}
  {2022}{\natexlab{a}})}\BibitemShut {NoStop}%
\bibitem [{\citenamefont {Reinhard}(1989)}]{Reinhard1989_RPP52-439}%
  \BibitemOpen
  \bibfield  {author} {\bibinfo {author} {\bibfnamefont {P.~G.}\ \bibnamefont
  {Reinhard}},\ }\href {\doibase 10.1088/0034-4885/52/4/002} {\bibfield
  {journal} {\bibinfo  {journal} {Rep. Prog. Phys.}\ }\textbf {\bibinfo
  {volume} {52}},\ \bibinfo {pages} {439} (\bibinfo {year} {1989})}\BibitemShut
  {NoStop}%
\bibitem [{\citenamefont {Ring}(1996)}]{Ring1996_PPNP37-193}%
  \BibitemOpen
  \bibfield  {author} {\bibinfo {author} {\bibfnamefont {P.}~\bibnamefont
  {Ring}},\ }\href {\doibase https://doi.org/10.1016/0146-6410(96)00054-3}
  {\bibfield  {journal} {\bibinfo  {journal} {Prog. Part. Nucl. Phys.}\
  }\textbf {\bibinfo {volume} {37}},\ \bibinfo {pages} {193} (\bibinfo {year}
  {1996})}\BibitemShut {NoStop}%
\bibitem [{\citenamefont {Meng}\ \emph {et~al.}(2006)\citenamefont {Meng},
  \citenamefont {Toki}, \citenamefont {Zhou}, \citenamefont {Zhang},
  \citenamefont {Long},\ and\ \citenamefont {Geng}}]{Meng2006_PPNP57-470}%
  \BibitemOpen
  \bibfield  {author} {\bibinfo {author} {\bibfnamefont {J.}~\bibnamefont
  {Meng}}, \bibinfo {author} {\bibfnamefont {H.}~\bibnamefont {Toki}}, \bibinfo
  {author} {\bibfnamefont {S.}~\bibnamefont {Zhou}}, \bibinfo {author}
  {\bibfnamefont {S.}~\bibnamefont {Zhang}}, \bibinfo {author} {\bibfnamefont
  {W.}~\bibnamefont {Long}}, \ and\ \bibinfo {author} {\bibfnamefont
  {L.}~\bibnamefont {Geng}},\ }\href {\doibase
  https://doi.org/10.1016/j.ppnp.2005.06.001} {\bibfield  {journal} {\bibinfo
  {journal} {Prog. Part. Nucl. Phys.}\ }\textbf {\bibinfo {volume} {57}},\
  \bibinfo {pages} {470} (\bibinfo {year} {2006})}\BibitemShut {NoStop}%
\bibitem [{\citenamefont {Paar}\ \emph {et~al.}(2007)\citenamefont {Paar},
  \citenamefont {Vretenar}, \citenamefont {Khan},\ and\ \citenamefont
  {Col\`{o}}}]{Paar2007_RPP70-R02}%
  \BibitemOpen
  \bibfield  {author} {\bibinfo {author} {\bibfnamefont {N.}~\bibnamefont
  {Paar}}, \bibinfo {author} {\bibfnamefont {D.}~\bibnamefont {Vretenar}},
  \bibinfo {author} {\bibfnamefont {E.}~\bibnamefont {Khan}}, \ and\ \bibinfo
  {author} {\bibfnamefont {G.}~\bibnamefont {Col\`{o}}},\ }\href {\doibase
  10.1088/0034-4885/70/5/R02} {\bibfield  {journal} {\bibinfo  {journal} {Rep.
  Prog. Phys.}\ }\textbf {\bibinfo {volume} {70}},\ \bibinfo {pages} {R02}
  (\bibinfo {year} {2007})}\BibitemShut {NoStop}%
\bibitem [{\citenamefont {Meng}\ and\ \citenamefont
  {Zhou}(2015)}]{Meng2015_NPP42-093101}%
  \BibitemOpen
  \bibfield  {author} {\bibinfo {author} {\bibfnamefont {J.}~\bibnamefont
  {Meng}}\ and\ \bibinfo {author} {\bibfnamefont {S.~G.}\ \bibnamefont
  {Zhou}},\ }\href {\doibase 10.1088/0954-3899/42/9/093101} {\bibfield
  {journal} {\bibinfo  {journal} {J. Phys. G: Nucl. Part. Phys.}\ }\textbf
  {\bibinfo {volume} {42}},\ \bibinfo {pages} {093101} (\bibinfo {year}
  {2015})}\BibitemShut {NoStop}%
\bibitem [{\citenamefont {Chen}\ \emph {et~al.}(2021)\citenamefont {Chen},
  \citenamefont {Sun}, \citenamefont {Li},\ and\ \citenamefont
  {Sun}}]{Chen2021_SCPMA64-282011}%
  \BibitemOpen
  \bibfield  {author} {\bibinfo {author} {\bibfnamefont {C.}~\bibnamefont
  {Chen}}, \bibinfo {author} {\bibfnamefont {Q.-K.}\ \bibnamefont {Sun}},
  \bibinfo {author} {\bibfnamefont {Y.-X.}\ \bibnamefont {Li}}, \ and\ \bibinfo
  {author} {\bibfnamefont {T.-T.}\ \bibnamefont {Sun}},\ }\href {\doibase
  https://doi.org/10.1007/s11433-021-1721-1} {\bibfield  {journal} {\bibinfo
  {journal} {Sci. China Phys. Mech. Astron.}\ }\textbf {\bibinfo {volume}
  {64}},\ \bibinfo {pages} {282011} (\bibinfo {year} {2021})}\BibitemShut
  {NoStop}%
\bibitem [{\citenamefont {Vretenar}\ \emph {et~al.}(1998)\citenamefont
  {Vretenar}, \citenamefont {P\"oschl}, \citenamefont {Lalazissis},\ and\
  \citenamefont {Ring}}]{Vretenar1998_PRC57-R1060}%
  \BibitemOpen
  \bibfield  {author} {\bibinfo {author} {\bibfnamefont {D.}~\bibnamefont
  {Vretenar}}, \bibinfo {author} {\bibfnamefont {W.}~\bibnamefont {P\"oschl}},
  \bibinfo {author} {\bibfnamefont {G.~A.}\ \bibnamefont {Lalazissis}}, \ and\
  \bibinfo {author} {\bibfnamefont {P.}~\bibnamefont {Ring}},\ }\href {\doibase
  10.1103/PhysRevC.57.R1060} {\bibfield  {journal} {\bibinfo  {journal} {Phys.
  Rev. C}\ }\textbf {\bibinfo {volume} {57}},\ \bibinfo {pages} {R1060}
  (\bibinfo {year} {1998})}\BibitemShut {NoStop}%
\bibitem [{\citenamefont {Lu}\ \emph {et~al.}(2011)\citenamefont {Lu},
  \citenamefont {Zhao},\ and\ \citenamefont {Zhou}}]{Lu2011_PRC84-014328}%
  \BibitemOpen
  \bibfield  {author} {\bibinfo {author} {\bibfnamefont {B.-N.}\ \bibnamefont
  {Lu}}, \bibinfo {author} {\bibfnamefont {E.-G.}\ \bibnamefont {Zhao}}, \ and\
  \bibinfo {author} {\bibfnamefont {S.-G.}\ \bibnamefont {Zhou}},\ }\href
  {\doibase 10.1103/PhysRevC.84.014328} {\bibfield  {journal} {\bibinfo
  {journal} {Phys. Rev. C}\ }\textbf {\bibinfo {volume} {84}},\ \bibinfo
  {pages} {014328} (\bibinfo {year} {2011})}\BibitemShut {NoStop}%
\bibitem [{\citenamefont {Ban}\ \emph {et~al.}(2004)\citenamefont {Ban},
  \citenamefont {Li}, \citenamefont {Zhang}, \citenamefont {Jia}, \citenamefont
  {Sang},\ and\ \citenamefont {Meng}}]{Ban2004_PRC69-045805}%
  \BibitemOpen
  \bibfield  {author} {\bibinfo {author} {\bibfnamefont {S.~F.}\ \bibnamefont
  {Ban}}, \bibinfo {author} {\bibfnamefont {J.}~\bibnamefont {Li}}, \bibinfo
  {author} {\bibfnamefont {S.~Q.}\ \bibnamefont {Zhang}}, \bibinfo {author}
  {\bibfnamefont {H.~Y.}\ \bibnamefont {Jia}}, \bibinfo {author} {\bibfnamefont
  {J.~P.}\ \bibnamefont {Sang}}, \ and\ \bibinfo {author} {\bibfnamefont
  {J.}~\bibnamefont {Meng}},\ }\href {\doibase 10.1103/PhysRevC.69.045805}
  {\bibfield  {journal} {\bibinfo  {journal} {Phys. Rev. C}\ }\textbf {\bibinfo
  {volume} {69}},\ \bibinfo {pages} {045805} (\bibinfo {year}
  {2004})}\BibitemShut {NoStop}%
\bibitem [{\citenamefont {Weber}\ \emph {et~al.}(2007)\citenamefont {Weber},
  \citenamefont {Negreiros}, \citenamefont {Rosenfield},\ and\ \citenamefont
  {Stejner}}]{Weber2007_PPNP59-94}%
  \BibitemOpen
  \bibfield  {author} {\bibinfo {author} {\bibfnamefont {F.}~\bibnamefont
  {Weber}}, \bibinfo {author} {\bibfnamefont {R.}~\bibnamefont {Negreiros}},
  \bibinfo {author} {\bibfnamefont {P.}~\bibnamefont {Rosenfield}}, \ and\
  \bibinfo {author} {\bibfnamefont {M.}~\bibnamefont {Stejner}},\ }\href
  {\doibase https://doi.org/10.1016/j.ppnp.2006.12.008} {\bibfield  {journal}
  {\bibinfo  {journal} {Prog. Part. Nucl. Phys.}\ }\textbf {\bibinfo {volume}
  {59}},\ \bibinfo {pages} {94} (\bibinfo {year} {2007})},\ \bibinfo {note}
  {international Workshop on Nuclear Physics 28th Course}\BibitemShut {NoStop}%
\bibitem [{\citenamefont {Long}\ \emph {et~al.}(2012)\citenamefont {Long},
  \citenamefont {Sun}, \citenamefont {Hagino},\ and\ \citenamefont
  {Sagawa}}]{Long2012_PRC85-025806}%
  \BibitemOpen
  \bibfield  {author} {\bibinfo {author} {\bibfnamefont {W.~H.}\ \bibnamefont
  {Long}}, \bibinfo {author} {\bibfnamefont {B.~Y.}\ \bibnamefont {Sun}},
  \bibinfo {author} {\bibfnamefont {K.}~\bibnamefont {Hagino}}, \ and\ \bibinfo
  {author} {\bibfnamefont {H.}~\bibnamefont {Sagawa}},\ }\href {\doibase
  10.1103/PhysRevC.85.025806} {\bibfield  {journal} {\bibinfo  {journal} {Phys.
  Rev. C}\ }\textbf {\bibinfo {volume} {85}},\ \bibinfo {pages} {025806}
  (\bibinfo {year} {2012})}\BibitemShut {NoStop}%
\bibitem [{\citenamefont {Sun}\ \emph {et~al.}(2012)\citenamefont {Sun},
  \citenamefont {Sun},\ and\ \citenamefont {Meng}}]{Sun2012_PRC86-014305}%
  \BibitemOpen
  \bibfield  {author} {\bibinfo {author} {\bibfnamefont {T.~T.}\ \bibnamefont
  {Sun}}, \bibinfo {author} {\bibfnamefont {B.~Y.}\ \bibnamefont {Sun}}, \ and\
  \bibinfo {author} {\bibfnamefont {J.}~\bibnamefont {Meng}},\ }\href {\doibase
  10.1103/PhysRevC.86.014305} {\bibfield  {journal} {\bibinfo  {journal} {Phys.
  Rev. C}\ }\textbf {\bibinfo {volume} {86}},\ \bibinfo {pages} {014305}
  (\bibinfo {year} {2012})}\BibitemShut {NoStop}%
\bibitem [{\citenamefont {Wang}\ \emph {et~al.}(2014)\citenamefont {Wang},
  \citenamefont {Zhang},\ and\ \citenamefont {Dong}}]{Wang2014_PRC90-055801}%
  \BibitemOpen
  \bibfield  {author} {\bibinfo {author} {\bibfnamefont {S.}~\bibnamefont
  {Wang}}, \bibinfo {author} {\bibfnamefont {H.~F.}\ \bibnamefont {Zhang}}, \
  and\ \bibinfo {author} {\bibfnamefont {J.~M.}\ \bibnamefont {Dong}},\ }\href
  {\doibase 10.1103/PhysRevC.90.055801} {\bibfield  {journal} {\bibinfo
  {journal} {Phys. Rev. C}\ }\textbf {\bibinfo {volume} {90}},\ \bibinfo
  {pages} {055801} (\bibinfo {year} {2014})}\BibitemShut {NoStop}%
\bibitem [{\citenamefont {Fedoseew}\ and\ \citenamefont
  {Lenske}(2015)}]{Fedoseew2015_PRC91-034307}%
  \BibitemOpen
  \bibfield  {author} {\bibinfo {author} {\bibfnamefont {A.}~\bibnamefont
  {Fedoseew}}\ and\ \bibinfo {author} {\bibfnamefont {H.}~\bibnamefont
  {Lenske}},\ }\href {\doibase 10.1103/PhysRevC.91.034307} {\bibfield
  {journal} {\bibinfo  {journal} {Phys. Rev. C}\ }\textbf {\bibinfo {volume}
  {91}},\ \bibinfo {pages} {034307} (\bibinfo {year} {2015})}\BibitemShut
  {NoStop}%
\bibitem [{\citenamefont {Gao}\ \emph {et~al.}(2017)\citenamefont {Gao},
  \citenamefont {Wang}, \citenamefont {Shan}, \citenamefont {Li},\ and\
  \citenamefont {Wang}}]{Gao2017_ApJ849-19}%
  \BibitemOpen
  \bibfield  {author} {\bibinfo {author} {\bibfnamefont {Z.-F.}\ \bibnamefont
  {Gao}}, \bibinfo {author} {\bibfnamefont {N.}~\bibnamefont {Wang}}, \bibinfo
  {author} {\bibfnamefont {H.}~\bibnamefont {Shan}}, \bibinfo {author}
  {\bibfnamefont {X.-D.}\ \bibnamefont {Li}}, \ and\ \bibinfo {author}
  {\bibfnamefont {W.}~\bibnamefont {Wang}},\ }\href {\doibase
  10.3847/1538-4357/aa8f49} {\bibfield  {journal} {\bibinfo  {journal}
  {Astrophys. J.}\ }\textbf {\bibinfo {volume} {849}},\ \bibinfo {pages} {19}
  (\bibinfo {year} {2017})}\BibitemShut {NoStop}%
\bibitem [{\citenamefont {{Baym}}\ \emph {et~al.}(1971)\citenamefont {{Baym}},
  \citenamefont {{Pethick}},\ and\ \citenamefont
  {{Sutherland}}}]{Baym1971_ApJ170-299}%
  \BibitemOpen
  \bibfield  {author} {\bibinfo {author} {\bibfnamefont {G.}~\bibnamefont
  {{Baym}}}, \bibinfo {author} {\bibfnamefont {C.}~\bibnamefont {{Pethick}}}, \
  and\ \bibinfo {author} {\bibfnamefont {P.}~\bibnamefont {{Sutherland}}},\
  }\href {\doibase 10.1086/151216} {\bibfield  {journal} {\bibinfo  {journal}
  {\apj}\ }\textbf {\bibinfo {volume} {170}},\ \bibinfo {pages} {299} (\bibinfo
  {year} {1971})}\BibitemShut {NoStop}%
\bibitem [{\citenamefont {Negele}\ and\ \citenamefont
  {Vautherin}(1973)}]{Negele1973_NPA207-298}%
  \BibitemOpen
  \bibfield  {author} {\bibinfo {author} {\bibfnamefont {J.}~\bibnamefont
  {Negele}}\ and\ \bibinfo {author} {\bibfnamefont {D.}~\bibnamefont
  {Vautherin}},\ }\href {\doibase https://doi.org/10.1016/0375-9474(73)90349-7}
  {\bibfield  {journal} {\bibinfo  {journal} {Nucl. Phys. A}\ }\textbf
  {\bibinfo {volume} {207}},\ \bibinfo {pages} {298} (\bibinfo {year}
  {1973})}\BibitemShut {NoStop}%
\bibitem [{\citenamefont {Ravenhall}\ \emph {et~al.}(1983)\citenamefont
  {Ravenhall}, \citenamefont {Pethick},\ and\ \citenamefont
  {Wilson}}]{Structure1983_PRL50-2066}%
  \BibitemOpen
  \bibfield  {author} {\bibinfo {author} {\bibfnamefont {D.~G.}\ \bibnamefont
  {Ravenhall}}, \bibinfo {author} {\bibfnamefont {C.~J.}\ \bibnamefont
  {Pethick}}, \ and\ \bibinfo {author} {\bibfnamefont {J.~R.}\ \bibnamefont
  {Wilson}},\ }\href {\doibase 10.1103/PhysRevLett.50.2066} {\bibfield
  {journal} {\bibinfo  {journal} {Phys. Rev. Lett.}\ }\textbf {\bibinfo
  {volume} {50}},\ \bibinfo {pages} {2066} (\bibinfo {year}
  {1983})}\BibitemShut {NoStop}%
\bibitem [{\citenamefont {Hashimoto}\ \emph {et~al.}(1984)\citenamefont
  {Hashimoto}, \citenamefont {Seki},\ and\ \citenamefont
  {Yamada}}]{Hashimoto1984_PTP71-320}%
  \BibitemOpen
  \bibfield  {author} {\bibinfo {author} {\bibfnamefont {M.-a.}\ \bibnamefont
  {Hashimoto}}, \bibinfo {author} {\bibfnamefont {H.}~\bibnamefont {Seki}}, \
  and\ \bibinfo {author} {\bibfnamefont {M.}~\bibnamefont {Yamada}},\ }\href
  {\doibase 10.1143/PTP.71.320} {\bibfield  {journal} {\bibinfo  {journal}
  {Prog. Theor. Phys.}\ }\textbf {\bibinfo {volume} {71}},\ \bibinfo {pages}
  {320} (\bibinfo {year} {1984})}\BibitemShut {NoStop}%
\bibitem [{\citenamefont {Williams}\ and\ \citenamefont
  {Koonin}(1985)}]{Williams1985_NPA435-844}%
  \BibitemOpen
  \bibfield  {author} {\bibinfo {author} {\bibfnamefont {R.}~\bibnamefont
  {Williams}}\ and\ \bibinfo {author} {\bibfnamefont {S.}~\bibnamefont
  {Koonin}},\ }\href {\doibase https://doi.org/10.1016/0375-9474(85)90191-5}
  {\bibfield  {journal} {\bibinfo  {journal} {Nucl. Phys. A}\ }\textbf
  {\bibinfo {volume} {435}},\ \bibinfo {pages} {844} (\bibinfo {year}
  {1985})}\BibitemShut {NoStop}%
\bibitem [{\citenamefont {Lenske}\ and\ \citenamefont
  {Fuchs}(1995)}]{Lenske1995_PLB345-4}%
  \BibitemOpen
  \bibfield  {author} {\bibinfo {author} {\bibfnamefont {H.}~\bibnamefont
  {Lenske}}\ and\ \bibinfo {author} {\bibfnamefont {C.}~\bibnamefont {Fuchs}},\
  }\href {\doibase https://doi.org/10.1016/0370-2693(94)01664-X} {\bibfield
  {journal} {\bibinfo  {journal} {Phys. Lett. B}\ }\textbf {\bibinfo {volume}
  {345}},\ \bibinfo {pages} {355} (\bibinfo {year} {1995})}\BibitemShut
  {NoStop}%
\bibitem [{\citenamefont {Cowling}(1941)}]{Cowling1941_MNRAS101-367}%
  \BibitemOpen
  \bibfield  {author} {\bibinfo {author} {\bibfnamefont {T.~G.}\ \bibnamefont
  {Cowling}},\ }\href {\doibase 10.1093/mnras/101.8.367} {\bibfield  {journal}
  {\bibinfo  {journal} {Mon. Not. R. Astron. Soc.}\ }\textbf {\bibinfo {volume}
  {101}},\ \bibinfo {pages} {367} (\bibinfo {year} {1941})}\BibitemShut
  {NoStop}%
\bibitem [{\citenamefont {Yoshida}\ and\ \citenamefont
  {Lee}(2002)}]{Yoshida2002_AAP395-201}%
  \BibitemOpen
  \bibfield  {author} {\bibinfo {author} {\bibfnamefont {S.}~\bibnamefont
  {Yoshida}}\ and\ \bibinfo {author} {\bibfnamefont {U.}~\bibnamefont {Lee}},\
  }\href {\doibase 10.1051/0004-6361:20021270} {\bibfield  {journal} {\bibinfo
  {journal} {Astron. Astrophys.}\ }\textbf {\bibinfo {volume} {395}},\ \bibinfo
  {pages} {201} (\bibinfo {year} {2002})}\BibitemShut {NoStop}%
\bibitem [{\citenamefont {Ranea-Sandoval}\ \emph {et~al.}(2018)\citenamefont
  {Ranea-Sandoval}, \citenamefont {Guilera}, \citenamefont {Mariani},\ and\
  \citenamefont {Orsaria}}]{Ranea-Sandoval2018_JCAR12-031}%
  \BibitemOpen
  \bibfield  {author} {\bibinfo {author} {\bibfnamefont {I.~F.}\ \bibnamefont
  {Ranea-Sandoval}}, \bibinfo {author} {\bibfnamefont {O.~M.}\ \bibnamefont
  {Guilera}}, \bibinfo {author} {\bibfnamefont {M.}~\bibnamefont {Mariani}}, \
  and\ \bibinfo {author} {\bibfnamefont {M.~G.}\ \bibnamefont {Orsaria}},\
  }\href {\doibase 10.1088/1475-7516/2018/12/031} {\bibfield  {journal}
  {\bibinfo  {journal} {JCAP}\ }\textbf {\bibinfo {volume} {12}},\ \bibinfo
  {pages} {031} (\bibinfo {year} {2018})}\BibitemShut {NoStop}%
\bibitem [{\citenamefont {Yoshida}\ and\ \citenamefont
  {Kojima}(1997)}]{Yoshida1997_MNRAS117-122-289}%
  \BibitemOpen
  \bibfield  {author} {\bibinfo {author} {\bibfnamefont {S.}~\bibnamefont
  {Yoshida}}\ and\ \bibinfo {author} {\bibfnamefont {Y.}~\bibnamefont
  {Kojima}},\ }\href {\doibase 10.1093/mnras/289.1.117} {\bibfield  {journal}
  {\bibinfo  {journal} {Mon. Not. R. Astron. Soc.}\
  }\textbf {\bibinfo {volume} {289}},\ \bibinfo {pages} {117} (\bibinfo {year}
  {1997})}\BibitemShut {NoStop}%
\bibitem [{\citenamefont {Sotani}\ and\ \citenamefont
  {Takiwaki}(2020)}]{Sotani2020_PRD102-063025}%
  \BibitemOpen
  \bibfield  {author} {\bibinfo {author} {\bibfnamefont {H.}~\bibnamefont
  {Sotani}}\ and\ \bibinfo {author} {\bibfnamefont {T.}~\bibnamefont
  {Takiwaki}},\ }\href {\doibase 10.1103/PhysRevD.102.063025} {\bibfield
  {journal} {\bibinfo  {journal} {Phys. Rev. D}\ }\textbf {\bibinfo {volume}
  {102}},\ \bibinfo {pages} {063025} (\bibinfo {year} {2020})}\BibitemShut
  {NoStop}%
\bibitem [{\citenamefont {Cox}(1980)}]{John1980_PUP}%
  \BibitemOpen
  \bibfield  {author} {\bibinfo {author} {\bibfnamefont {J.~P.}\ \bibnamefont
  {Cox}},\ }\href {http://www.jstor.org/stable/j.ctt1m32337} {{\bibinfo
  {title} {{Theory of Stellar Pulsation. (PSA-2)}}}}\ (\bibinfo  {publisher}
  {Princeton University Press},\ \bibinfo {year} {1980})\BibitemShut {NoStop}%
\bibitem [{\citenamefont {Counsell}\ \emph {et~al.}(2024)\citenamefont
  {Counsell}, \citenamefont {Gittins},\ and\ \citenamefont
  {Andersson}}]{Counsell2024_MNRAS531-1721-1729}%
  \BibitemOpen
  \bibfield  {author} {\bibinfo {author} {\bibfnamefont {A.~R.}\ \bibnamefont
  {Counsell}}, \bibinfo {author} {\bibfnamefont {F.}~\bibnamefont {Gittins}}, \
  and\ \bibinfo {author} {\bibfnamefont {N.}~\bibnamefont {Andersson}},\ }\href
  {\doibase 10.1093/mnras/stae1242} {\bibfield  {journal} {\bibinfo  {journal}
  {Mon. Not. R. Astron. Soc.}\ }\textbf {\bibinfo {volume} {531}},\ \bibinfo
  {pages} {1721} (\bibinfo {year} {2024})}\BibitemShut {NoStop}%
\bibitem [{\citenamefont {Pereira}\ \emph {et~al.}(2018)\citenamefont
  {Pereira}, \citenamefont {Flores},\ and\ \citenamefont
  {Lugones}}]{Pereira2018_ApJ860-12}%
  \BibitemOpen
  \bibfield  {author} {\bibinfo {author} {\bibfnamefont {J.~P.}\ \bibnamefont
  {Pereira}}, \bibinfo {author} {\bibfnamefont {C.~V.}\ \bibnamefont {Flores}},
  \ and\ \bibinfo {author} {\bibfnamefont {G.}~\bibnamefont {Lugones}},\ }\href
  {\doibase 10.3847/1538-4357/aabfbf} {\bibfield  {journal} {\bibinfo
  {journal} {Astrophys. J.}\ }\textbf {\bibinfo {volume} {860}},\ \bibinfo
  {pages} {12} (\bibinfo {year} {2018})}\BibitemShut {NoStop}%
\bibitem [{\citenamefont {Okamoto}\ \emph {et~al.}(2012)\citenamefont
  {Okamoto}, \citenamefont {Maruyama}, \citenamefont {Yabana},\ and\
  \citenamefont {Tatsumi}}]{Okamoto2012_PLB713-284}%
  \BibitemOpen
  \bibfield  {author} {\bibinfo {author} {\bibfnamefont {M.}~\bibnamefont
  {Okamoto}}, \bibinfo {author} {\bibfnamefont {T.}~\bibnamefont {Maruyama}},
  \bibinfo {author} {\bibfnamefont {K.}~\bibnamefont {Yabana}}, \ and\ \bibinfo
  {author} {\bibfnamefont {T.}~\bibnamefont {Tatsumi}},\ }\href {\doibase
  http://dx.doi.org/10.1016/j.physletb.2012.05.046} {\bibfield  {journal}
  {\bibinfo  {journal} {Phys. Lett. B}\ }\textbf {\bibinfo {volume} {713}},\
  \bibinfo {pages} {284} (\bibinfo {year} {2012})}\BibitemShut {NoStop}%
\bibitem [{\citenamefont {Okamoto}\ \emph {et~al.}(2013)\citenamefont
  {Okamoto}, \citenamefont {Maruyama}, \citenamefont {Yabana},\ and\
  \citenamefont {Tatsumi}}]{Okamoto2013_PRC88-025801}%
  \BibitemOpen
  \bibfield  {author} {\bibinfo {author} {\bibfnamefont {M.}~\bibnamefont
  {Okamoto}}, \bibinfo {author} {\bibfnamefont {T.}~\bibnamefont {Maruyama}},
  \bibinfo {author} {\bibfnamefont {K.}~\bibnamefont {Yabana}}, \ and\ \bibinfo
  {author} {\bibfnamefont {T.}~\bibnamefont {Tatsumi}},\ }\href {\doibase
  10.1103/PhysRevC.88.025801} {\bibfield  {journal} {\bibinfo  {journal} {Phys.
  Rev. C}\ }\textbf {\bibinfo {volume} {88}},\ \bibinfo {pages} {025801}
  (\bibinfo {year} {2013})}\BibitemShut {NoStop}%
\bibitem [{\citenamefont {Xia}\ \emph {et~al.}(2021)\citenamefont {Xia},
  \citenamefont {Maruyama}, \citenamefont {Yasutake}, \citenamefont {Tatsumi},\
  and\ \citenamefont {Zhang}}]{Xia2021_PRC103-055812}%
  \BibitemOpen
  \bibfield  {author} {\bibinfo {author} {\bibfnamefont {C.-J.}\ \bibnamefont
  {Xia}}, \bibinfo {author} {\bibfnamefont {T.}~\bibnamefont {Maruyama}},
  \bibinfo {author} {\bibfnamefont {N.}~\bibnamefont {Yasutake}}, \bibinfo
  {author} {\bibfnamefont {T.}~\bibnamefont {Tatsumi}}, \ and\ \bibinfo
  {author} {\bibfnamefont {Y.-X.}\ \bibnamefont {Zhang}},\ }\href {\doibase
  10.1103/PhysRevC.103.055812} {\bibfield  {journal} {\bibinfo  {journal}
  {Phys. Rev. C}\ }\textbf {\bibinfo {volume} {103}},\ \bibinfo {pages}
  {055812} (\bibinfo {year} {2021})}\BibitemShut {NoStop}%
\bibitem [{\citenamefont {Xia}\ \emph {et~al.}(2022{\natexlab{b}})\citenamefont
  {Xia}, \citenamefont {Sun}, \citenamefont {Maruyama}, \citenamefont {Long},\
  and\ \citenamefont {Li}}]{Xia2022_PRC105-045803}%
  \BibitemOpen
  \bibfield  {author} {\bibinfo {author} {\bibfnamefont {C.-J.}\ \bibnamefont
  {Xia}}, \bibinfo {author} {\bibfnamefont {B.~Y.}\ \bibnamefont {Sun}},
  \bibinfo {author} {\bibfnamefont {T.}~\bibnamefont {Maruyama}}, \bibinfo
  {author} {\bibfnamefont {W.-H.}\ \bibnamefont {Long}}, \ and\ \bibinfo
  {author} {\bibfnamefont {A.}~\bibnamefont {Li}},\ }\href {\doibase
  10.1103/PhysRevC.105.045803} {\bibfield  {journal} {\bibinfo  {journal}
  {Phys. Rev. C}\ }\textbf {\bibinfo {volume} {105}},\ \bibinfo {pages}
  {045803} (\bibinfo {year} {2022}{\natexlab{b}})}\BibitemShut {NoStop}%
\bibitem [{\citenamefont {Shlomo}\ \emph {et~al.}(2006)\citenamefont {Shlomo},
  \citenamefont {Kolomietz},\ and\ \citenamefont
  {Col{\`o}}}]{Shlomo2006_EPJA30-23}%
  \BibitemOpen
  \bibfield  {author} {\bibinfo {author} {\bibfnamefont {S.}~\bibnamefont
  {Shlomo}}, \bibinfo {author} {\bibfnamefont {V.~M.}\ \bibnamefont
  {Kolomietz}}, \ and\ \bibinfo {author} {\bibfnamefont {G.}~\bibnamefont
  {Col{\`o}}},\ }\href {\doibase 10.1140/epja/i2006-10100-3} {\bibfield
  {journal} {\bibinfo  {journal} {Eur. Phys. J. A}\ }\textbf {\bibinfo {volume}
  {30}},\ \bibinfo {pages} {23} (\bibinfo {year} {2006})}\BibitemShut {NoStop}%
\bibitem [{\citenamefont {Li}\ and\ \citenamefont
  {Han}(2013)}]{Li2013_PLB727-276}%
  \BibitemOpen
  \bibfield  {author} {\bibinfo {author} {\bibfnamefont {B.-A.}\ \bibnamefont
  {Li}}\ and\ \bibinfo {author} {\bibfnamefont {X.}~\bibnamefont {Han}},\
  }\href {\doibase http://dx.doi.org/10.1016/j.physletb.2013.10.006} {\bibfield
   {journal} {\bibinfo  {journal} {Phys. Lett. B}\ }\textbf {\bibinfo {volume}
  {727}},\ \bibinfo {pages} {276 } (\bibinfo {year} {2013})}\BibitemShut
  {NoStop}%
\bibitem [{\citenamefont {Oertel}\ \emph {et~al.}(2017)\citenamefont {Oertel},
  \citenamefont {Hempel}, \citenamefont {Kl\"ahn},\ and\ \citenamefont
  {Typel}}]{Oertel2017_RMP89-015007}%
  \BibitemOpen
  \bibfield  {author} {\bibinfo {author} {\bibfnamefont {M.}~\bibnamefont
  {Oertel}}, \bibinfo {author} {\bibfnamefont {M.}~\bibnamefont {Hempel}},
  \bibinfo {author} {\bibfnamefont {T.}~\bibnamefont {Kl\"ahn}}, \ and\
  \bibinfo {author} {\bibfnamefont {S.}~\bibnamefont {Typel}},\ }\href
  {\doibase 10.1103/RevModPhys.89.015007} {\bibfield  {journal} {\bibinfo
  {journal} {Rev. Mod. Phys.}\ }\textbf {\bibinfo {volume} {89}},\ \bibinfo
  {pages} {015007} (\bibinfo {year} {2017})}\BibitemShut {NoStop}%
\bibitem [{\citenamefont {Farine}\ \emph {et~al.}(1997)\citenamefont {Farine},
  \citenamefont {Pearson},\ and\ \citenamefont
  {Tondeur}}]{Farine1997_NPA615-135}%
  \BibitemOpen
  \bibfield  {author} {\bibinfo {author} {\bibfnamefont {M.}~\bibnamefont
  {Farine}}, \bibinfo {author} {\bibfnamefont {J.}~\bibnamefont {Pearson}}, \
  and\ \bibinfo {author} {\bibfnamefont {F.}~\bibnamefont {Tondeur}},\ }\href
  {\doibase https://doi.org/10.1016/S0375-9474(96)00453-8} {\bibfield
  {journal} {\bibinfo  {journal} {Nucl. Phys. A}\ }\textbf {\bibinfo {volume}
  {615}},\ \bibinfo {pages} {135} (\bibinfo {year} {1997})}\BibitemShut
  {NoStop}%
\bibitem [{\citenamefont {Zhang}\ \emph {et~al.}(2020)\citenamefont {Zhang},
  \citenamefont {Liu}, \citenamefont {Xia}, \citenamefont {Li},\ and\
  \citenamefont {Biswal}}]{Zhang2020_PRC101-034303}%
  \BibitemOpen
  \bibfield  {author} {\bibinfo {author} {\bibfnamefont {Y.}~\bibnamefont
  {Zhang}}, \bibinfo {author} {\bibfnamefont {M.}~\bibnamefont {Liu}}, \bibinfo
  {author} {\bibfnamefont {C.-J.}\ \bibnamefont {Xia}}, \bibinfo {author}
  {\bibfnamefont {Z.}~\bibnamefont {Li}}, \ and\ \bibinfo {author}
  {\bibfnamefont {S.~K.}\ \bibnamefont {Biswal}},\ }\href {\doibase
  10.1103/PhysRevC.101.034303} {\bibfield  {journal} {\bibinfo  {journal}
  {Phys. Rev. C}\ }\textbf {\bibinfo {volume} {101}},\ \bibinfo {pages}
  {034303} (\bibinfo {year} {2020})}\BibitemShut {NoStop}%
\bibitem [{\citenamefont {Essick}\ \emph {et~al.}(2021)\citenamefont {Essick},
  \citenamefont {Tews}, \citenamefont {Landry},\ and\ \citenamefont
  {Schwenk}}]{Essick2021_PRL127-192701}%
  \BibitemOpen
  \bibfield  {author} {\bibinfo {author} {\bibfnamefont {R.}~\bibnamefont
  {Essick}}, \bibinfo {author} {\bibfnamefont {I.}~\bibnamefont {Tews}},
  \bibinfo {author} {\bibfnamefont {P.}~\bibnamefont {Landry}}, \ and\ \bibinfo
  {author} {\bibfnamefont {A.}~\bibnamefont {Schwenk}},\ }\href {\doibase
  10.1103/PhysRevLett.127.192701} {\bibfield  {journal} {\bibinfo  {journal}
  {Phys. Rev. Lett.}\ }\textbf {\bibinfo {volume} {127}},\ \bibinfo {pages}
  {192701} (\bibinfo {year} {2021})}\BibitemShut {NoStop}%
\bibitem [{\citenamefont {Xie}\ and\ \citenamefont
  {Li}(2021)}]{Xie2021_JPG48-025110}%
  \BibitemOpen
  \bibfield  {author} {\bibinfo {author} {\bibfnamefont {W.-J.}\ \bibnamefont
  {Xie}}\ and\ \bibinfo {author} {\bibfnamefont {B.-A.}\ \bibnamefont {Li}},\
  }\href {\doibase 10.1088/1361-6471/abd25a} {\bibfield  {journal} {\bibinfo
  {journal} {J. Phys. G: Nucl. Part. Phys.}\ }\textbf {\bibinfo {volume}
  {48}},\ \bibinfo {pages} {025110} (\bibinfo {year} {2021})}\BibitemShut
  {NoStop}%
\bibitem [{\citenamefont {{PREX
  Collaboration}}(2021)}]{PREX2021_PRL126-172502}%
  \BibitemOpen
  \bibfield  {author} {\bibinfo {author} {\bibnamefont {{PREX
  Collaboration}}},\ }\href {\doibase 10.1103/PhysRevLett.126.172502}
  {\bibfield  {journal} {\bibinfo  {journal} {Phys. Rev. Lett.}\ }\textbf
  {\bibinfo {volume} {126}},\ \bibinfo {pages} {172502} (\bibinfo {year}
  {2021})}\BibitemShut {NoStop}%
\bibitem [{\citenamefont {{CREX
  Collaboration}}(2022)}]{CREX2022_PRL129-042501}%
  \BibitemOpen
  \bibfield  {author} {\bibinfo {author} {\bibnamefont {{CREX
  Collaboration}}},\ }\href {\doibase 10.1103/PhysRevLett.129.042501}
  {\bibfield  {journal} {\bibinfo  {journal} {Phys. Rev. Lett.}\ }\textbf
  {\bibinfo {volume} {129}},\ \bibinfo {pages} {042501} (\bibinfo {year}
  {2022})}\BibitemShut {NoStop}%
\bibitem [{\citenamefont {Salmi}\ \emph {et~al.}(2024)\citenamefont {Salmi},
  \citenamefont {Deneva}, \citenamefont {Ray}, \citenamefont {Watts},
  \citenamefont {Choudhury}, \citenamefont {Kini}, \citenamefont {Vinciguerra},
  \citenamefont {Cromartie}, \citenamefont {Wolff}, \citenamefont
  {Arzoumanian}, \citenamefont {Bogdanov}, \citenamefont {Gendreau},
  \citenamefont {Guillot}, \citenamefont {Ho}, \citenamefont {Morsink},
  \citenamefont {Cognard}, \citenamefont {Guillemot}, \citenamefont
  {Theureau},\ and\ \citenamefont {Kerr}}]{Salmi2024_ApJ976-58}%
  \BibitemOpen
  \bibfield  {author} {\bibinfo {author} {\bibfnamefont {T.}~\bibnamefont
  {Salmi}}, \bibinfo {author} {\bibfnamefont {J.~S.}\ \bibnamefont {Deneva}},
  \bibinfo {author} {\bibfnamefont {P.~S.}\ \bibnamefont {Ray}}, \bibinfo
  {author} {\bibfnamefont {A.~L.}\ \bibnamefont {Watts}}, \bibinfo {author}
  {\bibfnamefont {D.}~\bibnamefont {Choudhury}}, \bibinfo {author}
  {\bibfnamefont {Y.}~\bibnamefont {Kini}}, \bibinfo {author} {\bibfnamefont
  {S.}~\bibnamefont {Vinciguerra}}, \bibinfo {author} {\bibfnamefont {H.~T.}\
  \bibnamefont {Cromartie}}, \bibinfo {author} {\bibfnamefont {M.~T.}\
  \bibnamefont {Wolff}}, \bibinfo {author} {\bibfnamefont {Z.}~\bibnamefont
  {Arzoumanian}}, \bibinfo {author} {\bibfnamefont {S.}~\bibnamefont
  {Bogdanov}}, \bibinfo {author} {\bibfnamefont {K.}~\bibnamefont {Gendreau}},
  \bibinfo {author} {\bibfnamefont {S.}~\bibnamefont {Guillot}}, \bibinfo
  {author} {\bibfnamefont {W.~C.~G.}\ \bibnamefont {Ho}}, \bibinfo {author}
  {\bibfnamefont {S.~M.}\ \bibnamefont {Morsink}}, \bibinfo {author}
  {\bibfnamefont {I.}~\bibnamefont {Cognard}}, \bibinfo {author} {\bibfnamefont
  {L.}~\bibnamefont {Guillemot}}, \bibinfo {author} {\bibfnamefont
  {G.}~\bibnamefont {Theureau}}, \ and\ \bibinfo {author} {\bibfnamefont
  {M.}~\bibnamefont {Kerr}},\ }\href {\doibase 10.3847/1538-4357/ad81d2}
  {\bibfield  {journal} {\bibinfo  {journal} {Astrophys. J.}\ }\textbf
  {\bibinfo {volume} {976}},\ \bibinfo {pages} {58} (\bibinfo {year}
  {2024})}\BibitemShut {NoStop}%
\bibitem [{\citenamefont {{Vinciguerra}}\ \emph {et~al.}(2024)\citenamefont
  {{Vinciguerra}}, \citenamefont {{Salmi}}, \citenamefont {{Watts}},
  \citenamefont {{Choudhury}}, \citenamefont {{Riley}}, \citenamefont {{Ray}},
  \citenamefont {{Bogdanov}}, \citenamefont {{Kini}}, \citenamefont
  {{Guillot}}, \citenamefont {{Chakrabarty}}, \citenamefont {{Ho}},
  \citenamefont {{Huppenkothen}}, \citenamefont {{Morsink}}, \citenamefont
  {{Wadiasingh}},\ and\ \citenamefont {{Wolff}}}]{Vinciguerra2024_ApJ961-62}%
  \BibitemOpen
  \bibfield  {author} {\bibinfo {author} {\bibfnamefont {S.}~\bibnamefont
  {{Vinciguerra}}}, \bibinfo {author} {\bibfnamefont {T.}~\bibnamefont
  {{Salmi}}}, \bibinfo {author} {\bibfnamefont {A.~L.}\ \bibnamefont
  {{Watts}}}, \bibinfo {author} {\bibfnamefont {D.}~\bibnamefont
  {{Choudhury}}}, \bibinfo {author} {\bibfnamefont {T.~E.}\ \bibnamefont
  {{Riley}}}, \bibinfo {author} {\bibfnamefont {P.~S.}\ \bibnamefont {{Ray}}},
  \bibinfo {author} {\bibfnamefont {S.}~\bibnamefont {{Bogdanov}}}, \bibinfo
  {author} {\bibfnamefont {Y.}~\bibnamefont {{Kini}}}, \bibinfo {author}
  {\bibfnamefont {S.}~\bibnamefont {{Guillot}}}, \bibinfo {author}
  {\bibfnamefont {D.}~\bibnamefont {{Chakrabarty}}}, \bibinfo {author}
  {\bibfnamefont {W.~C.~G.}\ \bibnamefont {{Ho}}}, \bibinfo {author}
  {\bibfnamefont {D.}~\bibnamefont {{Huppenkothen}}}, \bibinfo {author}
  {\bibfnamefont {S.~M.}\ \bibnamefont {{Morsink}}}, \bibinfo {author}
  {\bibfnamefont {Z.}~\bibnamefont {{Wadiasingh}}}, \ and\ \bibinfo {author}
  {\bibfnamefont {M.~T.}\ \bibnamefont {{Wolff}}},\ }\href {\doibase
  10.3847/1538-4357/acfb83} {\bibfield  {journal} {\bibinfo  {journal}
  {Astrophys. J.}\ }\textbf {\bibinfo {volume} {961}},\ \bibinfo {eid} {62}
  (\bibinfo {year} {2024})}\BibitemShut {NoStop}%
\bibitem [{\citenamefont {Gondek}\ and\ \citenamefont
  {Zdunik}(1999)}]{Gondek1999_AAP344-117}%
  \BibitemOpen
  \bibfield  {author} {\bibinfo {author} {\bibfnamefont {D.}~\bibnamefont
  {Gondek}}\ and\ \bibinfo {author} {\bibfnamefont {J.~L.}\ \bibnamefont
  {Zdunik}},\ }\href {https://ui.adsabs.harvard.edu/abs/1999A&A...344..117G}
  {\bibfield  {journal} {\bibinfo  {journal} {Astron. Astrophys.}\ }\textbf
  {\bibinfo {volume} {344}},\ \bibinfo {pages} {117} (\bibinfo {year}
  {1999})}\BibitemShut {NoStop}%
\bibitem [{\citenamefont {Pratten}\ \emph {et~al.}(2020)\citenamefont
  {Pratten}, \citenamefont {Schmidt},\ and\ \citenamefont
  {Hinderer}}]{Pratten2020_NC11-1}%
  \BibitemOpen
  \bibfield  {author} {\bibinfo {author} {\bibfnamefont {G.}~\bibnamefont
  {Pratten}}, \bibinfo {author} {\bibfnamefont {P.}~\bibnamefont {Schmidt}}, \
  and\ \bibinfo {author} {\bibfnamefont {T.}~\bibnamefont {Hinderer}},\ }\href
  {\doibase 10.1038/s41467-020-15984-5} {\bibfield  {journal} {\bibinfo
  {journal} {Nat. Commun.}\ }\textbf {\bibinfo {volume} {11}},\ \bibinfo
  {pages} {2553} (\bibinfo {year} {2020})}\BibitemShut {NoStop}%
\end{thebibliography}

%

\end{document}